# DEVELOPMENT OF GRAPHICAL USER INTERFACE FOR MICROWAVE FILTER DESIGN

by

DJENGOMEMGOTO GERARD

Dissertation submitted to the Department of Electrical & Electronic
Engineering in partial fulfilment of the requirements
for the Bachelor of Engineering (Honours)
(Electrical & Electronic Engineering)

December 2013

Universiti Teknologi PETRONAS
Bandar Seri Iskandar
31750 Tronoh
Perak Darul Ridzuan

# CERTIFICATION OF APPROVAL

# DEVELOPMENT OF GRAPHICAL USER INTERFACE FOR MICROWAVE FILTER DESIGN

by

DJENGOMEMGOTO GERARD

A project dissertation submitted to the
Department of Electrical & Electronic Engineering
Universiti Teknologi PETRONAS
in partial fulfilment of the requirement for the
Bachelor of Engineering (Honours)
(Electrical & Electronic Engineering)

Approved by:

\_\_\_\_\_\_\_\_\_\_\_\_\_\_\_\_\_\_\_\_
Dr. Wong Peng Wen
Project Supervisor
Universiti Teknologi PETRONAS
Tronoh, Perak

December 2013

# CERTIFICATION OF ORIGINALITY

This is to certify that I am responsible for the work submitted in this project, that the original work is my own except as specified in the references and acknowledgements, and that the original work contained herein have not been undertaken or done by unspecified sources or persons.

______________________________
DJENGOMEMGOTO GERARD

# ABSTRACT

Microwave filters play a sterling role in wireless communication systems and they are used to discriminate between wanted and unwanted signal frequencies. Designing microwave filter is a long process in going from the given specification through the synthesis and simulation to the final prototype. However, due to the unavailability of synthesized theoretical values of reactive elements in the market, the result may disagree with the expected output.

Moreover, the existing software tools cover only conventional filter topologies; modern microwave filter topologies such as Ultra Wideband (UWB) filter are not yet incorporated into these software tools. In addition, they are costly and not easy to use.

This research project aims at developing a low-cost, time-effective and a stand-alone graphical user interface (GUI) that will be used to design microwave filters. Throughout the projects, the main theory behind the technology of microwave filters, their generalized mathematical equations and the analysis of their different circuit topologies have been reviewed. This review helps to extract the necessary information needed for the design of microwave filters. Besides, the guiding principles and the underlying engineering factors for a successful and information-oriented GUI were also highlighted.

To carry out the project, the High-Level GUI Development Environment (GUIDE) together with a structured programming approach have been used to design the GUI, and to program its related functionalities. The frequency responses are generated by using the generalized equation of each filter class and type; and by using their different circuit topologies (Shunt or Series topology). The GUI can also provide reactive element values from given specification. Moreover, the features for the design of ultra wideband (UWB) band-pass filter, capacitively coupled and combline filter are also incorporated into the stand-alone application. The finalized prototype will serve both industries and educational institutions.

# ACKNOLEDGEMENT

First and foremost, I would like to show and present my humble gratitude to Universiti Teknologi Petronas for putting in place this dynamic educational system for its students, to PETRONAS for its financial, physical and moral support throughout my degree programme in UTP.

Moreover, I would like to express my gratitude to my project supervisor Dr. Wong Peng Wen for his valuable support, guidance and encouragement throughout my final year project. His knowledge, brilliant ideas, advices and all his constructive criticisms helped me to make necessary improvements which helped in leading to a more effective project.

Similarly, I would like to thank Sovuthy Cheab and Sohail Khalid for sharing with me their expertise, for their endless help and technical assistance which helped me to have a vibrant practical experience in handling engineering projects.

I am also as ever, especially indebted to my parents, siblings and friends for their love and support throughout my life.

Last but not least, my greatest regards to The Almighty for bestowing upon me the courage to face the challenges and complexities of life; and complete this final year project successfully.

# TABLE OF CONTENTS

# LIST OF TABLES

# LIST OF FIGURES

# LIST OF ABBREVIATIONS

FCC     : Federal Communications Commission

UWB   : Ultra Wideband

GUI     : Graphical User Interface

MatLab:  Matrix Laboratory

LPF     : Low Pass Filter

HPF    : High Pass Filter

BPF     : Band Pass Filter

BSF     : Band Stop Filter

RF       : Radio Frequency

Radar   : Radio detection and ranging

SoC     : System on Chip

EMI     : Electromagnetic interference

CAD    : Computer Aided Design

ADS    : Advanced Design Systems

HFSS   : Ansoft High Frequency Structure Simulator

GUIDE: Graphical User Interface Development Environment

MCR   : MatLab Compiler Runtime

$Z_o$       : System impedance

UI        : User interface

LA      : Insertion Loss

LR      : Return Loss

N        : Order of the filter

L        : Inductor

C        : Capacitor

S        : Selectivity

$\omega_s$     : Stopband frequency

$\omega_p$     :  Passband frequency

dB     : Decibel

Hz     : Hertz

GHz    : Gigahertz

# CHAPTER 1
# INTRODUCTION

This chapter focuses mainly on presenting a brief background on microwave filters and graphical user interface. It states also the problem and the objectives of this research project.

## 1.1  Background

For the last few decades, the world of high-speed computing devices and systems have seen tremendous improvements. The latter technological advancement results from the endless contributions and efforts of researchers, scientists, system designers or industries from various engineering disciplines. Meanwhile, the demand for higher bandwidth systems and high performing devices that can handle multi-gigabits data transmission over communication media is also scaling rapidly. Consequently, this demand for high frequency and high bandwidth systems becomes the defining backbone of the emerging technologies such as high-speed system design, system on chip (SoC), radio frequency (RF) and microwave system design, telecommunication, wireless communication, satellite communication, radio detection and ranging (RADAR)... In designing these complex communication systems, it is crucial to ensure that the final product complies with the normal and targeted communication standards. Besides, it should also have the ability to handle signal interferences, to suppress electromagnetic interference (EMI) or the jittery interferences that might occurs during the data transmission; the system should also have a proper defined frequency spectrum for a given transmission.

In the same perspective, microwave filters are used as two-port network to control the rejection of unwanted signal frequencies while allowing the desired signal frequencies to pass via a system [25]. These microwave filters are generally classified as Low-Pass (LP), High-Pass (HP), Band-Pass (BP) and Band-Stop. The Band-Pass Filter can be broken down into narrow-band, wideband, ultra-wideband... Moreover, the latter type of BPF is still a novelty that has been seeing tremendous improvements.

Different techniques and methods have been used to design advanced microwave filters that help in maximizing the performance and effectiveness of high-speed devices used in modern communication technology [1]. In the same perspective, the Federal Communications Committee (FCC) released ultra-wideband (UWB) technology (with a bandwidth ranging from 3.1GHz to 10.6 GHz), a novelty in radio technology for the advancement in communication, networking, radar systems, imaging systems, and positioning systems, and gave the official authorization for the use of UWB technology and its commercialization in 2002 [2], [3]. Ever since, this new proposed technology has revolutionized the world of communication and networking industries and it has helped them to provide better services to their respective customers; UWB technology has become one of the major focus point and the center of research in microwave filters [4].

In line with this technological development, it is important to provide freshmen (in the area of microwave filter design), designers, innovators and researchers with new tools and platforms such as computer aided design (CAD) or graphical user interface (GUI) that will help in simplifying their work while boosting their efficiency but cutting down the design time. A GUI, with reference to [5], is defined as a pictorial interface to a program. The graphical user interface can help its user to better perform the data analysis and interpretation via the extraction of important design parameters and the display of data in both numerical value and graphical or pictorial form.

Throughout this research project, the emphasis will be put on the development of a graphical user interface (GUI) for the purpose of microwave filter design.

## 1.2 Problem statement

With the cutting-edge technology of RF and microwave devices used in high frequency applications, there are also powerful software and CAD tools available that help in the design of microwave filters or RF systems. In order to design RF systems or microwave filter for example, one or the combination of the following sophisticated platforms or simulation tools can be used for a computational design: Advance Design Systems (ADS) Agilent, Genesys RF and Microwave Design Software, Ansoft High Frequency Structure Simulator (HFSS), Microwave Office Design Suite, Cadence RF Design... However, designing microwave filters using these advanced tools requires a certain level of expertise, a strong programming skill or a very good knowledge in CAD.

It was reported that in order to master the functionalities and maximize the usage of ADS design tools, an intensive training and extensive exposure to the software's environment is required [14]. This makes it difficult to help beginners such as sophomores, freshmen or any starter to understand the fundamentals and basic theory behind microwave filter design before transitioning them to be familiar with high level microwave design and synthesis tools.

Similarly, due to time constraints in teaching microwave filters, as there are various materials to cover within a short period of time (approximately one semester), students are in need of software tools that will help them to easily practice computer simulations of microwave circuits for their homework and projects [30].

From professional designers' point of view, microwave filter design is a long and tiring journey towards reaching the final design and prototype. The process starts from the theoretical synthesis through schematic simulations to prototype fabrication and measurements. However, most of the time, the disagreement between the simulation and results occurs due to the unavailability of the synthesized theoretical values of the passive components in the market. Consequently, designers have to re-do the simulations due to undesired output.

Moreover, references [14] and [30] suggest that the existing software tools for microwave filter design are costly. In addition, modern filter topologies are not yet integrated in these filter design tools.

The proposed software tools in this project will incorporate both modern and conventional microwave filter topologies for industrial and educational applications. It will help users to get the 2D responses and their corresponding reactive element values from given specifications.

## 1.3   Objectives and Scope of Study

### 1.3.1   Objectives

The main objective of this research project is to develop a stand-alone graphical user interface (GUI) for microwave filter design. But most specifically, the intention is:

- ❖ To review the mathematical modeling of microwave filtering functions with their respective frequency responses.
- ❖ To design and build a graphical user interface that can stand-alone and help in the design of modern and conventional microwave filters.
- ❖ To provide a low-cost and time-effective software tool for industrial and educational applications.
- ❖ To generate responses using both 2D response emulation and simulation.

### 1.3.2   Scope of study

The scope of study covers basically all the necessary approaches and steps from the initial works to the final prototype resulting from this research project.

The primary works start with the identification of the design parameters and requirements. Throughout this phase, the main task is to find the input parameters and the basic valid design criteria for each type and class of microwave filter that will be incorporated into the GUI. This phase is followed by the initial drafts of the GUI layout design and its implementation in Graphical User Interface Development Environment (GUIDE). After a successful design of the GUI, the next major task will be the programming of the user interface (UI) functionalities such as save button, plot button, reset button, exit button...

The secondary works will focus on testing, debugging and running the finalized prototype of the GUI.

# CHAPTER 2
# LITTERATURE REVIEW

This chapter highlights briefly the technology and evolution of microwave systems; it covers also the key mathematical synthesis of microwave filtering functions, their respective transfer function using their scattering parameters (S-parameters), their ABCD matrix with relation to S-parameters through lumped elements and their design requirements. In addition, this section presents the previous works on GUI design and development for different applications and the fundamental design theory, principles or criteria behind GUI development using MATLAB software.

## 2.1 Theory of Microwave Filters

### 2.1.1 Basic Definition of terms relating to microwave filters

Before proceeding further, it is crucial to review the fundamentals resulting from the basic classification of microwave filters. In this generalized classification, each class of filter and its related concepts can be defined.

- ➢ Low Pass Filter (LPF): As the name of this filter class states "Low Pass", this filter passes only low frequency signal. That is to say, it allows signals with frequency from zero to cut-off frequency to pass and rejects all frequencies above its cut-off frequency.
- ➢ High Pass Filter (HPF): It blocks all signal frequencies lower than cut-off frequency while passing all signal frequencies higher than the cut-off frequency.
- ➢ Band Pass Filter (BPF): This filter passes only signal frequencies that are within its passband.
- ➢ Band Stop Filter (BSF): BSF rejects all frequency signals within the stopband and passes all signal frequencies that fall out of the stopband.
- ➢ Stopband refers to the frequency band that is occupied by the

unwanted signal frequencies.

➢ Passband refers to frequency band that is occupied by the desired signal frequencies.

➢ Transition band is the frequency band that allows the transition from passband to stopband or vise versa.

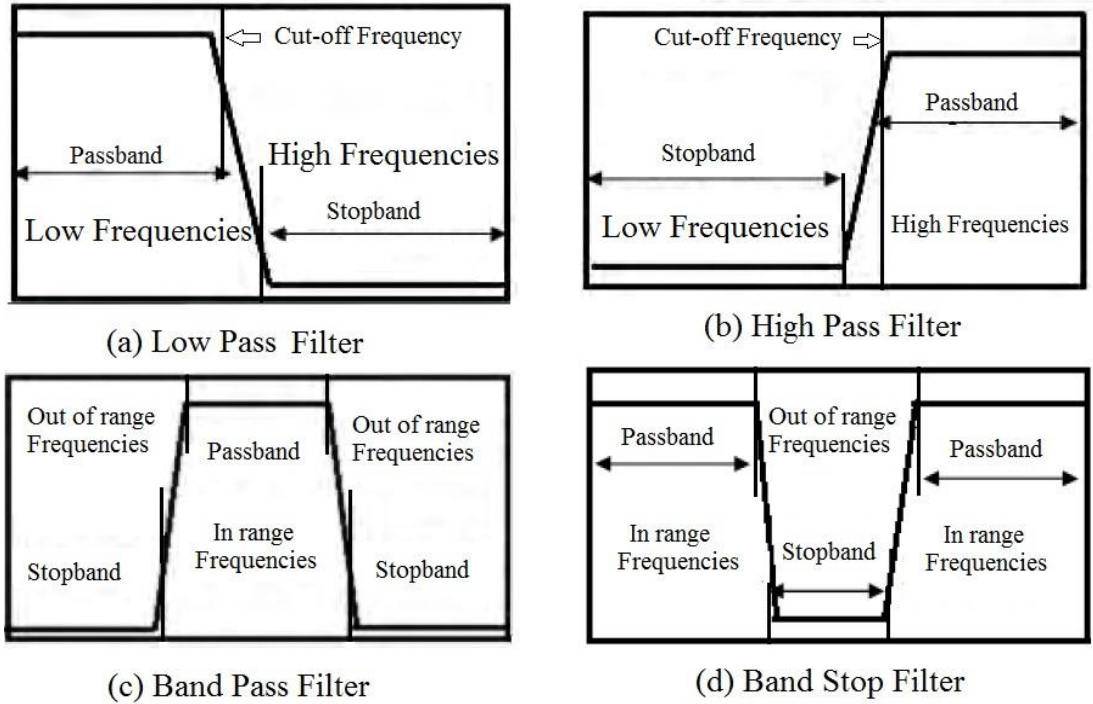

Figure 1    Frequency responses for the four classes of filters

### 2.1.2    Overview of Ultra Wideband (UWB) Filter

From the regulation issued by the FCC during the official release of UWB technology, three fundamental types of UWB systems have been defined [3]:

➢ Imaging systems.
➢ Communication and measurement systems.
➢ Vehicular and radar systems.

All UWB systems should be able to operate within the frequency band ranging from 3.1GHz to 10.6GHz. All signal being transmitted via UWB system should occupy a bandwidth (BW) of not less than 500MHz, the defined and selected signal BW is expected to be at least a quarter of the carrier or center frequency. With reference to this definition of UWB technology by the FCC, it is important to notice that UWB systems and principally UWB filters are designed by incorporating both low pass filter (LPF) and high pass filter (HPF), the cascaded equivalence of LPF and HPF helps to create a dynamic UWB band-pass filter (BPF) [3], [13].

The technology of UWB systems have shown remarkable and excellent performance in its application for in wireless communication. It is able to transfer high data rate (100Mbps to 1Gbps) at the lowest power consumption compared to WiFi, IEEE 802.16, 2G or 3G but only for short distance (not more than 30 meters) applications. Figure (2) presents the difference between different wireless communication standards in of their respective data rates and supported ranges.

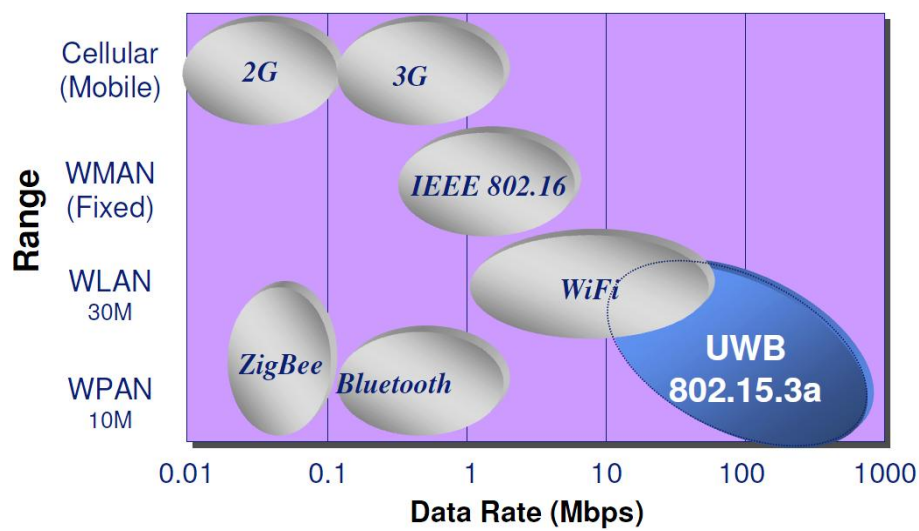

Figure 2   Data rate vs. Range for different wireless broadband standards [15].

## 2.2   Mathematical modeling of microwave filters

Microwave filters are mostly synthesized using different filter responses: maximally flat (Butterworth), equal ripple (Chebyshev), elliptic function, linear phase ... However, this project will mainly emphasize on Chebyshev and Butterworth filters as these two are the most commonly used attenuation characteristics in microwave filters [25].

### 2.2.1 Maximally Flat or Butterworth Filters

The generalized equation for Butterworth low pass prototype is specified by [25],[24]:

$$|S_{12}|^2 = \frac{1}{1+\omega^{2N}} \quad (2.1)$$

where $\omega \to \frac{\omega}{\omega_c}$, with $\omega_c$ as the cut-off frequency and $\omega$ as frequency variable. N is the filter order, and it is related to the selectivity (S), the return loss (LR) and the insertion loss (LA) by:

$$N = \frac{L_A + L_R}{20\log_{10}(S)} \quad (2.2)$$

with $S = \frac{\omega_s}{\omega_p} > 1 \quad (2.3)$

where $\omega_s$ refers to the stopband frequency and $\omega_p$ the passband frequency. LR and LA are in dB.

The transfer function related to $S_{11}$ can be expressed by using the following relationship:

$$|S_{12}|^2 + |S_{11}|^2 = 1 \quad (2.4)$$

#### 2.2.1.1 Element values for Butterworth filter

The N (where N is the number of reactive elements) element values for a two-port network ended at both ends with a resistor using maximally flat filter characteristics can be computed as follow [25],[27]:

$$g_o = g_{N+1} = 1 \quad (2.5)$$

$$g_k = 2\sin\left[\frac{(2k-1)\pi}{2N}\right] ; where \ k = 1, 2, 3, ..., N \quad (2.6)$$

The related reactive elements for both series and shunt N-section LC ladder network are obtained through:

$$L_k = \frac{Z_o g_k}{\omega_c} \ \text{(Series elements)} \quad (2.7)$$

$$C_k = \frac{g_k}{Z_o \omega_c} \ \text{(Shunt elements)} \quad (2.8)$$

where $k = 1, 2, 3, ..., N$

The reactive elements (L and C) alternate depending on the ladder type (shunt or series) as shown in figure (3). $Z_o$ is the system impedance.

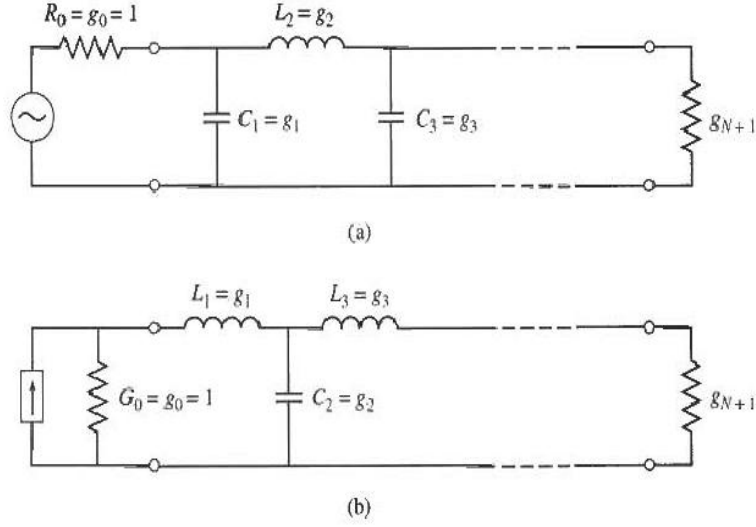

Figure 3  N-section LC ladder circuit with element definitions for LP prototype. (a) Prototype beginning with a shunt element. (b) Prototype beginning with a series element [27].

### 2.2.1.2 ABCD parameters and S-parameters

The ABCD parameters of a network are defined by the transfer matrix which is the result of a relationship between the input voltage and current to the output voltage and current [24].

$$T = \begin{vmatrix} A & B \\ C & D \end{vmatrix} \qquad (2.9)$$

The transfer matrix for series impedance and shunt impedance network is defined in Figure (4).

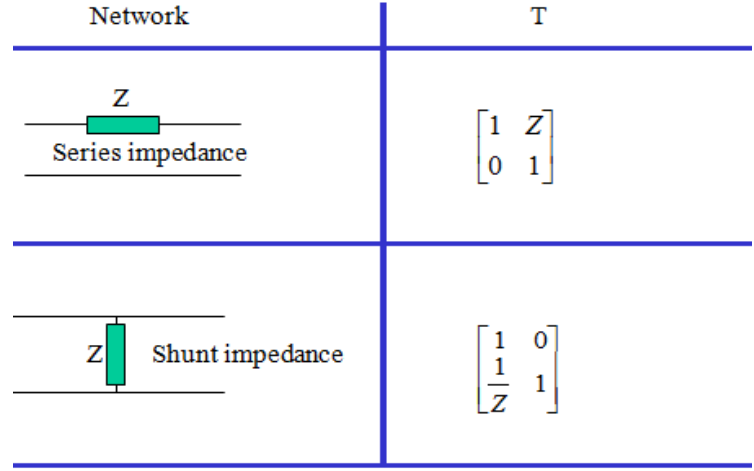

Figure 4    T matrix for shunt and series impedances [28]

For cascaded network like the N-section LC ladder network shown in figure (3), the total transfer matrix for both circuits can be obtained by matrix multiplication from the T matrix of the first reactive element till the T matrix of the last element.

By using the relationship that helps in converting the ABCD parameters to S-parameters [28], one can extract the expression of each S-parameter.

$$\begin{bmatrix} S_{11} & S_{12} \\ S_{21} & S_{22} \end{bmatrix} = \frac{1}{Z_o A + B + Z_o^2 C + Z_o D} \begin{bmatrix} Z_o A + B - Z_o^2 C - Z_o D & 2Z_o(AD-BC) \\ 2Z_o & -Z_o A + B - Z_o^2 C + Z_o D \end{bmatrix} \quad (2.10)$$

$$S_{21} = \frac{2}{A + B/Z_o + Z_o C + D} \quad (2.11)$$

$$|S_{21}|^2 = \frac{4}{|A + B/Z_o + Z_o C + D|^2} \quad (2.12)$$

Due to the symmetrical nature of maximally flat low pass prototype using its element values, deductively:

$$|S_{12}|^2 = |S_{21}|^2 = \frac{4}{|A + B/Z_o + Z_o C + D|^2} \quad (2.13)$$

Similarly, $S_{11}$ and $S_{22}$ can be expressed in term of ABCD parameters.

### 2.2.2    Equal Ripple or Chebyshev Filters

For Chebyshev response, the generalized filtering equation for low pass prototype is defined as [25]:

$$|S_{12}|^2 = \frac{1}{1+\varepsilon^2 T_N(\omega)^2} \tag{2.14}$$

$$\text{where } T_N(\omega) = \begin{cases} \cos(N\cos^{-1}\omega) \;,\; |\omega| \leq 1 \\ \cosh(N\cosh^{-1}\omega) \;,\; |\omega| \geq 1 \end{cases} \tag{2.15}$$

$$\omega \to \frac{\omega}{\omega_c} \tag{2.16}$$

$$\varepsilon = \sqrt{10^{0.1 L_{Ar}} - 1} \tag{2.17}$$

where $\omega_c$ referring to its cu-off frequency, $\omega$ the frequency variable and $\varepsilon$ is the ripple level.

The filter order (N) of this filter response is related to its insertion loss (LA), return loss (LR) and selectivity (S) by:

$$N = \frac{L_A + L_R + 6}{20 \log_{10}\left(S + \sqrt{S^2 - 1}\right)} \tag{2.18}$$

where $S$ is defined by equation (2.3).

The passband ripple ($L_{Ar}$) can be computed by using the following mathematical expressions:

$$L_{Ar} = -10 \log_{10}(1 - \Gamma^2) \tag{2.19}$$

where $\Gamma$ refers to the reflection coefficient and it is related to the return loss (LR) by:

$$\Gamma = 10^{(-RL/20)} \tag{2.20}$$

### 2.2.2.1 *Element values for Chebyshev filter*

For low pass prototype using Chebyshev response, the reactive elements of the N-section LC ladder circuit portrayed in figure (3) can be obtained through these equations [25]:

$$\begin{aligned} \beta &= \ln\left[\coth\left(\frac{L_{Ar}}{17.37}\right)\right] \\ \gamma &= \sin\left(\frac{\beta}{2N}\right) \\ a_k &= \sin\left[\frac{(2k-1)\pi}{2k}\right] \;;\; k = 1,2,3,...,N \\ b_k &= \gamma^2 + \sin^2\left(\frac{k\pi}{N}\right) \;;\; k = 1,2,3,...,N \end{aligned} \tag{2.21 - 2.24}$$

Therefore, the element values can be computed using:

$$g_o = 1$$

$$g_1 = \frac{2a_1}{\gamma}$$

$$g_k = \frac{4a_{k-1}a_k}{b_{k-1}g_{k-1}} \quad ; \quad k = 2, 3, ..., N \qquad (2.25 - 2.28)$$

$$g_{N+1} = \begin{cases} 1 & \text{for } N \text{ Odd} \\ \coth^2\left(\frac{\beta}{4}\right) & \text{for } N \text{ Odd} \end{cases}$$

where $L_{Ar}$ is the passband ripple (dB).

The associated reactive elements are similar to those described by equations (2.7) and (2.8); and the resulting ABCD matrix and S-parameters can be computed using equations listed in section (**2.2.1.2**).

### 2.2.3 Low Pass to High Pass, Band Pass or Band Stop Transformation.

In microwave filter design, with the use of filter prototype transformation, it is possible to convert a low pass prototype to either high pass, band pass or band stop prototype. This transformation is done through a frequency substitution [24],[25].

➤ For low pass to high pass transformation, frequency substitution is done as follow:

$$\omega \longrightarrow -\frac{\omega_c}{\omega} \qquad (2.29)$$

➤ When converting a low pass to band pass prototype, the low pass circuit undergoes the following frequency substitution:

$$\omega \longrightarrow \alpha\left(\frac{\omega}{\omega_o} - \frac{\omega_o}{\omega}\right) \qquad (2.30)$$

$$\alpha = \frac{\omega_o}{\omega_2 - \omega_1} \quad \text{and} \quad \omega_o = \sqrt{\omega_1 \omega_2} \qquad (2.31 - 2.32)$$

where $\omega_o$ is the mid-band frequency and $\alpha$ is the bandwidth scaling factor.

➤ The last transformation is the conversion of a low pass prototype to band pass prototype which is achieved via this frequency substitution:

$$\omega \longrightarrow -\cfrac{1}{\alpha\left(\cfrac{\omega}{\omega_o} - \cfrac{\omega_o}{\omega}\right)} \tag{2.33}$$

$$\alpha = \frac{\omega_o}{\omega_2 - \omega_1} \ , \ \omega_o = \sqrt{\omega_1 \omega_2} \tag{2.34 - 2.35}$$

where $\omega_o$ referring to the mid-band frequency and $\alpha$ the bandwidth scaling factor.

### 2.2.4  Extra classes of Chebyshev Band Pass Filter

#### 2.2.4.1  *Capacitively coupled Chebyshev Band Pass filter*

The circuit topology for this filter class is shown figure (5).

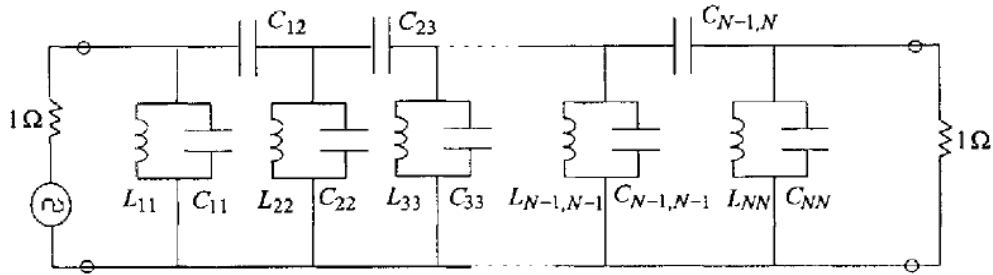

Figure 5   Circuit topology for capacitively coupled band pass filter [24].

With reference to [24], the reactive elements of this topology is derived using the following equations:

$$\eta = \sinh\left[\frac{1}{N}\sinh^{-1}\left(\frac{1}{\varepsilon}\right)\right]$$

$$C_r = \frac{2}{\eta}\sin\left[\frac{(2r-1)\pi}{2N}\right] \ ; \quad r = 1, 2, 3, ..., N \tag{2.36 - 2.38}$$

$$K_{r,\ r+1} = \frac{\sqrt{\eta^2 + \sin\left(\frac{r\pi}{N}\right)^2}}{\eta}$$

where the ripple level ($\varepsilon$) and the order of the filter (N) are described by equations (2.17) and (2.18) respectively.

By parameter values described in the above equations, the element values can be computed.

$$C_{01} = C_{N,N+1} = \frac{1}{\omega_o \sqrt{\alpha - 1}}$$

$$C_{r,r+1} = \frac{K_{r,r+1}}{\alpha \omega_o} \quad ; \quad r = 1, 2, 23, ..., N-1$$

$$C_{11} = \frac{C_r}{\omega_o} - \frac{\sqrt{\alpha - 1}}{\alpha \omega_o} - C_{12}$$

$$C_{NN} = \frac{C_N}{\omega_o} - \frac{\sqrt{\alpha - 1}}{\alpha \omega_o} - C_{N-1,N}$$

$$C_{rr} = \frac{C_r}{\omega_o} - C_{r-1,r} - C_{r,r+1} \quad ; \quad r = 2, 23, ..., N-1$$

$$L_{rr} = \frac{1}{C_r \omega_o}$$

(2.39 - 2.44)

The filter order (N), the mid-band frequency ($\omega_o$) and the bandwidth scaling factor ($\alpha$) are computed using equations (2.17), (2.34) and (2.33) respectively.

### 2.2.4.2 *Mathematical modeling of Chebyshev Ultra Wideband (UWB) BP Filter*

Throughout the gradual evolution of UWB technology, different design techniques, models and structures for microwave filters have been used to build prototypes that comply fully with the standards of the UWB systems [16]. Among these structures that exhibit reliably the qualitative and quantitative behaviour of a successful ultra wideband system, the structure involving microstrip made up of a microstrip multi-mode resonator (MMR) and a parallel-coupled line at each end of the network. The basic configuration of this category of microstrip is portrayed in figure (6).

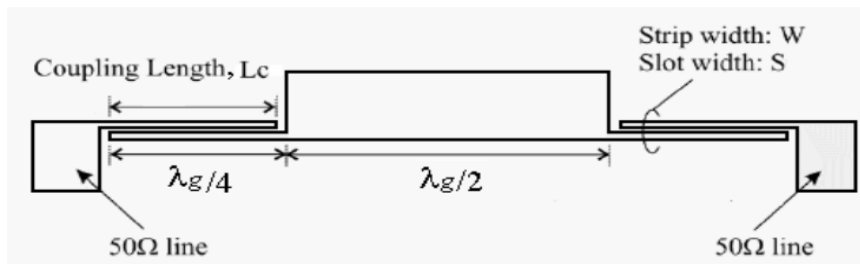

Figure 6    Basic schematic diagram of microstrip MMR with 2 parallel-couple lines at the two ends of the network [16].

Further improvements on the microstrip MMR based UWB systems in order to achieve better UWB passband behaviour have been done. Figure (7) depicts the layout of the MMR UWB band-pass filter (BPF) and its equivalent circuit diagram.

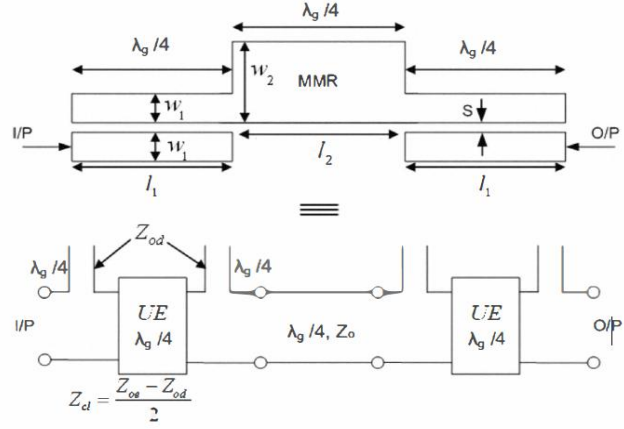

Figure 7  Layout and circuit diagram of MMR UWB band-pass filter.

Where $l_1$ refers to the coupling length, $w_1$ and $w_2$ the strip widths, $S$ and $\lambda_g$ refer to the slot width and the wavelength respectively. $Z_o$ defines the characteristic impedance while $Z_{oe}$ and $Z_{od}$ are the related even-mode and odd-mode characteristic impedances of microstrip MMR based UWB BPF.

The mathematical modeling of UWB BPFs, that is presented in the following sections of this sub-chapter and adapted from reference [13], [17], was derived using the above described topology.

### 2.2.4.2.1 Fourth order UWB BPF

The filtering function, $T(\theta)$, for fourth order UWB band-pass filter is defined as the Chebyshev function by:

$$T(\theta) = \frac{A\cos^4(\theta) + B\cos^2(\theta) + C}{\sin(\theta)} \qquad (2.45)$$

A reciprocal relationship can be established between the electrical length $\theta$, the center frequency $f_c$ (with $f_c = (f_{c(HPF)} + f_{c(LPF)})/2$) and the frequency $f$ of the filter.

$$\theta = \frac{\pi}{2} \times \frac{f}{f_c} \qquad (2.46)$$

Equation (1) presents the characteristic of a Chebyshev polynomial. For equal ripples in Chebyshev polynomial $T(\theta)$ with an inverse dependence in term of the filter frequency $f$, it is crucial to normalize $T(\theta)$. The restructured mathematical model of $T(\theta)$ is:

$$T(\theta) = \left(\frac{\varepsilon}{\overline{T}(\theta_{norm})}\right) \times \overline{T}(\theta) \qquad (2.47)$$

Here, $\theta_{norm} = \frac{\pi}{2}$, $\overline{T}(\theta)$ and $\varepsilon$ refers to the factor of the normalization of $T(\theta)$ and the ripple factor respectively. The ripple factor or ripple level ($\varepsilon$) is computed using equation (2.16). $\overline{T}(\theta)$ is given by:

$$\overline{T}(\theta) = \frac{\cos^4(\theta) + \alpha \cos^2(\theta) + \zeta}{\sin(\theta)} \qquad (2.48)$$

where

$$A = \frac{\varepsilon}{\overline{T}(\theta_{norm})} \qquad (2.49)$$

$$\alpha = \frac{B}{A} \qquad (2.50)$$

$$\zeta = \frac{C}{A} \qquad (2.51)$$

where $\theta_{norm} = \frac{\pi}{2}$, $\overline{T}(\theta_{norm}) = \zeta$, therefore the expression of A, B and C become: $A = \frac{\varepsilon}{\zeta}$, $B = \frac{\alpha \cdot \varepsilon}{\zeta}$ and $C = \varepsilon$. $\alpha$ and $\zeta$ components can be computed from their relationship with the bandwidth ($BW = f_{c(HPF)} - f_{c(LPF)}$) while complying with $\alpha^2 - 4\zeta > 0$ and $\alpha + \zeta + 1 > 0$ inequalities.

$$\alpha = \frac{3}{4}\left[\cos\left(\frac{BW}{2}\right)+\frac{1}{3}\right]^2 - \frac{4}{3} \tag{2.52}$$

$$\zeta = \frac{1}{4}\sin^2\left(\frac{BW}{2}\right)\left[1-\cos\left(\frac{BW}{2}\right)\right] \tag{2.53}$$

**Nota Bene**: The bandwidth (BW) value in GHz should be converted to its equivalent value in radian using equation (2.44) prior to the computation of $\alpha$ and $\zeta$ expressed in equation (2.50) and (2.51).

The following table summarizes the values of $\varepsilon$ for different return losses.

Table 1   Return Loss (LR) and its equivalent Ripple Factor or Ripple Level

| Return Loss (RL) (dB) | Reflection coefficient | Passband Ripple (dB) | Ripple Factor |
|---|---|---|---|
| 1 | 0.8913 | 6.8683 | 1.9652 |
| 2 | 0.7943 | 4.3292 | 1.3076 |
| 3 | 0.7079 | 3.0206 | 1.0024 |
| 4 | 0.6310 | 2.2048 | 0.8133 |
| 5 | 0.5623 | 1.6509 | 0.6801 |
| 6 | 0.5012 | 1.2563 | 0.5792 |
| 7 | 0.4467 | 0.9665 | 0.4993 |
| 8 | 0.3981 | 0.7494 | 0.4340 |
| 9 | 0.3548 | 0.5844 | 0.3795 |
| 10 | 0.3162 | 0.4576 | 0.3333 |
| 11 | 0.2818 | 0.3594 | 0.2937 |
| 12 | 0.2512 | 0.2830 | 0.2595 |
| 13 | 0.2239 | 0.2233 | 0.2297 |
| 14 | 0.1995 | 0.1764 | 0.2036 |
| 15 | 0.1778 | 0.1396 | 0.1807 |
| 16 | 0.1585 | 0.1105 | 0.1605 |
| 17 | 0.1413 | 0.0875 | 0.1427 |
| 18 | 0.1259 | 0.0694 | 0.1269 |
| 19 | 0.1122 | 0.0550 | 0.1129 |
| 20 | 0.1000 | 0.0436 | 0.1005 |

The transfer functions for UWB BPF are related to $T(\theta)$ by the following equations; their expressions in term of A and $\overline{T}(\theta)$ can also be derived using equation (2.45) and (2.47). Therefore:

$$\left|S_{11}(\theta)\right|^2 = \frac{T^2(\theta)}{1+T^2(\theta)} \Rightarrow \left|S_{11}(\theta)\right|^2 = \frac{\overline{T}^2(\theta)}{A^2+\overline{T}^2(\theta)} \tag{2.54}$$

$$|S_{21}(\theta)|^2 = \frac{1}{1+T^2(\theta)} \Rightarrow |S_{21}(\theta)|^2 = \frac{1}{1+A^2 \cdot \overline{T}^2(\theta)} \qquad (2.53)$$

From these equations, the dB expression of $|S_{11}(\theta)|^2$ and $|S_{21}(\theta)|^2$ for graphical analysis of their respective frequency response can be computed accordingly.

$$|S_{11}(\theta)|^2{}_{(dB)} = 20\log_{10}\overline{T}(\theta) - 10\log_{10}\left(A^2 + \overline{T}^2(\theta)\right) \qquad (2.54)$$

$$|S_{21}(\theta)|^2{}_{(dB)} = -10\log_{10}\left(1 + A^2 \cdot \overline{T}^2(\theta)\right) \qquad (2.55)$$

The graphical form of the $|S_{11}(\theta)|^2$ and $|S_{21}(\theta)|^2$ are portrayed in the following figure.

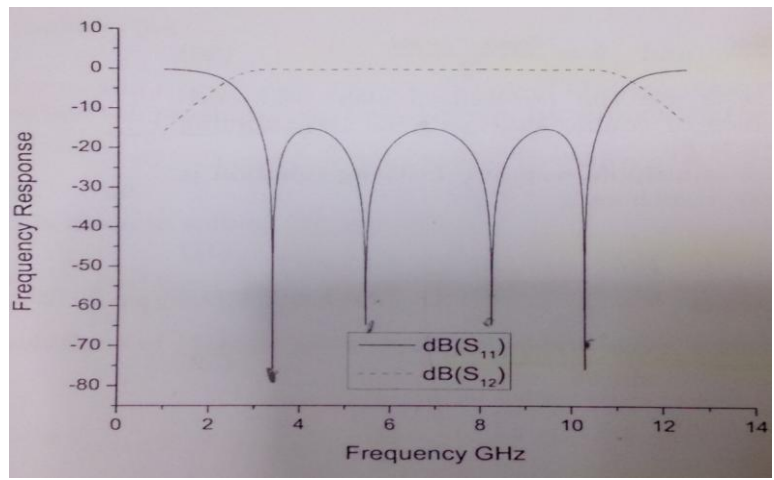

Figure 8    Frequency response of 4$^{\text{th}}$ order UWB BPF [17]

## 2.3   Graphical User Interface (GUI) Design

GUI plays an important role as the medium between the user and a computer program. The user interface provides the user with an environment integrated with interactive controls, commands, feedbacks ... with a predictive response. A simple example of human and computer interface is described in figure (9).

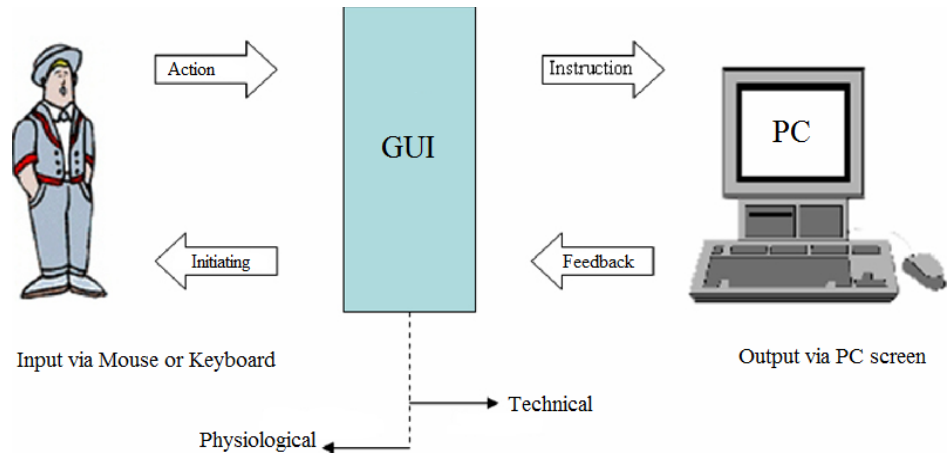

Figure 9　Human and GUI interaction [21].

### 2.3.1　Fundamental Principles for GUI design

According to a recent study conducted at the Xi'an University of science and technology, a number of design considerations and criteria such as dynamism, consistency and attractive visual appearance of the user interface, its layout and functions have been analyzed with respect to consumer's need, desire, taste and likes [20]. This important study was done using Kansei Engineering (KE), an engineering discipline with emphasis on implicit satisfaction and pleasure of consumer prior to product design, highlights the importance of the interaction between humans and the user interface and shows how the consideration of human's cognitive psychology and anthropology on the artificial intelligence of the GUI can help to boost the product's market.

According Martin [22], there are three fundamental design factors for a successful and information-oriented GUI: the development factors, the visibility factors and the acceptance factors. Similar to the study done by Sun [20], these three factors covers the basics requirements that are deemed fit in term of their compliance with human's anthropological and psychological perception. Moreover, Martin [22] has provided some guiding principles to be followed when designing a graphical user interface. For a successful GUI design, it is imperative to organize the layout design properly, to economize the window of the interface and to make a user friendly GUI. In figure (10) bellow are examples of a "good layout" and "bad layout" structures of a GUI.

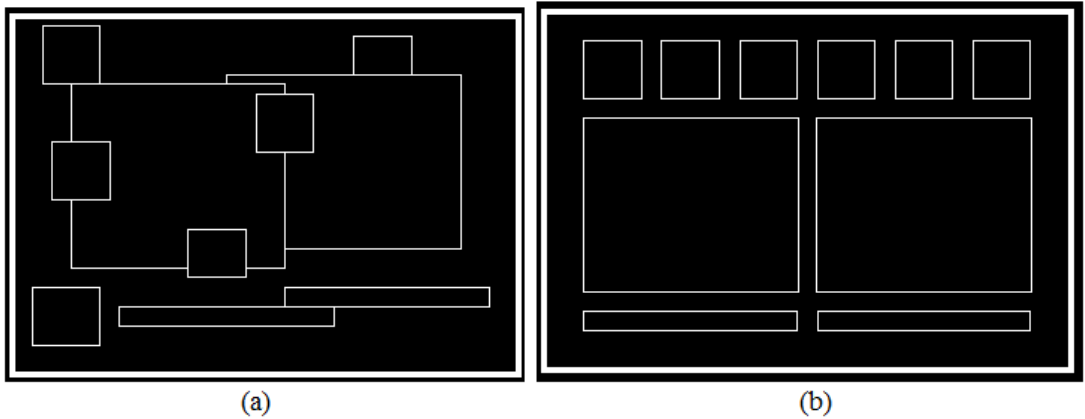

Figure 10 (a) Chaotic layout for GUI, (b) Ordered layout for GUI

### 2.3.2 MATLAB GUI

MATLAB GUI, like other graphical user interfaces, is to help people without a good command of programming language to be able to interact with the machine without the need to master it [23]. Marchand and Holland [23] continue on suggesting and defining the design stages involved in the development of a GUI using MATLAB software. Prior to the GUI design, the developer should make a proper analysis by identifying the end user and defining the user guide. This stage should be followed by the layout design of the user interface based on the analysis done during the first design stage and KE design considerations. Then, the designer should draft and sketch skillfully its layout design on a piece of paper before proceeding to the actual design and construction of the GUI.

When designing GUI using MATLAB, Marchand and Holland [23] present two different approaches: GUI Development Environment (GUIDE) technique and Programmatic approach.

❖ In Low-Level MATLAB GUI programming approach, the developer program the GUI using M-File, all the graphic elements of the interface are also programmed along with the GUI. In this approach, the designer should define all the functions or Callbacks involved.

❖ In High-Level GUI Development Environment (GUIDE) approach, compared to the programmatic approach which involve a lot of "hand-programming" for the creation of the GUI, this new technique uses the embedded feature (GUIDE) of MATLAB to create the complete GUI which can be saved as FIG-file. However, once the GUI is designed, all the Callback functions need to be programmed accordingly. The complete GUI layout editor shown in figure (11) is obtained by entering the **guide** at the command prompt.

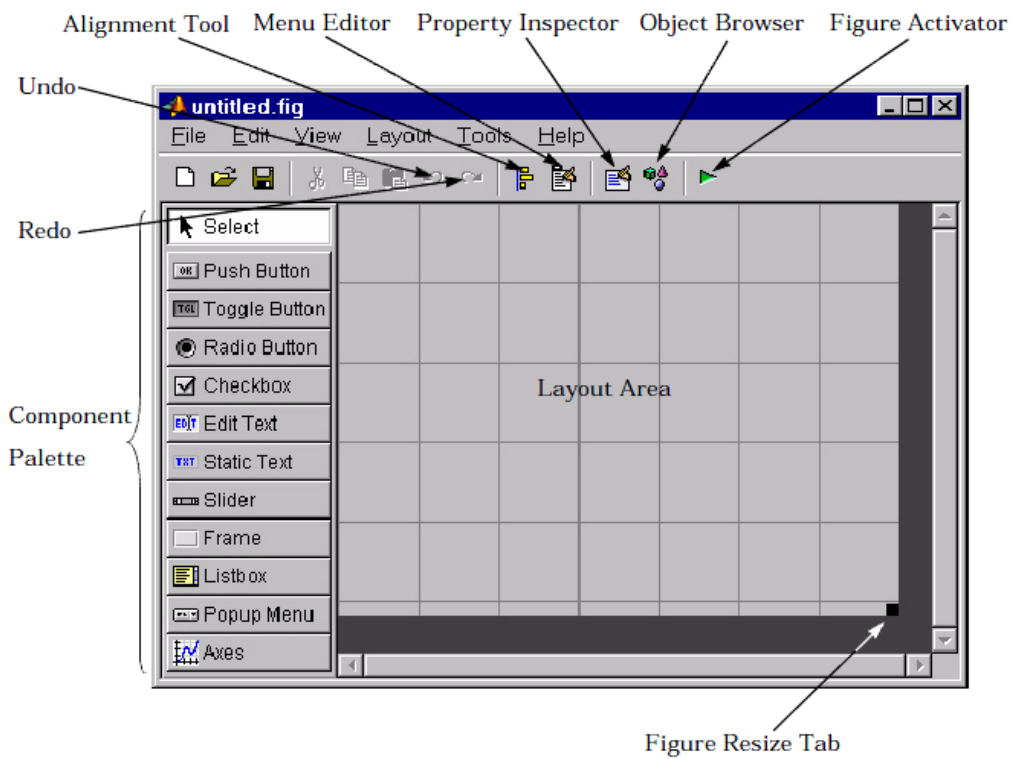

Figure 11    GUI Development Environment (GUIDE) Layout Editor [29].

# CHAPTER 3
# METHODOLOGY

This chapter focuses on the methodology that will be used in the conduct of this research project. It comprises of research methodology, project activities, key milestones, Gant Chart and tools used.

## 3.1 Research Methodology

Prior to stepping into the design phase of the user interface, a deeper exploration, research and review were conducted on the fundamental theories of microwave filters. The primary research was intended to help in grasping the behavior of each filter type and class; their respective frequency response characteristic and the identification of the major key elements needed for the design of a given class of filter. Moreover, some of the key factors, guiding principles and procedures for the design of a dynamic, user-friendly and marketable graphical user interface have been highlighted.

In order to carry out this project, which is to develop a stand-alone GUI for microwave filter design, the high-level GUI development environment (GUIDE) approach will be used to design the layout of the graphical user interface. The related M-file will then be used to program all the callback functions of the user interface. Moreover, all the filtering function, parameters, coefficients, frequency responses ... of the filter derived in **section 2.2** will be used to develop a proper algorithm to be integrated with the designed GUI. The finalized design will be tested for consistency, debugged for any designed error that might occur and submitted for educational and industrial use.

| ID | Project Activities | Start | Finish | Duration |
|----|--------------------|-------|--------|----------|
| 1 | Identifying the Design Requirements | 20/5/2013 | 31/5/2013 | 2w |
| 2 | Identifying the Design Parameters | 27/5/2013 | 7/6/2013 | 2w |
| 3 | Drafting the Layout Design of GUI | 7/6/2013 | 20/6/2013 | 2w |
| 4 | Designing the GUI in MatLab GUIDE | 21/6/2013 | 25/7/2013 | 5w |
| 5 | Programming the UI Controls | 1/7/2013 | 25/11/2013 | 21.2w |
| 6 | Saving and Running the GUI | 25/11/2013 | 29/11/2013 | 1w |
| 7 | Testing and Debugging the GUI | 29/11/2013 | 2/12/2013 | .4w |
| 8 | Finalizing the GUI Design | 2/12/2013 | 10/12/2013 | 1.4w |

Figure 12    Gantt Chart and main project activities

## 3.2  Microwave Filters Programming

The adopted algorithm, which is used to program the filtering functions, frequency responses and the related design parameters for microwave filter, is presented in the form of flow chart in figure (13.1). This flow chart shows also how the graphical user interface will operate. The user will need to select the desired type and class of microwave filter on the user interface. The GUI will take the valid input parameters (such as return loss, insertion loss, pass-band or stop-band frequencies, filter bandwidth ...) from the user. The next task of the GUI is to proceed to displaying the graphs and important data needed by the user on the user interface.

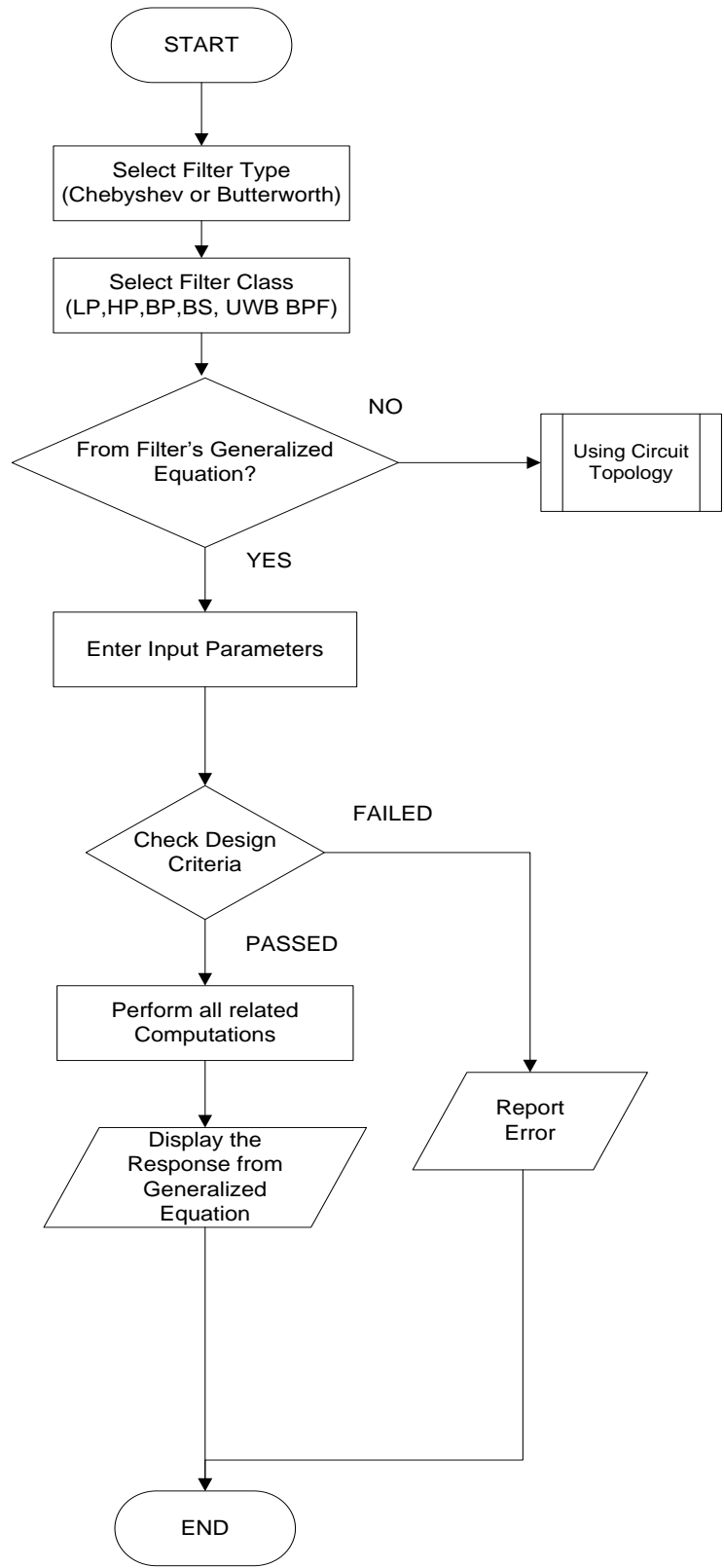

Figure 13   Flow Chart used in programming the generalized Chebyshev and Butterworth filter responses.

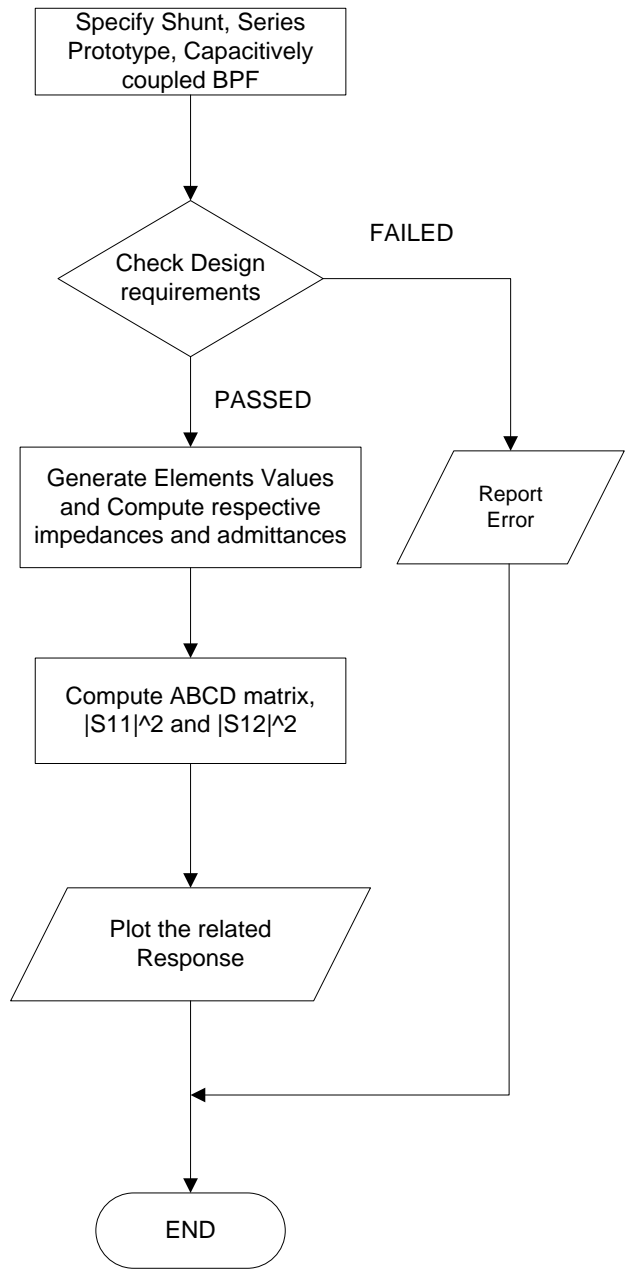

Figure 14   Flow Chart used in programming the Chebyshev and Butterworth filter responses from their respective circuit topologies.

## 3.3   Stand-alone GUI and MatLab Compiler Runtime (MCR)

The developed GUI (in MatLab) can be deployed and converted into an executable (.exe) file. The deployed MatLab GUI project can successfully run on computers with no MatLab software installed, however the stand-alone GUI can run on those computers with the help of MCR. The MCR is a set of libraries that help to run MatLab applications or components on computers with no MatLab installed, it is free to use, to distribute and it can be downloaded via MathWorks website. This MatLab Compiler Runtime is available for Windows, Linux and Mac operating systems.

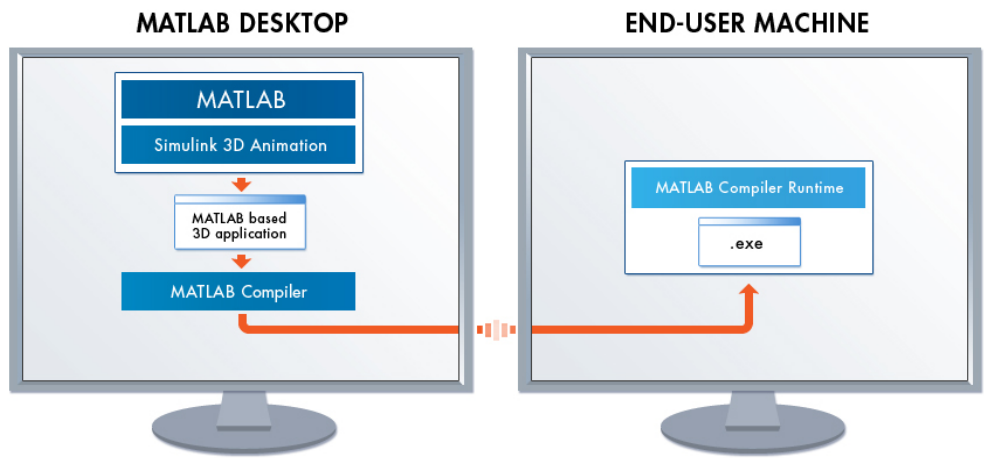

Figure 15    MCR and MatLab stand-alone applications

# CHAPTER 4
# RESULTS AND DISCUSSION

This chapter is objectively intended to provide the results of the research project. It presents a concise information on the GUI's artificial intelligence and user interface controls' operations. An ample discussion is also made on the subject matter.

## 4.1 Programming approach of the GUI and the Transfer Functions.

Structured programming approach was adopted for the design of the GUI, to write its related program codes for both transfer functions and user interface controls. Structured programming approach is the most commonly used approach in software development. By making use of the control structures (sequence, selection and repetition) that this approach provides, the application will be easier to debug, modify and its program code will be easier to understand.

### 4.1.1 Coding the Transfer Functions

The MatLab M-file has been used to program the transfer functions of all the classes of filters that are incorporated in the final prototype. The integrated microwave filters are listed as in table (2).

Table 2  Chebyshev and Butterworth Filters incorporated into the GUI.

| Total number of responses | Filter Types | Filter Classes | Filter Responses generated using |
|---|---|---|---|
| 15 | Chebyshev | LPF<br>HPF<br>BPF<br>BSF | • Chebyshev Generalized Equations<br>• ABCD Matrix |
| 12 | Butterworth | LPF<br>HPF<br>BPF<br>BSF | • Butterworth Generalized Equations<br>• ABCD Matrix |

In total, the GUI will enable its users to design microwave filters for four filter classes using both Chebyshev and Butterworth responses to generate up to twenty seven (27) responses using different techniques and approaches.

More importantly, the programmatic approach that has been adopted in handling each and every filter response differs with reference to each filter's prototype, topology, parameters, characteristics, definition and complexity. The following figure provides a fragment of the code used to generate Chebyshev low pass shunt prototype.

```
                    Tnew = T*[1 Z(t);0 1];
                    T = Tnew;
                elseif k==1
                    % Computing Z(C) values
                    Z(t) = j*Element(t).*w(i);
                    % T matrix
                    Tnew = T*[1 0;Z(t) 1];
                    T = Tnew;
                else
                end
            else
            end
        end
    T;
% Getting ABCD values by identification using 2X2 Matrix T=[A B;C D] by using
    A = T(1,1);
    B = T(1,2);
    C = T(2,1);
    D = T(2,2);
% Computing S12  and S11 values for different values ABCD resulting from the
% variation in w.
    Sa(i) = abs(2./(A+(B./Zo)+(C.*Zo)+D));
    Sb(i) = abs(1 - abs(2./(A+(B./Zo)+(C.*Zo)+D)).^2);
    end
```

Figure 16  Fragment code used to generate Frequency Response for Chebyshev Low Pass Shunt Prototype.

### 4.1.2 Design and implementation of the GUI in MatLab

The application has been designed in MatLab using its built-in tool (the Graphical User Interface Development Environment or GUIDE). The development of this GUI is characterized by the simplicity of the interface for the users to be easily familiar with it. It is intended to help users to perform a straight forward design by selecting the type and class of the filter, the next step will be just to key in the input parameters in order to obtain advanced graphical displays and numerical data for further analysis. The finalized user interface of the stand-alone application is portrayed in figure (17).

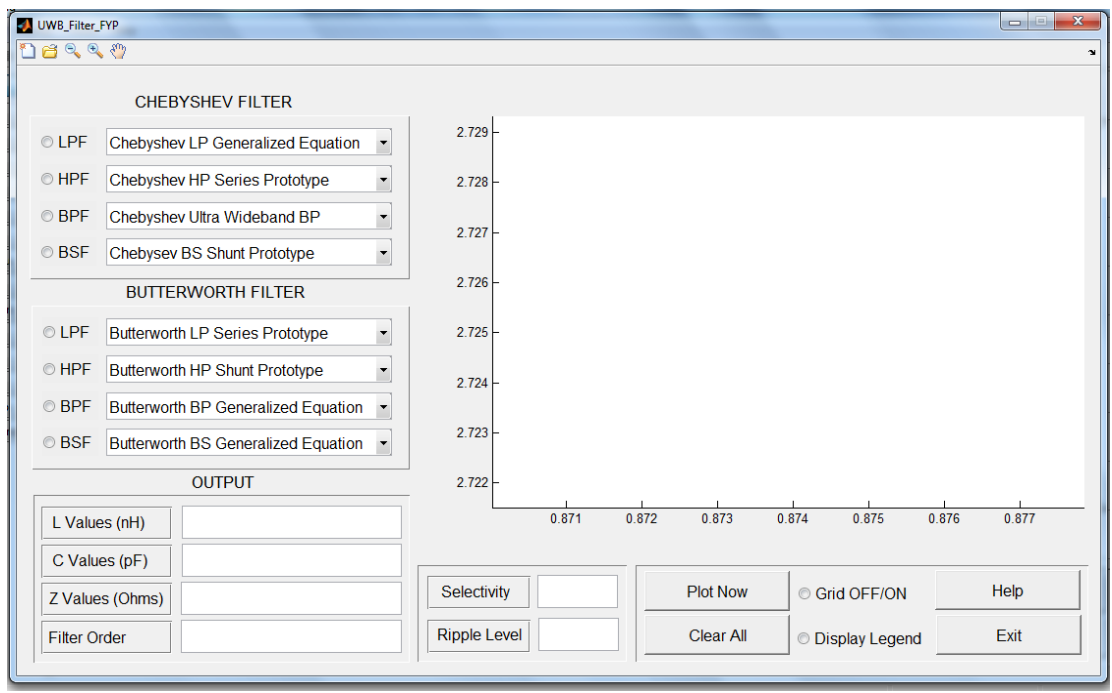

Figure 17    The user interface of the developed application for microwave filter design.

The related MatLab M-file has been used to write the program codes relating to the user interface controls (help dialog box, input dialog box, radio buttons for filter classes, exit button, save button...) of the GUI. The manipulation of this user interface controls helps in creating a dynamic artificial intelligence for this GUI that will make the user aware before making any unwanted, undesired or involuntary decision. The GUI can also inform the user through a pop-up dialog box in case the input parameters violate the design requirement.

In addition, the four classes of filters (LPF,HPF,BPF and BSF) for both Chebyshev and Butterworth filter type can be designed only one at a time. That is to say, if a radio button of Chebyshev low pass is selected, all the remaining buttons will be automatically inactive.

The finalized prototype is deployed and converted into an executable (.exe) format. The converted format can run as a stand-alone application by using the set of libraries provided in MatLab Compiler Runtime (MCR).

### 4.1.2.1  *Demonstration of microwave filter design using the GUI*

### 4.1.2.1.1  *Chebyshev Low Pass Filter*

In order to proceed to the design of Chebyshev LP filter, the user has three options (Chebyshev LP Generalized Equation, Chebyshev LP Shunt Prototype and Chebyshev Series Prototype) through the pop-up menu next to LPF under Chebyshev filter. The final step to displaying the response of the desired filter is to hit the plot button, a pop-up of input dialog box will be automatically displayed for the user to key in his/her data followed by a single click on the "OK" button of the pop-up input dialog box.

A step by step illustration of the situation described above is shown in the following figures.

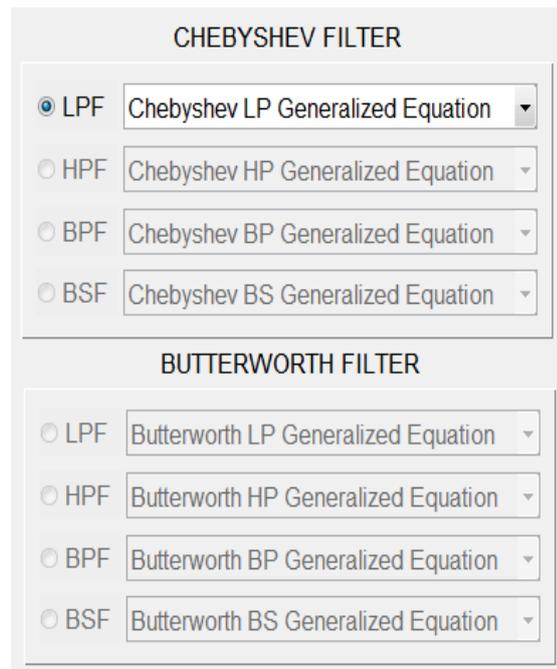

Figure 18    First Step: Select filter type and class

After hitting on the "Plot Now" button, this input dialog box will pop-up, the user will need to enter the input data in their respective field. The user will be required also to hit the OK button in order to proceed to the display of the filter's frequency response.

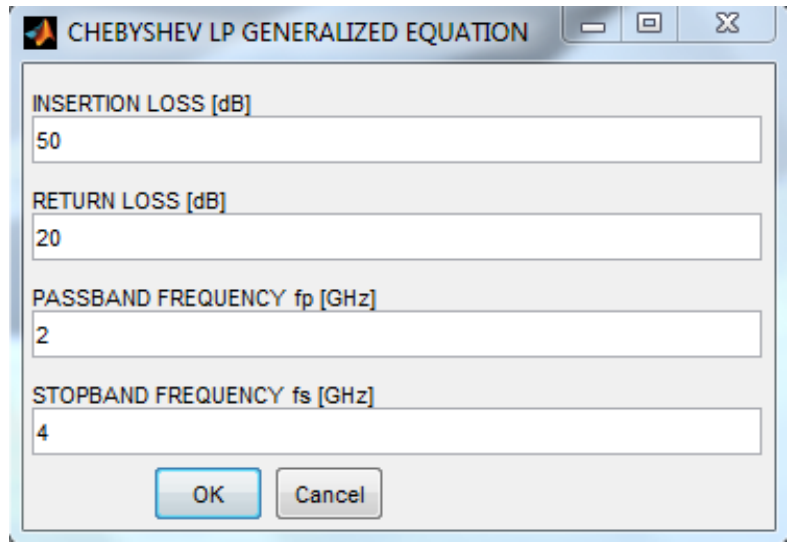

Figure 19    Second Step: Pop-up of the input dialog box, key in input data.

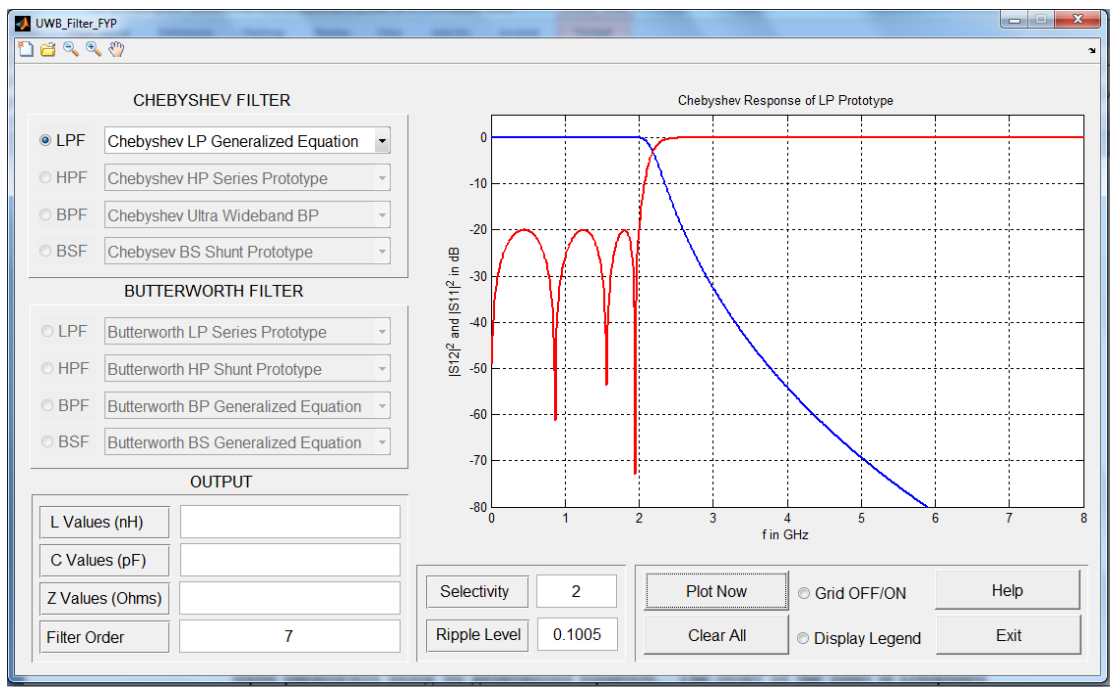

Figure 20    Final Step: Display of Frequency Response of the selected filter.

This Print Screen of the GUI shows the frequency response of the Chebyshev low pass filter with reference to the user's selection of filter type, filter class and input parameters using its generalized equation.  The order of the filter is computed using the input data and displayed in the field of "Filter Order" at the left side of the GUI; the selectivity and the ripple level are also displayed in their respective fields.

### 4.1.2.1.2  *Chebyshev High Pass Filter*

For the case of high-pass filter, responses can be generated. An example of 7th order Chebyshev HPF Series prototype is illustrated in the figures below. The first step is the same for all filter types and classes and it is similar to situation described in figure (18).

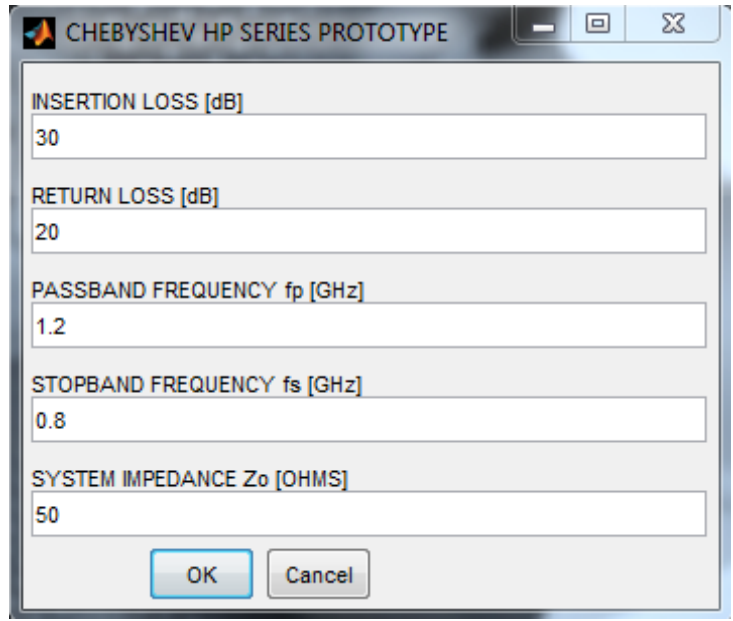

Figure 21   Second Step: Input data for Chebyshev HP Series Prototype.

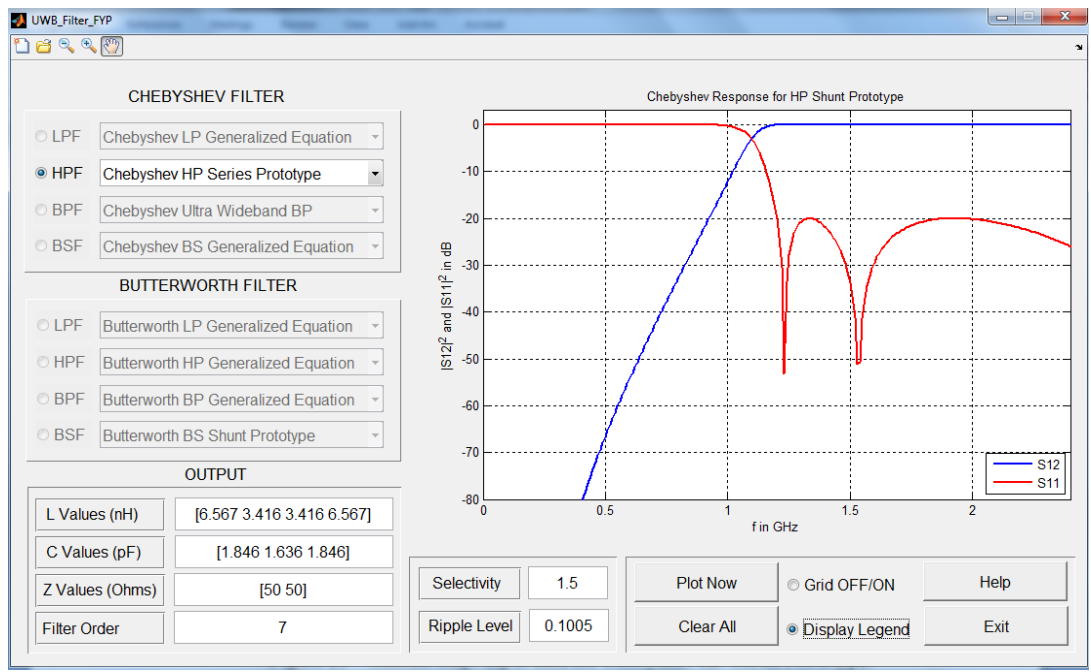

Figure 22   Final Step: Display of Frequency Response of Chebyshev HP Series

In figure (22), the corresponding values of reactive elements and the impedances (for both source and load) are displayed in their respective field. They are all displayed in array form. The user will need to identify the individual elements using this analogy:

$$L\ (nH) = [6.567\ 3.416\ 3.416\ 6.567]$$
$$= [L_1, L_3, L_5, L_7]$$

Thus $L_1 = 6.567\ nH$, $L_3 = 3.416\ nH$, $L_5 = 3.416\ nH$ and $L_7 = 6.567\ nH$.

$$C\ (pF) = [1.846\ 1.636\ 1.846]$$
$$= [C_2, C_4, C_6]$$

Hence $C_2 = 1.846\ pF$, $C_4 = 1.636\ pF$ and $C_6 = 1.846\ pF$.

And $Z = [50\ 50] = [Z_s, Z_L]$, thus $Z_s = 50\Omega$ and $Z_L = 50\Omega$

### 4.1.2.1.3 Chebyshev Band Pass Filter

As an illustration for BPF, the response and the important parameters can be obtained for capacitively coupled BPF. The following figures exemplify capacitively coupled circuit topologies.

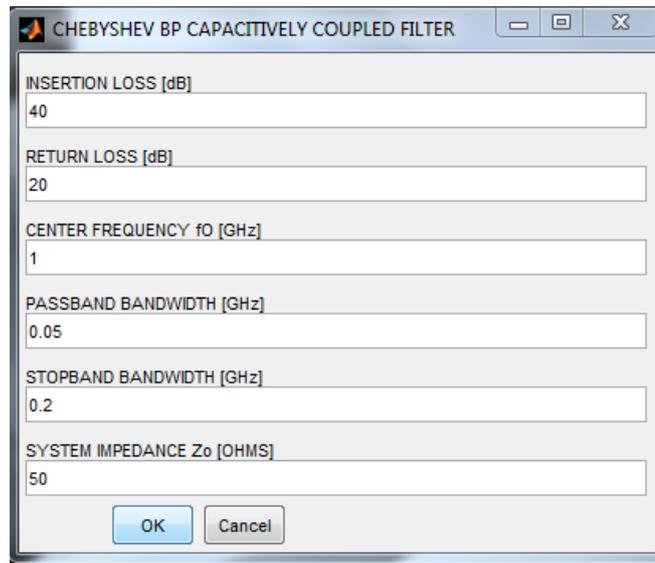

Figure 23  Second Step: Input data for Chebyshev Capacitively coupled BPF.

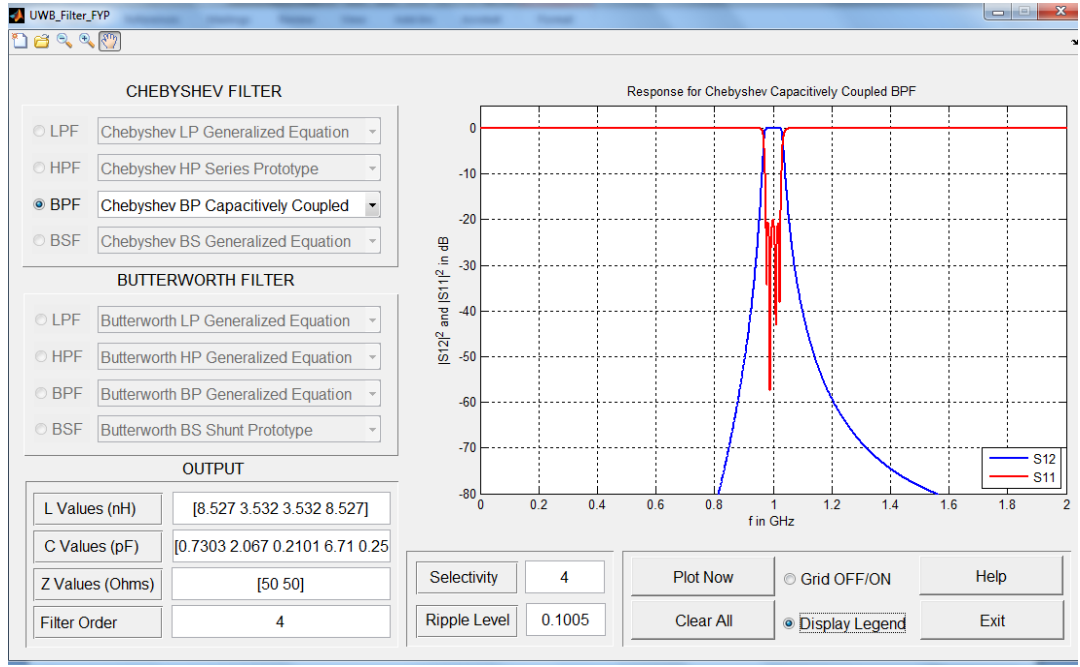

Figure 24  Final Step: Frequency Response of Chebyshev HP Series Prototype.

From figure (24), the impedances for both source and load are $Z_s = 50\Omega$ and $Z_L = 50\Omega$. The reactive elements are also portrayed in their respective field. By identification:

$$C\,(pF) = [0.7303\ 2.067\ 0.2101\ 6.71\ 0.251\ 6.71\ 0.2101\ 2.067\ 0.7303]$$
$$= [C_{01}, C_{11}, C_{12}, C_{22}, C_{23}, C_{33}, C_{34}, C_{44}, C_{45}]$$

Therefore

$$C_{01} = C_{45} = 0.7303\,pF$$
$$C_{11} = C_{44} = 2.067\,pF$$
$$C_{12} = C_{34} = 0.2101\,pF$$
$$C_{22} = C_{33} = 6.71\,pF$$
$$C_{23} = 0.251\,pF.$$

$$L\,(nH) = [8.527\ 3.532\ 3.532\ 8.527]$$
$$= [L_{11}, L_{22}, L_{33}, L_{44}]$$

Thus $L_{11} = L_{44} = 8.527\,nH$ and $L_{22} = L_{33} = 3.532\,nH$

These reactive element values, source and load impedances can be used to construct the corresponding microwave circuit for the filter.

It is also important to illustrate the case of UWB band-pass filter as it is one of the latest microwave filter topology that has been integrated for the first time and only in this filter design tool. The illustration for this case is presented as follow:

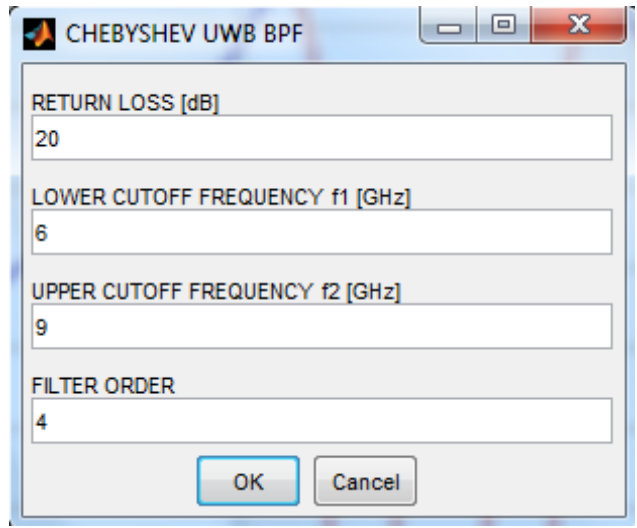

Figure 25    Second Step: Input data for Chebyshev Capacitively coupled BPF.

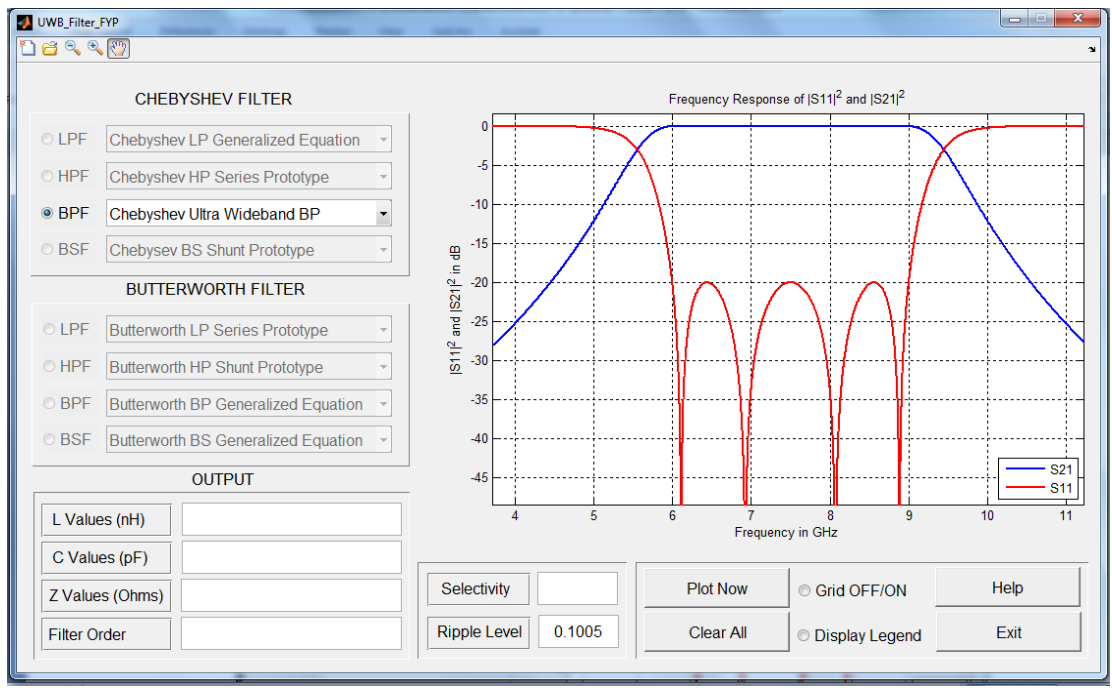

Figure 26    Final Step: Frequency Response of 4th order UWB BPF.

*4.1.2.1.4    Maximally Flat Band Pass Filter*

Likewise, when the user intend to design microwave filter using the maximally flat response, the procedures are almost the same with equal-ripple response demonstrations. For example, to design a band-stop filter using Butterworth response, the following steps are followed (assuming that the filter type and class were selected).

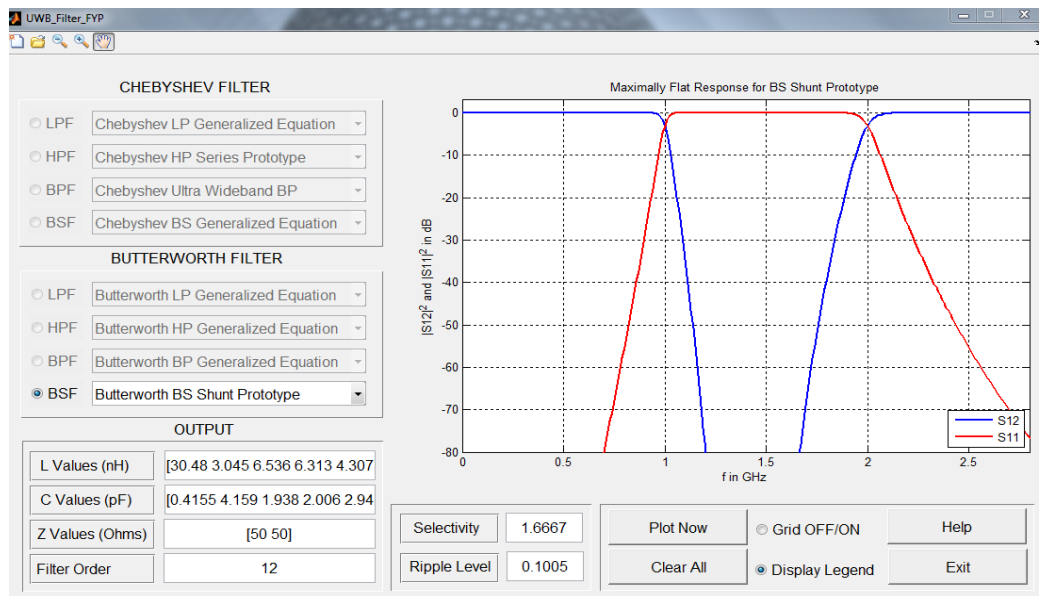

Figure 27    Second Step: Input data for Butterworth BSF.

Figure 28    Final Step: Frequency Response of Maximally flat BSF.

The individual reactive element values for the linear microwave circuit can be obtained by using this following identification.

$$L\,(nH) = [30.48\ 3.045\ 6.536\ 6.313\ 4.307\ 7.89\ 4.013\ 7.352\ 5.015\ 4.844\ 10.4\ 1.039]$$
$$= [L_1, L_2, L_3, L_4, L_5, L_6, L_7, L_8, L_9, L_{10}, L_{11}, L_{12}]$$

Thus $L_1 = 30.48\ nH$, $L_2 = 3.045\ nH$, $L_3 = 6.536\ nH$, $L_4 = 6.313\ nH$,

$L_5 = 4.307\ nH$, $L_6 = 7.89\ nH$, $L_7 = 4.013\ nH$, $L_8 = 7.352\ nH$,

$L_9 = 5.015\ nH$, $L_{10} = 4.844\ nH$, $L_{11} = 10.4\ nH$ and $L_{12} = 1.039\ nH$.

$$C(pF) = [0.4155\ 4.159\ 1.938\ 2.006\ 2.94\ 1.605\ 3.156\ 1.723\ 2.525\ 2.614\ 1.218\ 12.19]$$
$$= [C_1, C_2, C_3, C_4, C_5, C_6, C_7, C_8, C_9, C_{10}, C_{11}, C_{12}]$$

Hence $C_1 = 0.4155\ pF$, $C_2 = 4.159\ pF$, $C_3 = 1.938\ pF$, $C_4 = 2.006\ pF$,

$C_5 = 2.94\ pF$, $C_6 = 1.605\ pF$, $C_7 = 3.156\ pF$, $C_8 = 1.723\ pF$,

$C_9 = 2.525\ pF$, $C_{10} = 2.614\ pF$, $C_{11} = 1.218\ pF$ and $C_{12} = 12.19\ pF$.

And $Z = [50\ 50] = [Z_s, Z_L]$, thus $Z_s = 50\Omega$ and $Z_L = 50\Omega$

### *4.1.2.2 Hallmarks of the graphical user interface*

The major feature of this GUI is its simplicity. It is a straightforward software tool for microwave filter design that can help students and learners to grasp the concept of microwave engineering. In addition, it takes less than a minute to display an impeccable filter response that can be the basis and the framework for a scientific and technical discussion.

The key characteristics and features of the proposed GUI are presented in the table below.

Table 3  Major features of the GUI and its comparison with the existing design tools.

| Features | Classical tools | Proposed tool |
|---|---|---|
| Conventional filter topologies | ✓ | ✓ |
| Modern filter topology |  | ✓ |
| 2-D response simulation | ✓ | ✓ |
| 2-D response emulation |  | ✓ |
| Easy to use, good for sophomores |  | ✓ |
| Stand-alone capability | ✓ | ✓ |

# CHAPTER 5
# CONCLUSION & RECOMMENDATIONS

The developed graphical user interface for microwave filter design can successfully run on both computers with MatLab installed and on PCs with no MatLab software. For the latter case, the stand-alone GUI can run with the help of MatLab Compiler Runtime (MCR) that carries the necessary library required to run stand-alone MatLab applications. The GUI environment is easier for users to comprehend and familiarize themselves with. This modest GUI design can help the end-users to create impeccable and professional microwave filters even if they do not have a high level of expertise in the area of filter design tools, simulator or advanced design systems.

The proposed graphical user interface is integrated with both modern and conventional microwave filter topologies. This software tool can significantly help in reducing the training time and cost. It is the right and appropriate software tools for microwave education for both industries and educational institutions.

For future research works in this area, the graphical user interface can be developed using another approach. This approach can be directed to the use of object oriented programming such as java GUI programming, QT5 or VB.Net to create a dynamic synthesis tool for microwave filter design. From this approach, the possibility of integrating circuits and systems (such as schematic diagrams, functional blocks, nonlinear components, symbolic variables ...) can also be explored.

# APPENDICES
# APPENDIX A : COMPLETE GUI CODE

```matlab
% |======================================================================|
% | The following codes help in building the GUI for the generation of   |
% | Chebyshev (Equi-ripple) and Butterworth (Maximally flate) LP, HP, BP |
% |or BS responses for any order. These codes were developed and completed|
% |on 10th December 2013 by DJENGOMEMGOTO GERARD, for the purpose of Final|
% |Year Project under the supervision of Dr Wong Peng Wen.               |
% | Version 1.0 , December 2013 Universiti Teknologi PETRONAS.           |
% | © Copyright 2013, Universiti Teknologi PETRONAS                      |
% |======================================================================|

function varargout = MW_Filter_FYP2(varargin)
%MW_FILTER_FYP2 M-file for MW_Filter_FYP2.fig

% Last Modified by GUIDE v2.5 10-Dec-2013 11:35:30

% Begin initialization code - DO NOT EDIT
gui_Singleton = 1;
gui_State = struct('gui_Name',       mfilename, ...
                   'gui_Singleton',  gui_Singleton, ...
                   'gui_OpeningFcn', @MW_Filter_FYP2_OpeningFcn, ...
                   'gui_OutputFcn',  @MW_Filter_FYP2_OutputFcn, ...
                   'gui_LayoutFcn',  [], ...
                   'gui_Callback',   []);
if nargin && ischar(varargin{1})
   gui_State.gui_Callback = str2func(varargin{1});
end

if nargout
    [varargout{1:nargout}] = gui_mainfcn(gui_State, varargin{:});
else
    gui_mainfcn(gui_State, varargin{:});
end
% End initialization code - DO NOT EDIT

% --- Executes just before MW_Filter_FYP2 is made visible.
function MW_Filter_FYP2_OpeningFcn(hObject, eventdata, handles, varargin)
% This function has no output args, see OutputFcn.
% hObject    handle to figure
% eventdata  reserved - to be defined in a future version of MATLAB
% handles    structure with handles and user data (see GUIDATA)
% varargin   unrecognized PropertyName/PropertyValue pairs from the
%            command line (see VARARGIN)

% Choose default command line output for MW_Filter_FYP2
handles.output = hObject;

% Update handles structure
guidata(hObject, handles);

% --- Outputs from this function are returned to the command line.
function varargout = MW_Filter_FYP2_OutputFcn(hObject, eventdata, handles)
% varargout  cell array for returning output args (see VARARGOUT);

% Get default command line output from handles structure
varargout{1} = handles.output;

% --- Executes during object creation, after setting all properties.
function pushbuton_plot_CreateFcn(hObject, eventdata, handles)
% hObject    handle to pushbuton_plot (see GCBO)
% eventdata  reserved - to be defined in a future version of MATLAB
% handles    empty - handles not created until after all CreateFcns called
```

```matlab
% --- Executes on button press in pushbutton_resetAll.
function pushbutton_resetAll_Callback(hObject, eventdata, handles)

reset_button =questdlg('Are you sure you want to clear all data? Make sure to save your current data.','Clear All','Yes','No','No');

switch reset_button
    
    % if user anser is yes, then reset all data and figure.
    case 'Yes'
        set(handles.InductorValues1,'String','');      % to clear the field for Inductor values
        set(handles.CapacitorValues1,'String','');     % to clear the field for capacitor values
        set(handles.ImpedanceValues1,'String','');     % to clear the field for imepdance values
        set(handles.edit14_selectivity,'String','');   % to clear the field for selectivity
        set(handles.edit15_ripple,'String','');        % to clear the field for ripple level
        set(handles.FilterOrder,'String','');          % to clear the field for filter order
        legend('hide');                                %To clear the legend
        % Clear all the previous stored variables in the workspace
        clear all;
         % Clear all the lines currently in the MatLab Command Window
        clc;
         % this command clears the current plotted figure in the axe window.
        cla;
    % Otherwise, return to the current window with its current data and
    % figure
    case 'No'
        return
        
end

% --- Executes on button press in pushbutton_help.
function pushbutton_help_Callback(hObject, eventdata, handles)
%Providing help to users
helpdlg('To design your microwave filter, follow these three steps: 1) Select the filter type and class using the radio buttons at the top left corner.  2) Using your mouse, hit the "Plot Now" button.   3) Key in your input data in the pop-up input dialog box. Click on "ON" button of the input dialog box to validate your data entry.')

% --- Executes on button press in pushbutton_exit.
function pushbutton_exit_Callback(hObject, eventdata, handles)

% To exit the current GUI, select Yes to exit and No to return to the
% current window
exit_button =questdlg('Do you want to close this program? Save your current data before closing this application.','Exit Program','Yes','No','No');

switch exit_button
    
    case 'Yes'
        % Close all previously generated figures
        close all;
    case 'No'
        return
        
end

% --- Executes on button press in radiobutton_legend.
function radiobutton_legend_Callback(hObject, eventdata, handles)

% Getting the user's choice for either displaying the legend or not.
radiobutton_legend = get(hObject,'Value');

% Show legend if the radiobutton is selected
if radiobutton_legend == get(hObject,'Max')
    
    % Inserting legend for the rest of plots other than UWB BPF
        legend('S12','S11');
        % Moving the legend to the lower right corner of the graph
        legend('location','SouthEast');
```

```matlab
    % Hide legend otherwise.
    elseif radiobutton_legend == get(hObject,'Min')
        % Radio button is not selected-take appropriate action
        legend('hide');

end

% --- Executes on button press in radiobutton_gridOFF.
function radiobutton_gridOFF_Callback(hObject, eventdata, handles)

% Getting the user's choice for either displaying the grid or not.
radiobutton_grid = get(hObject,'Value');

% Hide grid if the radiobutton is selected
if radiobutton_grid == get(hObject,'Max')
   grid off;

%  Show grid on the figure in case the radiobutton for Grid OFF is not selected.
else
    % Radio button is not selected, show grid
    grid on;

end

% --- Executes during object creation, after setting all properties.
function ChebyshevFilter_CreateFcn(hObject, eventdata, handles)
% hObject    handle to ChebyshevFilter (see GCBO)
% eventdata  reserved - to be defined in a future version of MATLAB
% handles    empty - handles not created until after all CreateFcns called

% ------------------------------------------------------------------
function Figure_Callback(hObject, eventdata, handles)
% hObject    handle to Figure (see GCBO)
% eventdata  reserved - to be defined in a future version of MATLAB
% handles    structure with handles and user data (see GUIDATA)

% ------------------------------------------------------------------
function Tools_Callback(hObject, eventdata, handles)
% hObject    handle to Tools (see GCBO)
% eventdata  reserved - to be defined in a future version of MATLAB
% handles    structure with handles and user data (see GUIDATA)

function InductorValues1_Callback(hObject, eventdata, handles)
% hObject    handle to InductorValues1 (see GCBO)
%

% --- Executes during object creation, after setting all properties.
function InductorValues1_CreateFcn(hObject, eventdata, handles)
%       See ISPC and COMPUTER.
if ispc && isequal(get(hObject,'BackgroundColor'), get(0,'defaultUicontrolBackgroundColor'))
    set(hObject,'BackgroundColor','white');
end

function CapacitorValues1_Callback(hObject, eventdata, handles)
% hObject    handle to CapacitorValues1 (see GCBO)
% eventdata  reserved - to be defined in a future version of MATLAB
% handles    structure with handles and user data (see GUIDATA)

% --- Executes during object creation, after setting all properties.
function CapacitorValues1_CreateFcn(hObject, eventdata, handles)
% Hint: edit controls usually have a white background on Windows.
%       See ISPC and COMPUTER.
if ispc && isequal(get(hObject,'BackgroundColor'), get(0,'defaultUicontrolBackgroundColor'))
    set(hObject,'BackgroundColor','white');
end

function FilterOrder_Callback(hObject, eventdata, handles)
```

```matlab
% hObject    handle to FilterOrder (see GCBO)
% eventdata  reserved - to be defined in a future version of MATLAB
% handles    structure with handles and user data (see GUIDATA)

% --- Executes during object creation, after setting all properties.
function FilterOrder_CreateFcn(hObject, eventdata, handles)

% Hint: edit controls usually have a white background on Windows.
%       See ISPC and COMPUTER.
if ispc && isequal(get(hObject,'BackgroundColor'), get(0,'defaultUicontrolBackgroundColor'))
    set(hObject,'BackgroundColor','white');
end

% --- Executes on selection change in ButterworthLPF.
function ButterworthLPF_Callback(hObject, eventdata, handles)

% Hints: contents = cellstr(get(hObject,'String')) returns ButterworthLPF contents as cell array
%        contents{get(hObject,'Value')} returns selected item from ButterworthLPF

ButterworthLP = get(handles.ButterworthLPF,'Value'); %get currently selected option from menu

% --- Executes during object creation, after setting all properties.
function ButterworthLPF_CreateFcn(hObject, eventdata, handles)

% Hint: popupmenu controls usually have a white background on Windows.
%       See ISPC and COMPUTER.
if ispc && isequal(get(hObject,'BackgroundColor'), get(0,'defaultUicontrolBackgroundColor'))
    set(hObject,'BackgroundColor','white');
end

% --- Executes on button press in radiobtnButterLP.
function radiobtnButterLP_Callback(hObject, eventdata, handles)

% Hint: get(hObject,'Value') returns toggle state of radiobtnButterLP

if get(hObject,'Value') == 1
        set(handles.radiobtnButterLP,'Value');
      % Chebyshev Section
        set(handles.radiobtnChebHP,'Enable','off')  %disable radio button for HPF
        set(handles.ChebyshevHPF,'Enable','off')  %disable Chebyshev HP pop-up menu
        set(handles.radiobtnChebLP,'Enable','off')  %disable radio button for BPF
        set(handles.ChebyshevLPF,'Enable','off')  %disable Chebyshev BP pop-up menu
        set(handles.radiobtnChebBS,'Enable','off')  %disable radio button for BSF
        set(handles.ChebyshevBSF,'Enable','off')  %disable Chebyshev BS pop-up menu
        set(handles.radiobtnChebBP,'Enable','off')  %disable radio button for BPF
        set(handles.ChebyshevBPF,'Enable','off')  %disable Chebyshev BP pop-up menu

       % Butterworth Section
        set(handles.radiobtnButterHP,'Enable','off')  %disable radio button for HPF in Butterworth section
        set(handles.ButterworthHPF,'Enable','off')  %disable Butterworth HP pop-up menu
        set(handles.radiobtnButterBP,'Enable','off')  %disable radio button for BPF in Butterworth section
        set(handles.ButterworthBPF,'Enable','off')  %disable Butterworth BP pop-up menu
        set(handles.radiobtnButterBS,'Enable','off')  %disable radio button for BSF in Butterworth section
        set(handles.ButterworthBSF,'Enable','off')  %disable Buterworth BS pop-up menu

elseif get(hObject, 'Value')==0

       % Chebyshev Section
        set(handles.radiobtnChebHP,'Enable','on')  %Enable radio button for HPF
        set(handles.ChebyshevHPF,'Enable','on')  %Enable Chebyshev HP pop-up menu
        set(handles.radiobtnChebLP,'Enable','on')  %Enable radio button for BPF
        set(handles.ChebyshevLPF,'Enable','on')  %Enable Chebyshev BP pop-up menu
        set(handles.radiobtnChebBS,'Enable','on')  %Enable radio button for BSF
        set(handles.ChebyshevBSF,'Enable','on')  %Enable Chebyshev BS pop-up menu
        set(handles.radiobtnChebBP,'Enable','on')  %disable radio button for BPF
        set(handles.ChebyshevBPF,'Enable','on')  %disable Chebyshev BP pop-up menu

       % Butterworth Section
        set(handles.radiobtnButterHP,'Enable','on')  %Enable radio button for HPF in Butterworth section
```

```matlab
            set(handles.ButterworthHPF,'Enable','on')   %Enable Butterworth HP pop-up menu
            set(handles.radiobtnButterBP,'Enable','on')   %Enable radio button for BPF in Butterworth section
            set(handles.ButterworthBPF,'Enable','on')   %Enable Butterworth BP pop-up menu
            set(handles.radiobtnButterBS,'Enable','on')   %Enable radio button for BSF in Butterworth section
            set(handles.ButterworthBSF,'Enable','on')   %Enable Buterworth BS pop-up menu
            set(handles.radiobtnButterLP,'Enable','on')   %Enable radio button for LPF in Butterworth section
            set(handles.ButterworthLPF,'Enable','on')   %Enable Buterworth LP pop-up menu
    else
            % Chebyshev Section
            set(handles.radiobtnChebHP,'Enable','on')   %Enable radio button for HPF
            set(handles.ChebyshevHPF,'Enable','on')   %Enable Chebyshev HP pop-up menu
            set(handles.radiobtnChebLP,'Enable','on')   %Enable radio button for BPF
            set(handles.ChebyshevLPF,'Enable','on')   %Enable Chebyshev BP pop-up menu
            set(handles.radiobtnChebBS,'Enable','on')   %Enable radio button for BSF
            set(handles.ChebyshevBSF,'Enable','on')   %Enable Chebyshev BS pop-up menu
            set(handles.radiobtnChebBP,'Enable','on')   %disable radio button for BPF
            set(handles.ChebyshevBPF,'Enable','on')   %disable Chebyshev BP pop-up menu

            % Butterworth Section
            set(handles.radiobtnButterHP,'Enable','on')   %Enable radio button for HPF in Butterworth section
            set(handles.ButterworthHPF,'Enable','on')   %Enable Butterworth HP pop-up menu
            set(handles.radiobtnButterBP,'Enable','on')   %Enable radio button for BPF in Butterworth section
            set(handles.ButterworthBPF,'Enable','on')   %Enable Butterworth BP pop-up menu
            set(handles.radiobtnButterBS,'Enable','on')   %Enable radio button for BSF in Butterworth section
            set(handles.ButterworthBSF,'Enable','on')   %Enable Buterworth BS pop-up menu
            set(handles.radiobtnButterLP,'Enable','on')   %Enable radio button for LPF in Butterworth section
            set(handles.ButterworthLPF,'Enable','on')   %Enable Butterworth LP pop-up menu
end

% --- Executes on selection change in ButterworthBPF.
function ButterworthBPF_Callback(hObject, eventdata, handles)
% Hints: contents = cellstr(get(hObject,'String')) returns ButterworthBPF contents as cell array
%        contents{get(hObject,'Value')} returns selected item from ButterworthBPF

ButterworthBP = get(handles.ButterworthBPF,'Value'); %get currently selected option from menu

% --- Executes during object creation, after setting all properties.
function ButterworthBPF_CreateFcn(hObject, eventdata, handles)
% Hint: popupmenu controls usually have a white background on Windows.
%       See ISPC and COMPUTER.
if ispc && isequal(get(hObject,'BackgroundColor'), get(0,'defaultUicontrolBackgroundColor'))
    set(hObject,'BackgroundColor','white');
end

% --- Executes on button press in radiobtnButterBP.
function radiobtnButterBP_Callback(hObject, eventdata, handles)

if get(hObject,'Value') == 1
        set(handles.radiobtnButterBP,'Value');
      % Chebyshev Section
        set(handles.radiobtnChebHP,'Enable','off')   %disable radio button for HPF
        set(handles.ChebyshevHPF,'Enable','off')   %disable Chebyshev HP pop-up menu
        set(handles.radiobtnChebLP,'Enable','off')   %disable radio button for BPF
        set(handles.ChebyshevLPF,'Enable','off')   %disable Chebyshev BP pop-up menu
        set(handles.radiobtnChebBS,'Enable','off')   %disable radio button for BSF
        set(handles.ChebyshevBSF,'Enable','off')   %disable Chebyshev BS pop-up menu
        set(handles.radiobtnChebBP,'Enable','off')   %disable radio button for BPF
        set(handles.ChebyshevBPF,'Enable','off')   %disable Chebyshev BP pop-up menu

       % Butterworth Section
        set(handles.radiobtnButterHP,'Enable','off')   %disable radio button for HPF in Butterworth section
        set(handles.ButterworthHPF,'Enable','off')   %disable Butterworth HP pop-up menu
        set(handles.radiobtnButterLP,'Enable','off')   %disable radio button for LPF in Butterworth section
        set(handles.ButterworthLPF,'Enable','off')   %disable Butterworth LP pop-up menu
        set(handles.radiobtnButterBS,'Enable','off')   %disable radio button for BSF in Butterworth section
        set(handles.ButterworthBSF,'Enable','off')   %disable Buterworth BS pop-up menu

elseif get(hObject, 'Value')==0
```

```matlab
        % Chebyshev Section
        set(handles.radiobtnChebHP,'Enable','on')  %Enable radio button for HPF
        set(handles.ChebyshevHPF,'Enable','on')  %Enable Chebyshev HP pop-up menu
        set(handles.radiobtnChebLP,'Enable','on')  %Enable radio button for BPF
        set(handles.ChebyshevLPF,'Enable','on')  %Enable Chebyshev BP pop-up menu
        set(handles.radiobtnChebBS,'Enable','on')  %Enable radio button for BSF
        set(handles.ChebyshevBSF,'Enable','on')  %Enable Chebyshev BS pop-up menu
        set(handles.radiobtnChebBP,'Enable','on')  %disable radio button for BPF
        set(handles.ChebyshevBPF,'Enable','on')  %disable Chebyshev BP pop-up menu

         % Butterworth Section
         set(handles.radiobtnButterHP,'Enable','on')  %Enable radio button for HPF in Butterworth section
         set(handles.ButterworthHPF,'Enable','on')  %Enable Butterworth HP pop-up menu
         set(handles.radiobtnButterLP,'Enable','on')  %Enable radio button for LPF in Butterworth section
         set(handles.ButterworthLPF,'Enable','on')  %Enable Butterworth LP pop-up menu
         set(handles.radiobtnButterBS,'Enable','on')   %Enable radio button for BSF in Butterworth section
         set(handles.ButterworthBSF,'Enable','on')  %Enable Buterworth BS pop-up menu
         set(handles.radiobtnButterBP,'Enable','on')   %Enable radio button for BPF in Butterworth section
         set(handles.ButterworthBPF,'Enable','on')  %Enable Buterworth BP pop-up menu
else
end

% --- Executes on selection change in ButterworthHPF.
function ButterworthHPF_Callback(hObject, eventdata, handles)

ButterworthHP = get(handles.ButterworthHPF,'Value'); %get currently selected option from menu

% --- Executes during object creation, after setting all properties.
function ButterworthHPF_CreateFcn(hObject, eventdata, handles)

if ispc && isequal(get(hObject,'BackgroundColor'), get(0,'defaultUicontrolBackgroundColor'))
    set(hObject,'BackgroundColor','white');
end

% --- Executes on button press in radiobtnButterHP.
function radiobtnButterHP_Callback(hObject, eventdata, handles)

if get(hObject,'Value') == 1
        set(handles.radiobtnButterHP,'Value');
      % Chebyshev Section
        set(handles.radiobtnChebHP,'Enable','off')  %disable radio button for HPF
        set(handles.ChebyshevHPF,'Enable','off')  %disable Chebyshev HP pop-up menu
        set(handles.radiobtnChebLP,'Enable','off')  %disable radio button for BPF
        set(handles.ChebyshevLPF,'Enable','off')  %disable Chebyshev BP pop-up menu
        set(handles.radiobtnChebBS,'Enable','off')  %disable radio button for BSF
        set(handles.ChebyshevBSF,'Enable','off')  %disable Chebyshev BS pop-up menu
        set(handles.radiobtnChebBP,'Enable','off')  %disable radio button for BPF
        set(handles.ChebyshevBPF,'Enable','off')  %disable Chebyshev BP pop-up menu

         % Butterworth Section
         set(handles.radiobtnButterHP,'Enable','on')  %Enable radio button for HPF in Butterworth section
         set(handles.ButterworthHPF,'Enable','on')  %Enable Buterworth HP pop-up menu
         set(handles.radiobtnButterLP,'Enable','off')   %disable radio button for LPF in Butterworth section
         set(handles.ButterworthLPF,'Enable','off')  %disable Butterworth LP pop-up menu
         set(handles.radiobtnButterBP,'Enable','off')   %disable radio button for BPF in Butterworth section
         set(handles.ButterworthBPF,'Enable','off')  %disable Butterworth BP pop-up menu
         set(handles.radiobtnButterBS,'Enable','off')   %disable radio button for BSF in Butterworth section
         set(handles.ButterworthBSF,'Enable','off')  %disable Buterworth BS pop-up menu

    elseif get(hObject, 'Value')==0

         % Chebyshev Section
        set(handles.radiobtnChebHP,'Enable','on')  %Enable radio button for HPF
        set(handles.ChebyshevHPF,'Enable','on')  %Enable Chebyshev HP pop-up menu
        set(handles.radiobtnChebLP,'Enable','on')  %Enable radio button for BPF
        set(handles.ChebyshevLPF,'Enable','on')  %Enable Chebyshev BP pop-up menu
        set(handles.radiobtnChebBS,'Enable','on')  %Enable radio button for BSF
        set(handles.ChebyshevBSF,'Enable','on')  %Enable Chebyshev BS pop-up menu
        set(handles.radiobtnChebBP,'Enable','on')  %disable radio button for BPF
```

```matlab
            set(handles.ChebyshevBPF,'Enable','on')  %disable Chebyshev BP pop-up menu

        % Butterworth Section
            set(handles.radiobtnButterLP,'Enable','on')  %Enable radio button for LPF in Butterworth section
            set(handles.ButterworthLPF,'Enable','on')  %Enable Butterworth LP pop-up menu
            set(handles.radiobtnButterBP,'Enable','on')  %Enable radio button for BPF in Butterworth section
            set(handles.ButterworthBPF,'Enable','on')  %Enable Butterworth BP pop-up menu
            set(handles.radiobtnButterBS,'Enable','on')  %Enable radio button for BSF in Butterworth section
            set(handles.ButterworthBSF,'Enable','on')  %Enable Buterworth BS pop-up menu
else
end

% --- Executes on selection change in ButterworthBSF.
function ButterworthBSF_Callback(hObject, eventdata, handles)

ButterworthBS = get(handles.ButterworthBSF,'Value'); %get currently selected option from menu

% --- Executes during object creation, after setting all properties.
function ButterworthBSF_CreateFcn(hObject, eventdata, handles)

if ispc && isequal(get(hObject,'BackgroundColor'), get(0,'defaultUicontrolBackgroundColor'))
    set(hObject,'BackgroundColor','white');
end

% --- Executes on button press in radiobtnButterBS.
function radiobtnButterBS_Callback(hObject, eventdata, handles)

if get(hObject,'Value') == 1
        set(handles.radiobtnButterBS,'Value');
       % Chebyshev Section
        set(handles.radiobtnChebHP,'Enable','off')  %disable radio button for HPF
        set(handles.ChebyshevHPF,'Enable','off')  %disable Chebyshev HP pop-up menu
        set(handles.radiobtnChebLP,'Enable','off')  %disable radio button for BPF
        set(handles.ChebyshevLPF,'Enable','off')  %disable Chebyshev BP pop-up menu
        set(handles.radiobtnChebBS,'Enable','off')  %disable radio button for BSF
        set(handles.ChebyshevBSF,'Enable','off')  %disable Chebyshev BS pop-up menu
        set(handles.radiobtnChebBP,'Enable','off')  %disable radio button for BPF
        set(handles.ChebyshevBPF,'Enable','off')  %disable Chebyshev BP pop-up menu

         % Butterworth Section
        set(handles.radiobtnButterHP,'Enable','off')  %disable radio button for HPF in Butterworth section
        set(handles.ButterworthHPF,'Enable','off')  %disable Butterworth HP pop-up menu
        set(handles.radiobtnButterBP,'Enable','off')  %disable radio button for BPF in Butterworth section
        set(handles.ButterworthBPF,'Enable','off')  %disable Butterworth BP pop-up menu
        set(handles.radiobtnButterLP,'Enable','off')  %disable radio button for LPF in Butterworth section
        set(handles.ButterworthLPF,'Enable','off')  %disable Buterworth LP pop-up menu

elseif get(hObject, 'Value')==0

       % Chebyshev Section
        set(handles.radiobtnChebHP,'Enable','on')  %Enable radio button for HPF
        set(handles.ChebyshevHPF,'Enable','on')  %Enable Chebyshev HP pop-up menu
        set(handles.radiobtnChebLP,'Enable','on')  %Enable radio button for BPF
        set(handles.ChebyshevLPF,'Enable','on')  %Enable Chebyshev BP pop-up menu
        set(handles.radiobtnChebBS,'Enable','on')  %Enable radio button for BSF
        set(handles.ChebyshevBSF,'Enable','on')  %Enable Chebyshev BS pop-up menu
        set(handles.radiobtnChebBP,'Enable','on')  %disable radio button for BPF
        set(handles.ChebyshevBPF,'Enable','on')  %disable Chebyshev BP pop-up menu

       % Butterworth Section
        set(handles.radiobtnButterHP,'Enable','on')  %Enable radio button for HPF in Butterworth section
        set(handles.ButterworthHPF,'Enable','on')  %Enable Butterworth HP pop-up menu
        set(handles.radiobtnButterBP,'Enable','on')  %Enable radio button for BPF in Butterworth section
        set(handles.ButterworthBPF,'Enable','on')  %Enable Butterworth BP pop-up menu
        set(handles.radiobtnButterLP,'Enable','on')  %Enable radio button for LPF in Butterworth section
        set(handles.ButterworthLPF,'Enable','on')  %Enable Buterworth LP pop-up menu
        set(handles.radiobtnButterBS,'Enable','on')  %Enable radio button for BSF in Butterworth section
        set(handles.ButterworthBSF,'Enable','on')  %Enable Buterworth BS pop-up menu
else
```

```matlab
end

% --- Executes on selection change in ChebyshevLPF.
function ChebyshevLPF_Callback(hObject, eventdata, handles)

ChebyshevLP = get(handles.ChebyshevLPF,'Value'); %get currently selected option from menu

% --- Executes during object creation, after setting all properties.
function ChebyshevLPF_CreateFcn(hObject, eventdata, handles)

if ispc && isequal(get(hObject,'BackgroundColor'), get(0,'defaultUicontrolBackgroundColor'))
    set(hObject,'BackgroundColor','white');
end

% --- Executes on button press in radiobtnChebLP.
function radiobtnChebLP_Callback(hObject, eventdata, handles)

if get(hObject,'Value') == 1
        set(handles.radiobtnChebLP,'Value');
       % Chebyshev Section
        set(handles.radiobtnChebHP,'Enable','off')  %disable radio button for HPF
        set(handles.ChebyshevHPF,'Enable','off')   %disable Chebyshev HP pop-up menu
        set(handles.radiobtnChebBP,'Enable','off')  %disable radio button for BPF
        set(handles.ChebyshevBPF,'Enable','off')   %disable Chebyshev BP pop-up menu
        set(handles.radiobtnChebBS,'Enable','off')  %disable radio button for BSF
        set(handles.ChebyshevBSF,'Enable','off')   %disable Chebyshev BS pop-up menu
        set(handles.radiobtnChebBS,'Enable','off')  %disable radio button for BSF
        % Butterworth Section
        set(handles.radiobtnButterLP,'Enable','off')  %disable radio button for LPF in Butterworth section
        set(handles.ButterworthLPF,'Enable','off')   %disable Butterworth LP pop-up menu
        set(handles.radiobtnButterHP,'Enable','off')  %disable radio button for HPF in Butterworth section
        set(handles.ButterworthHPF,'Enable','off')   %disable Butterworth HP pop-up menu
        set(handles.radiobtnButterBP,'Enable','off')  %disable radio button for BPF in Butterworth section
        set(handles.ButterworthBPF,'Enable','off')   %disable Butterworth BP pop-up menu
        set(handles.radiobtnButterBS,'Enable','off')  %disable radio button for BSF in Butterworth section
        set(handles.ButterworthBSF,'Enable','off')   %disable Buterworth BS pop-up menu

else
        % Chebyshev Section
        set(handles.radiobtnChebLP,'Enable','on')  %Enable radio button for LPF
        set(handles.ChebyshevLPF,'Enable','on')   %Enable Chebyshev LP pop-up menu
        set(handles.radiobtnChebHP,'Enable','on')  %Enable radio button for HPF
        set(handles.ChebyshevHPF,'Enable','on')   %Enable Chebyshev HP pop-up menu
        set(handles.radiobtnChebBP,'Enable','on')  %Enable radio button for BPF
        set(handles.ChebyshevBPF,'Enable','on')   %Enable Chebyshev BP pop-up menu
        set(handles.radiobtnChebBS,'Enable','on')  %Enable radio button for BSF
        set(handles.ChebyshevBSF,'Enable','on')   %Enable Chebyshev BS pop-up menu
        set(handles.radiobtnChebBS,'Enable','on')  %Enable radio button for BSF
         % Butterworth Section
        set(handles.radiobtnButterLP,'Enable','on')  %Enable radio button for LPF in Butterworth section
        set(handles.ButterworthLPF,'Enable','on')   %Enable Butterworth LP pop-up menu
        set(handles.radiobtnButterHP,'Enable','on')  %Enable radio button for HPF in Butterworth section
        set(handles.ButterworthHPF,'Enable','on')   %Enable Butterworth HP pop-up menu
        set(handles.radiobtnButterBP,'Enable','on')  %Enable radio button for BPF in Butterworth section
        set(handles.ButterworthBPF,'Enable','on')   %Enable Butterworth BP pop-up menu
        set(handles.radiobtnButterBS,'Enable','on')  %Enable radio button for BSF in Butterworth section
        set(handles.ButterworthBSF,'Enable','on')   %Enable Buterworth BS pop-up menu
end
% --- Executes on selection change in ChebyshevBPF.
function ChebyshevBPF_Callback(hObject, eventdata, handles)

ChebyshevBP = get(handles.ChebyshevBPF,'Value'); %get currently selected option from menu

% --- Executes during object creation, after setting all properties.
function ChebyshevBPF_CreateFcn(hObject, eventdata, handles)

if ispc && isequal(get(hObject,'BackgroundColor'), get(0,'defaultUicontrolBackgroundColor'))
    set(hObject,'BackgroundColor','white');
end
```

```matlab
% --- Executes on button press in radiobtnChebBP.
function radiobtnChebBP_Callback(hObject, eventdata, handles)

if get(hObject,'Value') == 1
         set(handles.radiobtnChebBP,'Value');
       % Chebyshev Section
         set(handles.radiobtnChebHP,'Enable','off')  %disable radio button for HPF
         set(handles.ChebyshevHPF,'Enable','off')   %disable Chebyshev HP pop-up menu
         set(handles.radiobtnChebLP,'Enable','off')  %disable radio button for LPF
         set(handles.ChebyshevLPF,'Enable','off')   %disable Chebyshev LP pop-up menu
         set(handles.radiobtnChebBS,'Enable','off')  %disable radio button for BSF
         set(handles.ChebyshevBSF,'Enable','off')   %disable Chebyshev BS pop-up menu

       % Butterworth Section
         set(handles.radiobtnButterLP,'Enable','off')  %disable radio button for LPF in Butterworth section
         set(handles.ButterworthLPF,'Enable','off')   %disable Butterworth LP pop-up menu
         set(handles.radiobtnButterHP,'Enable','off')  %disable radio button for HPF in Butterworth section
         set(handles.ButterworthHPF,'Enable','off')   %disable Butterworth HP pop-up menu
         set(handles.radiobtnButterBP,'Enable','off')  %disable radio button for BPF in Butterworth section
         set(handles.ButterworthBPF,'Enable','off')   %disable Butterworth BP pop-up menu
         set(handles.radiobtnButterBS,'Enable','off')  %disable radio button for BSF in Butterworth section
         set(handles.ButterworthBSF,'Enable','off')   %disable Buterworth BS pop-up menu

elseif get(hObject, 'Value')==0

       % Chebyshev Section
          set(handles.radiobtnChebBP,'Enable','on')  %Enable radio button for BPF
         set(handles.ChebyshevBPF,'Enable','on')   %Enable Chebyshev BP pop-up menu
         set(handles.radiobtnChebHP,'Enable','on')  %Enable radio button for HPF
         set(handles.ChebyshevHPF,'Enable','on')   %Enable Chebyshev HP pop-up menu
         set(handles.radiobtnChebLP,'Enable','on')  %Enable radio button for LPF
         set(handles.ChebyshevLPF,'Enable','on')   %Enable Chebyshev LP pop-up menu
         set(handles.radiobtnChebBS,'Enable','on')  %Enable radio button for BSF
         set(handles.ChebyshevBSF,'Enable','on')   %Enable Chebyshev BS pop-up menu

       % Butterworth Section
         set(handles.radiobtnButterLP,'Enable','on')  %Enable radio button for LPF in Butterworth section
         set(handles.ButterworthLPF,'Enable','on')   %Enable Butterworth LP pop-up menu
         set(handles.radiobtnButterHP,'Enable','on')  %Enable radio button for HPF in Butterworth section
         set(handles.ButterworthHPF,'Enable','on')   %Enable Butterworth HP pop-up menu
         set(handles.radiobtnButterBP,'Enable','on')  %Enable radio button for BPF in Butterworth section
         set(handles.ButterworthBPF,'Enable','on')   %Enable Butterworth BP pop-up menu
         set(handles.radiobtnButterBS,'Enable','on')  %Enable radio button for BSF in Butterworth section
         set(handles.ButterworthBSF,'Enable','on')   %Enable Buterworth BS pop-up menu
else
end

% --- Executes on selection change in ChebyshevHPF.
function ChebyshevHPF_Callback(hObject, eventdata, handles)

ChebyshevHP = get(handles.ChebyshevHPF,'Value'); %get currently selected option from menu

% --- Executes during object creation, after setting all properties.
function ChebyshevHPF_CreateFcn(hObject, eventdata, handles)
if ispc && isequal(get(hObject,'BackgroundColor'), get(0,'defaultUicontrolBackgroundColor'))
    set(hObject,'BackgroundColor','white');
end

% --- Executes on button press in radiobtnChebHP.
function radiobtnChebHP_Callback(hObject, eventdata, handles)

if get(hObject,'Value') == 1
         set(handles.radiobtnChebHP,'Value');
       % Chebyshev Section
         set(handles.radiobtnChebLP,'Enable','off')  %disable radio button for HPF
         set(handles.ChebyshevLPF,'Enable','off')   %disable Chebyshev HP pop-up menu
         set(handles.radiobtnChebBP,'Enable','off')  %disable radio button for BPF
         set(handles.ChebyshevBPF,'Enable','off')   %disable Chebyshev BP pop-up menu
```

```matlab
            set(handles.radiobtnChebBS,'Enable','off')   %disable radio button for BSF
            set(handles.ChebyshevBSF,'Enable','off')   %disable Chebyshev BS pop-up menu
            set(handles.radiobtnChebBS,'Enable','off')   %disable radio button for BSF
          % Butterworth Section
            set(handles.radiobtnButterLP,'Enable','off')   %disable radio button for LPF in Butterworth section
            set(handles.ButterworthLPF,'Enable','off')   %disable Butterworth LP pop-up menu
            set(handles.radiobtnButterHP,'Enable','off')   %disable radio button for HPF in Butterworth section
            set(handles.ButterworthHPF,'Enable','off')   %disable Butterworth HP pop-up menu
            set(handles.radiobtnButterBP,'Enable','off')   %disable radio button for BPF in Butterworth section
            set(handles.ButterworthBPF,'Enable','off')   %disable Butterworth BP pop-up menu
            set(handles.radiobtnButterBS,'Enable','off')   %disable radio button for BSF in Butterworth section
            set(handles.ButterworthBSF,'Enable','off')   %disable Buterworth BS pop-up menu

    elseif get(hObject, 'Value')==0

           % Chebyshev Section
            set(handles.radiobtnChebHP,'Enable','on')   %Enable radio button for HPF
            set(handles.ChebyshevHPF,'Enable','on')   %Enable Chebyshev HP pop-up menu
            set(handles.radiobtnChebLP,'Enable','on')   %Enable radio button for LPF
            set(handles.ChebyshevLPF,'Enable','on')   %Enable Chebyshev LP pop-up menu
            set(handles.radiobtnChebBP,'Enable','on')   %Enable radio button for BPF
            set(handles.ChebyshevBPF,'Enable','on')   %Enable Chebyshev BP pop-up menu
            set(handles.radiobtnChebBS,'Enable','on')   %Enable radio button for BSF
            set(handles.ChebyshevBSF,'Enable','on')   %Enable Chebyshev BS pop-up menu
            set(handles.radiobtnChebBS,'Enable','on')   %Enable radio button for BSF
          % Butterworth Section
            set(handles.radiobtnButterLP,'Enable','on')   %Enable radio button for LPF in Butterworth section
            set(handles.ButterworthLPF,'Enable','on')   %Enable Butterworth LP pop-up menu
            set(handles.radiobtnButterHP,'Enable','on')   %Enable radio button for HPF in Butterworth section
            set(handles.ButterworthHPF,'Enable','on')   %Enable Butterworth HP pop-up menu
            set(handles.radiobtnButterBP,'Enable','on')   %Enable radio button for BPF in Butterworth section
            set(handles.ButterworthBPF,'Enable','on')   %Enable Butterworth BP pop-up menu
            set(handles.radiobtnButterBS,'Enable','on')   %Enable radio button for BSF in Butterworth section
            set(handles.ButterworthBSF,'Enable','on')   %Enable Buterworth BS pop-up menu
    else
    end

% --- Executes on selection change in ChebyshevBSF.
function ChebyshevBSF_Callback(hObject, eventdata, handles)

ChebyshevBS = get(handles.ChebyshevBSF,'Value'); %get currently selected option from menu

% --- Executes during object creation, after setting all properties.
function ChebyshevBSF_CreateFcn(hObject, eventdata, handles)

if ispc && isequal(get(hObject,'BackgroundColor'), get(0,'defaultUicontrolBackgroundColor'))
    set(hObject,'BackgroundColor','white');
end

% --- Executes on button press in radiobtnChebBS.
function radiobtnChebBS_Callback(hObject, eventdata, handles)

if get(hObject,'Value') == 1
        set(handles.radiobtnChebBS,'Value');
       % Chebyshev Section
        set(handles.radiobtnChebHP,'Enable','off')   %disable radio button for HPF
        set(handles.ChebyshevHPF,'Enable','off')   %disable Chebyshev HP pop-up menu
        set(handles.radiobtnChebLP,'Enable','off')   %disable radio button for LPF
        set(handles.ChebyshevLPF,'Enable','off')   %disable Chebyshev LP pop-up menu
        set(handles.radiobtnChebBP,'Enable','off')   %disable radio button for BSF
        set(handles.ChebyshevBPF,'Enable','off')   %disable Chebyshev BP pop-up menu

        % Butterworth Section
        set(handles.radiobtnButterLP,'Enable','off')   %disable radio button for LPF in Butterworth section
        set(handles.ButterworthLPF,'Enable','off')   %disable Butterworth LP pop-up menu
        set(handles.radiobtnButterHP,'Enable','off')   %disable radio button for HPF in Butterworth section
        set(handles.ButterworthHPF,'Enable','off')   %disable Butterworth HP pop-up menu
        set(handles.radiobtnButterBP,'Enable','off')   %disable radio button for BPF in Butterworth section
        set(handles.ButterworthBPF,'Enable','off')   %disable Butterworth BP pop-up menu
```

```matlab
            set(handles.radiobtnButterBS,'Enable','off')  %disable radio button for BSF in Butterworth section
            set(handles.ButterworthBSF,'Enable','off')  %disable Buterworth BS pop-up menu

elseif get(hObject, 'Value')==0

             % Chebyshev Section
            set(handles.radiobtnChebBS,'Enable','on')  %Enable radio button for BSF
            set(handles.ChebyshevBSF,'Enable','on')  %Enable Chebyshev BS pop-up menu
            set(handles.radiobtnChebHP,'Enable','on')  %Enable radio button for HPF
            set(handles.ChebyshevHPF,'Enable','on')  %Enable Chebyshev HP pop-up menu
            set(handles.radiobtnChebLP,'Enable','on')  %Enable radio button for LPF
            set(handles.ChebyshevLPF,'Enable','on')  %Enable Chebyshev LP pop-up menu
            set(handles.radiobtnChebBP,'Enable','on')  %Enable radio button for BSF
            set(handles.ChebyshevBPF,'Enable','on')  %Enable Chebyshev BP pop-up menu
            set(handles.radiobtnChebBP,'Enable','on')  %Enable radio button for BPF
         % Butterworth Section
            set(handles.radiobtnButterLP,'Enable','on')  %Enable radio button for LPF in Butterworth section
            set(handles.ButterworthLPF,'Enable','on')  %Enable Butterworth LP pop-up menu
            set(handles.radiobtnButterHP,'Enable','on')  %Enable radio button for HPF in Butterworth section
            set(handles.ButterworthHPF,'Enable','on')  %Enable Butterworth HP pop-up menu
            set(handles.radiobtnButterBP,'Enable','on')  %Enable radio button for BPF in Butterworth section
            set(handles.ButterworthBPF,'Enable','on')  %Enable Butterworth BP pop-up menu
            set(handles.radiobtnButterBS,'Enable','on')  %Enable radio button for BSF in Butterworth section
            set(handles.ButterworthBSF,'Enable','on')  %Enable Buterworth BS pop-up menu
else
end

% --------------------------------------------------------------------

%For the display of slectivity value
function edit14_selectivity_Callback(hObject, eventdata, handles)

% Hints: get(hObject,'String') returns contents of edit14_selectivity as text
%        str2double(get(hObject,'String')) returns contents of edit14_selectivity as a double

% --- Executes during object creation, after setting all properties.
function edit14_selectivity_CreateFcn(hObject, eventdata, handles)

if ispc && isequal(get(hObject,'BackgroundColor'), get(0,'defaultUicontrolBackgroundColor'))
    set(hObject,'BackgroundColor','white');
end

%To display the ripple factor
function edit15_ripple_Callback(hObject, eventdata, handles)

% Hints: get(hObject,'String') returns contents of edit15_ripple as text
%        str2double(get(hObject,'String')) returns contents of edit15_ripple as a double

% --- Executes during object creation, after setting all properties.
function edit15_ripple_CreateFcn(hObject, eventdata, handles)

if ispc && isequal(get(hObject,'BackgroundColor'), get(0,'defaultUicontrolBackgroundColor'))
    set(hObject,'BackgroundColor','white');
end

% --- Executes on button press in radiobutton_gridOFF.
function radiobutton__Callback(hObject, eventdata, handles)
% hObject    handle to radiobutton_gridOFF (see GCBO)
% eventdata  reserved - to be defined in a future version of MATLAB
% handles    structure with handles and user data (see GUIDATA)
% Hint: get(hObject,'Value') returns toggle state of radiobutton_gridOFF

% --- Executes on button press in pushbutton_plot.
function pushbutton_plot_Callback(hObject, eventdata, handles)
ChebyshevLP = get(handles.ChebyshevLPF,'Value'); %getting currently selected option from menu for Chebyshev LP
ChebyshevHP = get(handles.ChebyshevHPF,'Value'); %getting currently selected option from menu for Chebyshev HP
ChebyshevBP = get(handles.ChebyshevBPF,'Value'); %getting currently selected option from menu for Chebyshev BP
ChebyshevBS = get(handles.ChebyshevBSF,'Value'); %getting currently selected option from menu for Chebyshev BS
```

```matlab
ButterworthLP = get(handles.ButterworthLPF,'Value'); %getting currently selected option from menu for Butterworth LP
ButterworthHP = get(handles.ButterworthHPF,'Value'); %getting currently selected option from menu for Butterworth HP
ButterworthBP = get(handles.ButterworthBPF,'Value'); %getting currently selected option from menu for Butterworth BP
ButterworthBS = get(handles.ButterworthBSF,'Value'); %getting currently selected option from menu for Butterworth BS
selected_radio1 = get(handles.radiobtnChebLP,'Value'); %getting currently selected option from radio button for Chebyshev LP
selected_radio2 = get(handles.radiobtnChebHP,'Value'); %getting currently selected option from radio button for Chebyshev HP
selected_radio3 = get(handles.radiobtnChebBP,'Value'); %getting currently selected option from radio button for Chebyshev BP
selected_radio4 = get(handles.radiobtnChebBS,'Value');  %getting currently selected option from radio button for Chebyshev BS
selected_radio5 = get(handles.radiobtnButterLP,'Value'); % getting currently selected option from radio button for Butterworth LP
selected_radio6 = get(handles.radiobtnButterHP,'Value'); % getting currently selected option from radio button for Butterworth HP
selected_radio7 = get(handles.radiobtnButterBP,'Value'); % getting currently selected option from radio button for Butterworth BP
selected_radio8 = get(handles.radiobtnButterBS,'Value'); % getting currently selected option from radio button for Butterworth BS

% LP radio button is selected

if selected_radio1 ==1

if ChebyshevLP == 1     % Chebyshev LP from its Generalized Equation

    % Generation an input dialog box for Chebyshev LP Generalized Equation
    prompt={'INSERTION LOSS [dB]','RETURN LOSS [dB]','PASSBAND FREQUENCY fp [GHz]','STOPBAND FREQUENCY fs [GHz]'};
    title1='CHEBYSHEV LP GENERALIZED EQUATION';
    numlines=1;
    defaultanswer={'','','',''};
    options.Resize='on';
    options.WindowStyle='normal';
    options.Interpreter='tex';
    input_array = inputdlg(prompt,title1,numlines,defaultanswer, options); % Complete input dialog box

    LA = str2num(input_array{1}); % Accepting first element of the input as LA
    LR = str2num(input_array{2}); % Accepting 2nd element of the input as LR
    fc = str2num(input_array{3}); % Accepting 3rd element of the input as fp
    fs = str2num(input_array{4});% Accepting 4th element of the input as fs

   if (LA>LR) && (LR>0) && (fc>=0.3) && (fs>fc) && (fs<=300) % Microwave frequence range: 300MHz - 300GHz.

% |====================================================================|
% |                                      1                             |
% | Plotting Chebyshev LP Prototype from |S12|^2 = ----------------------|
% |                                      1 + [epsillon*T(w)]^2         |
% | Here: w <-------- w/wc                                             |
% |====================================================================|

% Cut-off frequency Wc and Stopband frequency Ws:
    wc = 2*pi*fc;
    ws = 2*pi*fs;

% Selectivity
    S = ws/wc;
    % Displaying the selectivity
    Selectivity=num2str(S); % converting number to sder:tring
    set(handles.edit14_selectivity,'String',Selectivity); % Display the converted string

% Degree for Chebyshev Filter
```

```matlab
        Degree = (LA + LR + 6)/(20*log10(S + sqrt(S^2 - 1)));
        Degree_rounded = round(Degree);
        if Degree_rounded<Degree
            N = Degree_rounded + 1;
        elseif Degree_rounded>=Degree
            N = Degree_rounded;
        else
        end
        N;
% Displaying the filter order/degree
    filter_order=num2str(N); % converting number to sder:tring
    set(handles.FilterOrder,'String',filter_order); % Display the converted string

%    Reflection Coefficient
     Ref = 10^(-LR/20);
%    dB pass-band ripple
     Lar = -10*log10(1-Ref^2);
% Ripple level epsillon
    epsillon = sqrt(10^(0.1*Lar) - 1);
    % Displaying the ripple level or factor
    ripple_level=num2str(epsillon); % converting number to sder:tring
    set(handles.edit15_ripple,'String',ripple_level); % Display the converted string

%  Computation of S12 and its Plotting
     f = 0:0.01:30;
     w = 0:(2*pi*0.01):(30*2*pi);

        for i=1:1:3001
     if (w(i)<=wc)
          Deno = 1 + (epsillon*cos(N*acos(w(i)./wc))).^2;

     elseif (w(i)>=wc)
          Deno = 1 + (epsillon*cosh(N*acosh(w(i)./wc))).^2;

     else
     end

 % Computing S12 values for different values ABCD resulting from the variation in w.
    S1(i) = abs(1./Deno);
    S10(i) = 1 - abs((1./Deno));
   end

     S12 = 10*log10(S1);
     S11 = 10*log10(S10);

    plot(f,S12,'blue', f,S11,'red')
    graph = plot(f,S12,'blue', f,S11,'red');
    % Setting the line width or thickness for both curves
    set (graph, 'LineWidth',1.5);
    title('Chebyshev Response of LP Prototype');
    ylabel('|S12|^2 and |S11|^2 in dB');
    xlabel('f in GHz');
    axis([0 2*fs -80 5]);
    grid on;

   else
      helpdlg('Please check your input parameters. Make sure that fp>=0.3GHz, fs>fp, fs<=300GHz, LA>LR and LR>0 dB.');
   end
% |====================================================================|
% |                                                                    |
% |     Plotting Chebyshev LP Prototype using A,B,C,D parameters:      |
% |                                 4                                  |
%                    S12|^2 = -------------------                      |
% |                            |A+(B/Zo)+(C*Zo)+D|^2                   |
% |====================================================================|
elseif ChebyshevLP == 2   % Chebyshev LP Shunt Prototype to plot with its ABCD Parameters

     % Generation an input dialog box for Chebyshev LP Shunt Prototype
```

```matlab
    prompt={'INSERTION LOSS [dB]','RETURN LOSS [dB]','PASSBAND FREQUENCY fp [GHz]','STOPBAND FREQUENCY fs [GHz]','SYSTEM IMPEDANCE Zo [OHMS]'};
    title1='CHEBYSHEV LP SHUNT PROTOTYPE';
    numlines=1;
    defaultanswer={'','','','',''};
    options.Resize='on';
    options.WindowStyle='normal';
    options.Interpreter='tex';
    input_array = inputdlg(prompt,title1,numlines,defaultanswer, options); % Complete input dialog box

    LA = str2num(input_array{1}); % Accepting first element of the input as LA
    LR = str2num(input_array{2}); % Accepting 2nd element of the input as LR
    fc = str2num(input_array{3}); % Accepting 3rd element of the input as fp
    fs = str2num(input_array{4});% Accepting 4th element of the input as fs
    Zo = str2num(input_array{5}); % Accepting last element of the input as Zo

   if (LA>LR) && (LR>0) && (fc>=0.3) && (fs>fc) && (fs<=300)

% |**********************************************************************|
% |                Chebyshev - Shunt Low Pass Prototype                  |
% |**********************************************************************|
% Cut-off frequency Wc and Stopband frequency Ws:
    wc = 2*pi*fc;
    ws = 2*pi*fs;

% Selectivity must be > 1.
    S = ws/wc;
    % Displaying the selectivity
    Selectivity=num2str(S); % converting number to sder:tring
    set(handles.edit14_selectivity,'String',Selectivity); % Display the converted string

% Degree for Chebyshev Filter
    Degree = (LA + LR + 6)/(20*log10(S + sqrt(S^2 - 1)));
    Degree_rounded = round(Degree);
    if Degree_rounded<Degree
        N = Degree_rounded + 1;
    elseif Degree_rounded>=Degree
        N = Degree_rounded;
    else
    end
    N;
    % Displaying the filter order/degree
        filter_order=num2str(N); % converting number to sder:tring
        set(handles.FilterOrder,'String',filter_order); % Display the converted string

%     Reflection Coefficient
    Ref = 10^(-LR/20);
%     dB pass-band ripple
    Lar = -10*log10(1-Ref^2);
% Ripple level epsillon
    epsillon = sqrt(10^(0.1*Lar) - 1);
    % Displaying the ripple level or factor
    ripple_level=num2str(epsillon); % converting number to sder:tring
    set(handles.edit15_ripple,'String',ripple_level); % Display the converted string

% Computation of element values,  Ref. G. Mattahaei MW Filters, Pg 99
R = coth(Lar/17.37);
beta = log(R);
gamma = sinh(beta/(2*N));
k = N+1;

for i=1:1:k
    a(i) = sin(((2*i-1).*pi)./(2*N));
    b(i) = gamma^2 + [sin((i.*pi)./N)]^2;

    if i==1
    g(1)= 2.*a(1)/gamma;
    elseif (i>1) && (i<=N)
        g(i) = (4.*a(i-1).*a(i))./(b(i-1).*g(i-1));
```

```matlab
        else
            A = mod (N,2);
            if A==0
                g(k)=[coth(beta/4)]^2;

            elseif A==1
                g(k)=1;
            end
        end
end

%  Plotting with element values for different filter order

% Generatng L and C elements for LP Prototype - Shunt

for i=1:1:N
    if i==1
        % Computing C(1) value
        Element(1) = g(1)/(wc*Zo);
        Capacitor(1)=Element(1);
    elseif (i>1) && (i<=N)
        k=mod(i,2);
        if k==0
            % Computing L values for odd values of i
            Element(i) = (g(i)*Zo)/wc;
            Inductor(i)=Element(i);
        elseif k==1
            % Computing C values for even values of i
            Element(i) = g(i)/(wc*Zo);
            Capacitor(i)=Element(i);
        else
        end
    else
    end
end
%Scaled reactive element values and scaled source Zs and load ZL impedances, Z = [Zs ZL]
Element;
Inductor(Inductor==0)=[];
Inductor;                        %L remains in nH.
Capacitor(Capacitor==0)=[];
Capacitor2 = 1000.*Capacitor;   %to convert from nF to pF.
Zs = 1*Zo;                       % Source impedance
ZL = g(N+1)*Zo;                  % Load impedance in Ohm
Impedance = [Zs ZL];             %Z are in Ohms. in Ohm

% Displaying the inductor, capacitor and impedance values.
Inductor1 = mat2str(Inductor,4);
set(handles.InductorValues1,'String',Inductor1); % Display the converted string
Capacitor1 = mat2str(Capacitor2,4);
set(handles.CapacitorValues1,'String',Capacitor1); % Display the converted string
Impedance1 = mat2str(Impedance,4);
set(handles.ImpedanceValues1,'String',Impedance1); % Display the converted string

f = 0:0.01:30;                  % Range of the frequency in GHz
w = 0:(2*pi*0.01):(2*pi*30);   % Range of w = 2*pi*f.

 % Computing Z, Y and T = [A B;C D] values for different frequency values
 % This is for Chebyshev LP Prototype - Series

    for i=1:1:3001
        for t=1:1:N
            if t==1
                % Computing Z(C1) values
                Z(1) = j*Element(1).*w(i);
                T = [1 0;Z(1) 1];
            elseif (t>1) && (t<=N)
                    k=mod(t,2);
                if k==0
                    % Computing Z(L) values
```

```matlab
                        Z(t) = j*Element(t).*w(i);
                        Tnew = T*[1 Z(t);0 1];
                        T = Tnew;
                    elseif k==1
                        % Computing Z(C) values
                        Z(t) = j*Element(t).*w(i);
                        Tnew = T*[1 0;Z(t) 1];
                        T = Tnew;
                    else
                    end
                else
                end
        end

    T;

% Getting ABCD values by identification using 2X2 Matrix T=[A B;C D] by using
    A = T(1,1);
    B = T(1,2);
    C = T(2,1);
    D = T(2,2);

% Computing S12 values for different values ABCD resulting from the
% variation in w.
    Sa(i) = abs(2./(A+(B./Zo)+(C.*Zo)+D));
    Sb(i) = abs(1 - abs(2./(A+(B./Zo)+(C.*Zo)+D)).^2);
   end

%                                                  4
% Plotting with A,B,C,D parameters,  |S12|^2 = -------------------
%                                      |A+(B/Zo)+(C*Zo)+D|^2

    S12 = 20*log10(Sa);  % Computing the dB values of |S12|^2
    S11 = 10*log10(Sb);

    plot(f,S12,'blue', f,S11,'red')
    graph = plot(f,S12,'blue', f,S11,'red');
    % Setting the line width or thickness for both curves
    set (graph, 'LineWidth',1.5);
    title('Chebyshev Response for LP Shunt Prototype');
    ylabel('|S12|^2 and |S11|^2 in dB');
    xlabel('f in GHz');
    axis([0 2*fc -80 3]);
    grid on;

% |**********************************************************************|
% |           End Plotting Chebyshev LP shunt Prototype                   |
% |**********************************************************************|
    else
        helpdlg('Please check your input parameters. Make sure that fp>=0.3GHz, fs>fp, fs<=300GHz, LA>LR and LR>0 dB. It is also advised to set Zo=50 Ohms');
    end

elseif ChebyshevLP == 3   % Chebyshev LP Series Prototype to plot with its ABCD Parameters

    % Generation an input dialog box for Chebyshev LP Series Prototype
    prompt={'INSERTION LOSS [dB]','RETURN LOSS [dB]','PASSBAND FREQUENCY fp [GHz]','STOPBAND FREQUENCY fs [GHz]','SYSTEM IMPEDANCE Zo [OHMS]'};
    title1='CHEBYSHEV LP SERIES PROTOTYPE';
    numlines=1;
    defaultanswer={'','','','',''};
    options.Resize='on';
    options.WindowStyle='normal';
    options.Interpreter='tex';
    input_array = inputdlg(prompt,title1,numlines,defaultanswer, options); % Complete input dialog box

    LA = str2num(input_array{1}); % Accepting first element of the input as LA
    LR = str2num(input_array{2}); % Accepting 2nd element of the input as LR
    fc = str2num(input_array{3}); % Accepting 3rd element of the input as fp
```

```matlab
        fs = str2num(input_array{4});% Accepting 4th element of the input as fs
        Zo = str2num(input_array{5}); % Accepting last element of the input as Zo

    if (LA>LR) && (LR>0) && (fc>=0.3) && (fs>fc) && (fs<=300)

% |************************************************************************|
% |                 Chebyshev - Shunt Low Pass Prototype                    |
% |************************************************************************|
% Cut-off frequency Wc and Stopband frequency Ws:
    wc = 2*pi*fc;
    ws = 2*pi*fs;

% Selectivity must be > 1.
    S = ws/wc;
    % Displaying the selectivity
    Selectivity=num2str(S); % converting number to sder:tring
    set(handles.edit14_selectivity,'String',Selectivity); % Display the converted string

% Degree for Chebyshev Filter
    Degree = (LA + LR + 6)/(20*log10(S + sqrt(S^2 - 1)));
    Degree_rounded = round(Degree);
    if Degree_rounded<Degree
        N = Degree_rounded + 1;
    elseif Degree_rounded>=Degree
        N = Degree_rounded;
    else
    end
    N;
    % Displaying the filter order/degree
        filter_order=num2str(N); % converting number to sder:tring
        set(handles.FilterOrder,'String',filter_order); % Display the converted string

%    Reflection Coefficient
    Ref = 10^(-LR/20);
%    dB pass-band ripple
    Lar = -10*log10(1-Ref^2);
% Ripple level epsillon
    epsillon = sqrt(10^(0.1*Lar) - 1);
    % Displaying the ripple level or factor
    ripple_level=num2str(epsillon); % converting number to sder:tring
    set(handles.edit15_ripple,'String',ripple_level); % Display the converted string

% Computation of element values,  Ref. G. Mattahaei MW Filters, Pg 99
R = coth(Lar/17.37);
beta = log(R);
gamma = sinh(beta/(2*N));
k = N+1;

for i=1:1:k
    a(i) = sin(((2*i-1).*pi)./(2*N));
    b(i) = gamma^2 + [sin((i.*pi)./N)]^2;

    if i==1

        g(1)= 2.*a(1)/gamma;

    elseif (i>1) && (i<=N)
        g(i) = (4.*a(i-1).*a(i))./(b(i-1).*g(i-1));

    else

        A = mod (N,2);
        if A==0
            g(k)=[coth(beta/4)]^2;

        elseif A==1

            g(k)=1;
```

```matlab
            end

        end

end

%  Plotting with element values for different filter order

% Generatng L and C elements for LP Prototype - Series

for i=1:1:N
    if i==1
        Element(1) = (g(1)*Zo)/wc; % Computing L1 values
        Inductor(1) = Element(1);
    elseif (i>1) && (i<=N)
        k=mod(i,2);
        if k==0
            Element(i) = g(i)/(wc*Zo); % Computing C values
            Capacitor(i)=Element(i);
        elseif k==1
            Element(i) = (g(i)*Zo)/wc; % Computing L values
            Inductor(i)=Element(i);
        else
        end
    else
    end
end
%Scaled reactive element values and scaled source Zs and load ZL impedances, Z = [Zs ZL]
Element;
Inductor(Inductor==0)=[];
Inductor;                      %L remains in nH.
Capacitor(Capacitor==0)=[];
Capacitor2 = 1000.*Capacitor;   %to convert from nF to pF.
Zs = 1*Zo;                     % Source impedance
ZL = g(N+1)*Zo;                % Load impedance in Ohm
Impedance = [Zs ZL];           %Z are in Ohms. in Ohm

% Displaying the inductor, capacitor and impedance values.
Inductor1 = mat2str(Inductor,4);
set(handles.InductorValues1,'String',Inductor1); % Display the converted string
Capacitor1 = mat2str(Capacitor2,4);
set(handles.CapacitorValues1,'String',Capacitor1); % Display the converted string
Impedance1 = mat2str(Impedance,4);
set(handles.ImpedanceValues1,'String',Impedance1); % Display the converted string

f = 0:0.01:30;                  % Range of the frequency in GHz
w = 0:(2*pi*0.01):(2*pi*30);    % Range of w = 2*pi*f.

% Computing Z, Y and T = [A B;C D] values for different frequency values
% This is for Chebyshev LP Prototype - Series

    for i=1:1:3001
        for t=1:1:N
            if t==1
                Z(1) = j*Element(1).*w(i);  % Computing Z(L1) values
                T = [1 Z(1);0 1];
            elseif (t>1) && (t<=N)
                    k=mod(t,2);
                if k==0
                    Z(t) = j*Element(t).*w(i); % Computing Zc values
                    Tnew = T*[1 0;Z(t) 1];
                    T = Tnew;
                elseif k==1
                    Z(t) = j*Element(t).*w(i); % Computing Z(L) values
                    Tnew = T*[1 Z(t);0 1];
                    T = Tnew;
                else
                end
```

```matlab
            else
            end
        end

    T;

% Getting ABCD values by identification using 2X2 Matrix T=[A B;C D] by using
    A = T(1,1);
    B = T(1,2);
    C = T(2,1);
    D = T(2,2);

 % Computing S12 values for different values ABCD resulting from the
 % variation in w.
    Sa(i) = abs(2./(A+(B./Zo)+(C.*Zo)+D));
    Sb(i) = abs(1 - abs((2./(A+(B./Zo)+(C.*Zo)+D)).^2));
   end

%                                           4
% Plotting with A,B,C,D parameters, |S12|^2 = -------------------
%                                           |A+(B/Zo)+(C*Zo)+D|^2

    S12 = 20*log10(Sa);  % Computing the dB values of |S12|^2
    S11 = 10*log10(Sb);

    plot(f,S12,'blue', f,S11,'red')
    graph = plot(f,S12,'blue', f,S11,'red');
    % Setting the line width or thickness for both curves
    set (graph, 'LineWidth',1.5);
    title('Chebyshev Response of LP Series Prototype');
    ylabel('|S12|^2 and |S11|^2 in dB');
    xlabel('f in GHz');
    axis([0 2*fc -80 3]);
    grid on

% |************************************************************************|
% |              End Plotting Chebyshev LP Series Prototype                |
% |************************************************************************|

    else
        helpdlg('Please check your input parameters. Make sure that fp>=0.3GHz, fs>fp, fs<=300GHz, LA>LR and LR>0 dB. It is also advised to set Zo=50 Ohms');
    end
else
end
% |************************************************************************|
% |              End Plotting Chebyshev LP  Prototypes                     |
% |************************************************************************|

%=========================================================================

% |************************************************************************|
% |              Begin Plotting Chebyshev HP  Prototypes                   |
% |************************************************************************|

% HP radio button is selected
elseif selected_radio2 == 1
if ChebyshevHP == 1     % Chebyshev HP from its Generalized Equation
% |========================================================================|
% | Chebyshev BP Prototype using LP to BP transformation:                  |
% |   w <----- -wc/w   frequency substitution.                             |
% |                                                   1                    |
% | Plotting Chebyshev BP Prototype from   |S12|^2 = ----------------      |
% |                                                  1 + w^2N              |
% |========================================================================|
    % Generation an input dialog box for Chebyshev HP Generalized Equation
```

```matlab
    prompt={'INSERTION LOSS [dB]','RETURN LOSS [dB]','PASSBAND FREQUENCY fp [GHz]','STOPBAND FREQUENCY fs [GHz]'};
    title1='CHEBYSHEV HP GENERALIZED EQUATION';
    numlines=1;
    defaultanswer={'','','',''};
    options.Resize='on';
    options.WindowStyle='normal';
    options.Interpreter='tex';
    input_array = inputdlg(prompt,title1,numlines,defaultanswer, options); % Complete input dialog box

    LA = str2num(input_array{1}); % Accepting first element of the input as LA
    LR = str2num(input_array{2}); % Accepting 2nd element of the input as LR
    fc = str2num(input_array{3}); % Accepting 3rd element of the input as fp
    fs = str2num(input_array{4});% Accepting 4th element of the input as fs

   if (LA>LR) && (LR>0) && (fs>=0.3) && (fc>fs) && (fc<=300)
    % Cut-off frequency Wc and Stopband frequency Ws:
    wc = 2*pi*fc;
    ws = 2*pi*fs;
% Selectivity
        S = wc/ws;
        % Displaying the selectivity
        Selectivity=num2str(S); % converting number to sder:tring
        set(handles.edit14_selectivity,'String',Selectivity); % Display the converted string

% Degree for Chebyshev Filter
    Degree = (LA + LR + 6)/(20*log10(S + sqrt(S^2 - 1)));
    Degree_rounded = round(Degree);
    if Degree_rounded<Degree
       N = Degree_rounded + 1;
    elseif Degree_rounded>=Degree
       N = Degree_rounded;
    else
    end
    N;
     % Displaying the filter order/degree
     filter_order=num2str(N); % converting number to sder:tring
     set(handles.FilterOrder,'String',filter_order); % Display the converted string
%    Reflection Coefficient
     Ref = 10^(-LR/20);
%    dB pass-band ripple
     Lar = -10*log10(1-Ref^2);
% Ripple level epsillon
    epsillon = sqrt(10^(0.1*Lar) - 1);
    ripple_level=num2str(epsillon); % converting number to sder:tring
    set(handles.edit15_ripple,'String',ripple_level); % Display the converted string

    f = 0:0.01:30;                  % Range of the frequency in GHz
    w = 0:(2*pi*0.01):(2*pi*30);    % Range of w = 2*pi*f.

   for i=1:1:3001
    if (w(i)>=wc)
       Deno = 1 + (epsillon*cos(N*acos(-wc./w(i)))).^2;

    elseif (w(i)<=wc)
       Deno = 1 + (epsillon*cosh(N*acosh(-wc./w(i)))).^2;

    else
    end

 % Computing S12 values for different values ABCD resulting from the variation in w.
    S1(i) = abs(1./Deno);
    S10(i)= 1 - (abs(1./Deno));
  end

    S12 = 10*log10(S1); % Computating |S12|^2 in dB
    S11 = 10*log10(S10);
```

```matlab
% Plotting |S12|^2 and |S11|^2
        plot(f,S12,'blue', f,S11,'red')
        graph = plot(f,S12,'blue', f,S11,'red');
        % Setting the line width or thickness for both curves
        set (graph, 'LineWidth',1.5);
        title('Chebyshev Response HP');
        ylabel('|S12|^2 and |S11|^2 in dB');
        xlabel('f in GHz');
        axis([0 2*fc -80 5]);
        grid on;
% |***********************************************************************|
% |              End Plotting Chebyshev HP Series Prototype               |
% |***********************************************************************|
    else
        helpdlg('Please check your input parameters. Make sure that fs>=0.3GHz, fs<fp, LA>LR and LR>0 dB.');
    end
% |***********************************************************************|
% |             Begin Plotting Chebyshev HP Series Prototype              |
% |***********************************************************************|

% |=======================================================================|
% |                                                                       |
% |     Plotting Chebyshev HP Prototypes using A,B,C,D parameters:        |
% |                              4                                        |
% |                  |S12|^2 = -------------------                        |
% |                            |A+(B/Zo)+(C*Zo)+D|^2                      |
% |=======================================================================|
elseif ChebyshevHP == 2   % Chebyshev HP Shunt Prototype to plot with its ABCD Parameters

    % Generation an input dialog box for Chebyshev HP Shunt Prototype
    prompt={'INSERTION LOSS [dB]','RETURN LOSS [dB]','PASSBAND FREQUENCY fp [GHz]','STOPBAND FREQUENCY fs [GHz]','SYSTEM IMPEDANCE Zo [OHMS]'};
    title1='CHEBYSHEV HP SHUNT PROTOTYPE';
    numlines=1;
    defaultanswer={'','','','',''};
    options.Resize='on';
    options.WindowStyle='normal';
    options.Interpreter='tex';
    input_array = inputdlg(prompt,title1,numlines,defaultanswer, options); % Complete input dialog box

    LA = str2num(input_array{1}); % Accepting first element of the input as LA
    LR = str2num(input_array{2}); % Accepting 2nd element of the input as LR
    fc = str2num(input_array{3}); % Accepting 3rd element of the input as fp
    fs = str2num(input_array{4});% Accepting 4th element of the input as fs
    Zo = str2num(input_array{5}); % Accepting last element of the input as Zo

    if (LA>LR) && (LR>=0) && (fs>=0.3) && (fs<fc) && (fc<=300) && (Zo==50)
    % Cut-off frequency Wc and Stopband frequency Ws:
    wc = 2*pi*fc;
    ws = 2*pi*fs;
% Selectivity
    S = wc/ws;
       % Displaying the selectivity
       Selectivity=num2str(S); % converting number to sder:tring
       set(handles.edit14_selectivity,'String',Selectivity); % Display the converted string

% Degree for Chebyshev Filter
    Degree = (LA + LR + 6)/(20*log10(S + sqrt(S^2 - 1)));
    Degree_rounded = round(Degree);
    if Degree_rounded<Degree
        N = Degree_rounded + 1;
    elseif Degree_rounded>=Degree
        N = Degree_rounded;
    else
    end
    N;
    % Displaying the filter order/degree
    filter_order=num2str(N); % converting number to sder:tring
    set(handles.FilterOrder,'String',filter_order); % Display the converted string
```

```matlab
%     Reflection Coefficient
     Ref = 10^(-LR/20);
%     dB pass-band ripple
     Lar = -10*log10(1-Ref^2);
% Ripple level epsillon
    epsillon = sqrt(10^(0.1*Lar) - 1);
    ripple_level=num2str(epsillon); % converting number to sder:tring
    set(handles.edit15_ripple,'String',ripple_level); % Display the converted string
 % Computation of element values,  Ref. G. Mattahaei MW Filters, Pg 99
R = coth(Lar/17.37);
beta = log(R);
gamma = sinh(beta/(2*N));
k = N+1;

for i=1:1:k
    a(i) = sin(((2*i-1).*pi)./(2*N));
    b(i) = gamma^2 + [sin((i.*pi)./N)]^2;
    
    if i==1
    g(1)= 2.*a(1)/gamma;
    elseif (i>1) && (i<=N)
        g(i) = (4.*a(i-1).*a(i))./(b(i-1).*g(i-1));
    else
        A = mod (N,2);
        if A==0
            g(k)=[coth(beta/4)]^2;
            
        elseif A==1
            g(k)=1;
        end
    end
end

%  Plotting with element values for different filter order

% Generatng L and C elements for HP Prototype - Series

for i=1:1:N
    if i==1
        Element(1) = (g(1)*Zo)*wc; % Computing L1 values
        %LP to HP transformation for element value
        Capacitor(1)=1/Element(1);
    elseif (i>1) && (i<=N)
        k=mod(i,2);
        if k==0
            Element(i) = (g(i)*wc)/Zo; % Computing C values
            %LP to HP transformation for element values
            Inductor(i)=1./Element(i);
        elseif k==1
            Element(i) = (g(i)*Zo)*wc; % Computing L values
            %LP to HP transformation for element values
            Capacitor(i)=1./Element(i);
        else
        end
    else
    end
end
%Scaled reactive element values and scaled source Zs and load ZL impedances, Z = [Zs ZL]
Element;
Inductor(Inductor==0)=[];
Inductor;                        %L remains in nH.
Capacitor(Capacitor==0)=[];
Capacitor2 = 1000.*Capacitor;  %to convert from nF to pF.
Zs = 1*Zo;                    % Source impedance
ZL = g(N+1)*Zo;               % Load impedance in Ohm
Impedance = [Zs ZL];          %Z are in Ohms. in Ohm

% Displaying the inductor, capacitor and impedance values.
Inductor1 = mat2str(Inductor,4);
```

```matlab
set(handles.InductorValues1,'String',Inductor1); % Display the converted string
Capacitor1 = mat2str(Capacitor2,4);
set(handles.CapacitorValues1,'String',Capacitor1); % Display the converted string
Impedance1 = mat2str(Impedance,4);
set(handles.ImpedanceValues1,'String',Impedance1); % Display the converted string

f = 0:0.01:30;                % Range of the frequency in GHz
w = 0:(2*pi*0.01):(2*pi*30);  % Range of w = 2*pi*f.

% Computing Z, Y and T = [A B;C D] values for different frequency values
% This is for Chebyshev HP Prototype - Series

    for i=1:1:3001
        for t=1:1:N
            if t==1
                Z(1) = -j*Element(1)./w(i);  % Computing Z(L1) values
                T = [1 Z(1);0 1];
            elseif (t>1) && (t<=N)
                    k=mod(t,2);
                if k==0
                    Z(t) = -j*Element(t)./w(i); % Computing Zc values
                    Tnew = T*[1 0;Z(t) 1];
                    T = Tnew;
                elseif k==1
                    Z(t) = -j*Element(t)./w(i); % Computing Z(L) values
                    Tnew = T*[1 Z(t);0 1];
                    T = Tnew;
                else
                end
            else
            end
        end

    T;

% Getting ABCD values by identification using 2X2 Matrix T=[A B;C D] by using
    A = T(1,1);
    B = T(1,2);
    C = T(2,1);
    D = T(2,2);

% Computing S12 values for different values ABCD resulting from the
% variation in w.
    Sa(i) = abs(2./(A+(B./Zo)+(C.*Zo)+D));
    Sb(i) = 1 - (abs((2./(A+(B./Zo)+(C.*Zo)+D)))).^2;
    end
%                                              4
% Plotting with A,B,C,D parameters, |S12|^2 = -------------------
%                                             |A+(B/Zo)+(C*Zo)+D|^2

    S12 = 20*log10(Sa);  % Computing the dB values of |S12|^2
    S11 = 10*log10(Sb);

    plot(f,S12,'blue', f,S11,'red')
    graph = plot(f,S12,'blue', f,S11,'red');
    % Setting the line width or thickness for both curves
    set (graph, 'LineWidth',1.5);
    title('Chebyshev Response for HP Shunt Prototype');
    ylabel('|S12|^2 and |S11|^2 in dB');
    xlabel('f in GHz');
    axis([0 2*fc -80 3]);
    grid on;

% |************************************************************|
% |          End Plotting Chebyshev HP shunt Prototype         |
% |************************************************************|

    else
```

```matlab
        helpdlg('Please check your input parameters. Make sure that fs>=0.3GHz, fs<fp, LA>LR, LR>0 dB and it is recommended to keep Zo=50 Ohms');
    end

elseif ChebyshevHP == 3   % Chebyshev HP Series Prototype to plot with its ABCD Parameters

    % Generation an input dialog box for Chebyshev HP Series Prototype
    prompt={'INSERTION LOSS [dB]','RETURN LOSS [dB]','PASSBAND FREQUENCY fp [GHz]','STOPBAND FREQUENCY fs [GHz]','SYSTEM IMPEDANCE Zo [OHMS]'};
    title1='CHEBYSHEV HP SERIES PROTOTYPE';
    numlines=1;
    defaultanswer={'','','','',''};
    options.Resize='on';
    options.WindowStyle='normal';
    options.Interpreter='tex';
    input_array = inputdlg(prompt,title1,numlines,defaultanswer, options); % Complete input dialog box

    LA = str2num(input_array{1}); % Accepting first element of the input as LA
    LR = str2num(input_array{2}); % Accepting 2nd element of the input as LR
    fc = str2num(input_array{3}); % Accepting 3rd element of the input as fp
    fs = str2num(input_array{4});% Accepting 4th element of the input as fs
    Zo = str2num(input_array{5}); % Accepting last element of the input as Zo

    if (LA>LR) && (LR>=0) && (fs>=0.3) && (fs<fc) && (fc<=300) && (Zo==50)
    % Cut-off frequency Wc and Stopband frequency Ws:
    wc = 2*pi*fc;
    ws = 2*pi*fs;
% Selectivity
        S = wc/ws;
        % Displaying the selectivity
        Selectivity=num2str(S); % converting number to sder:tring
        set(handles.edit14_selectivity,'String',Selectivity); % Display the converted string

% Degree for Chebyshev Filter
    Degree = (LA + LR + 6)/(20*log10(S + sqrt(S^2 - 1)));
    Degree_rounded = round(Degree);
    if Degree_rounded<Degree
        N = Degree_rounded + 1;
    elseif Degree_rounded>=Degree
        N = Degree_rounded;
    else
    end
    N;
     % Displaying the filter order/degree
     filter_order=num2str(N); % converting number to sder:tring
     set(handles.FilterOrder,'String',filter_order); % Display the converted string
%    Reflection Coefficient
    Ref = 10^(-LR/20);
%    dB pass-band ripple
    Lar = -10*log10(1-Ref^2);
% Ripple level epsillon
    epsillon = sqrt(10^(0.1*Lar) - 1);
    ripple_level=num2str(epsillon); % converting number to sder:tring
    set(handles.edit15_ripple,'String',ripple_level); % Display the converted string

% Computation of element values,  Ref. G. Mattahaei MW Filters, Pg 99
R = coth(Lar/17.37);
beta = log(R);
gamma = sinh(beta/(2*N));
k = N+1;

for i=1:1:k
    a(i) = sin(((2*i-1).*pi)./(2*N));
    b(i) = gamma^2 + [sin((i.*pi)./N)]^2;

    if i==1
    g(1)= 2.*a(1)/gamma;
    elseif (i>1) && (i<=N)
        g(i) = (4.*a(i-1).*a(i))./(b(i-1).*g(i-1));
```

```matlab
        else
            A = mod (N,2);
            if A==0
                g(k)=[coth(beta/4)]^2;
            elseif A==1
                g(k)=1;
            end
        end
end

% Plotting with element values for different filter order

% Generatng L and C elements for HP Prototype - Shunt

for i=1:1:N
    if i==1
        % Computing C(1) value
        Element(1) = (g(1)*wc)/Zo;
        %LP to HP transformation
        Inductor(1)=1/Element(1);
    elseif (i>1) && (i<=N)
        k=mod(i,2);
        if k==0
            % Computing L values for odd values of i
            Element(i) = (g(i)*Zo)*wc;
            %LP to HP transformation
            Capacitor(i)=1./Element(i);
        elseif k==1
            % Computing C values for even values of i
            Element(i) = (g(i)*wc)/Zo;
            %LP to HP transformation
            Inductor(i)=1./Element(i);
        else
        end
    else
    end
end
%Scaled reactive element values and scaled source Zs and load ZL impedances, Z = [Zs ZL]
Element;
Inductor(Inductor==0)=[];
Inductor;                       %L remains in nH.
Capacitor(Capacitor==0)=[];
Capacitor2 = 1000.*Capacitor;   %to convert from nF to pF.
Zs = 1*Zo;                      % Source impedance
ZL = g(N+1)*Zo;                 % Load impedance in Ohm
Impedance = [Zs ZL];            %Z are in Ohms. in Ohm

% Displaying the inductor, capacitor and impedance values.
Inductor1 = mat2str(Inductor,4);
set(handles.InductorValues1,'String',Inductor1); % Display the converted string
Capacitor1 = mat2str(Capacitor2,4);
set(handles.CapacitorValues1,'String',Capacitor1); % Display the converted string
Impedance1 = mat2str(Impedance,4);
set(handles.ImpedanceValues1,'String',Impedance1); % Display the converted string

f = 0:0.01:30;                  % Range of the frequency in GHz
w = 0:(2*pi*0.01):(2*pi*30);    % Range of w = 2*pi*f.

 % Computing Z, Y and T = [A B;C D] values for different frequency values
 % This is for Chebyshev HP Prototype - Series

    for i=1:1:3001
        for t=1:1:N
            if t==1
                % Computing Z(C1) values
                Z(1) = -j*Element(1)./w(i);
                T = [1 0;Z(1) 1];
            elseif (t>1) && (t<=N)
                k=mod(t,2);
```

```matlab
            if k==0
                % Computing Z(L) values
                Z(t) = -j*Element(t)./w(i);
                Tnew = T*[1 Z(t);0 1];
                T = Tnew;
            elseif k==1
                % Computing Z(C) values
                Z(t) = -j*Element(t)./w(i);
                Tnew = T*[1 0;Z(t) 1];
                T = Tnew;
            else
            end
        else
        end
    end

    T;

% Getting ABCD values by identification using 2X2 Matrix T=[A B;C D] by using
    A = T(1,1);
    B = T(1,2);
    C = T(2,1);
    D = T(2,2);

 % Computing S12 values for different values ABCD resulting from the
 % variation in w.
    Sa(i) = abs(2./(A+(B./Zo)+(C.*Zo)+D));
    Sb(i) = 1 - (abs((2./(A+(B./Zo)+(C.*Zo)+D)))).^2;
   end
%                                                    4
% Plotting with A,B,C,D parameters, |S12|^2 = -------------------
%                                              |A+(B/Zo)+(C*Zo)+D|^2

    S12 = 20*log10(Sa);  % Computing the dB values of |S12|^2
    S11 = 10*log10(Sb);
    plot(f,S12,'blue', f,S11,'red')
    graph = plot(f,S12,'blue', f,S11,'red');
    % Setting the line width or thickness for both curves
    set (graph, 'LineWidth',1.5);
    title('Chebyshev Response of HP Series Prototype');
    ylabel('|S12|^2 and |S11|^2 in dB');
    xlabel('f in GHz');
    axis([0 2*fc -80 3]);
    grid on;

% |**********************************************************************|
% |              End Plotting Chebyshev HP Series Prototype              |
% |**********************************************************************|

    else
        helpdlg('Please check your input parameters. Make sure that fp>=0.3 GHz, fs<fp, LA>LR, LR>=0 dB and  it is advised to keep Zo=50 Ohms');
    end
else
end

% |**********************************************************************|
% |              End Plotting Chebyshev HP  Prototypes                   |
% |**********************************************************************|

%=========================================================================

% |**********************************************************************|
% |              Begin Plotting Chebyshev BP  Prototypes                 |
% |**********************************************************************|

% Chebyshev BP radio button is selected
elseif selected_radio3 == 1
```

```matlab
if ChebyshevBP == 1     % Chebyshev BP from its Generalized Equation
% |===========================================================================|
% | Chebyshev BP Prototype using LP to BP transformation:                     |
% |   w <----- [wo/(w1-w1)]*[(w./wo)-(wo./w)]   with wo = sqrt(w1*w2)         |
% |                                             1                             |
% | Plotting Chebyshev BP Prototype from   |S12|^2 = ----------------         |
% |                                             1 + w^2N                      |
% |===========================================================================|
    % Generation an input dialog box for Chebyshev BP Generalized Equation
    prompt={'INSERTION LOSS [dB]','RETURN LOSS [dB]','LOWER SIDE CUTOFF FREQUENCY f1 [GHz]','UPPER SIDE CUTOFF FREQUENCY f2 [GHz]','STOPBAND FREQUENCY AT HIGHER SIDE fs [GHz]'};
    title1='CHEBYSHEV BP GENERALIZED EQUATION';
    numlines=1;
    defaultanswer={'','','','',''};
    options.Resize='on';
    options.WindowStyle='normal';
    options.Interpreter='tex';
    input_array = inputdlg(prompt,title1,numlines,defaultanswer, options); % Complete input dialog box

    LA = str2num(input_array{1}); % Accepting first element of the input as LA
    LR = str2num(input_array{2}); % Accepting 2nd element of the input as LR
    f1 = str2num(input_array{3}); % Accepting 3rd element of the input as f1
    f2 = str2num(input_array{4});% Accepting 4th element of the input as f2
    fs = str2num(input_array{5});% Accepting 4th element of the input as fs

   if (LA>LR) && (LR>0) && (f1>=0.3) && (f2>f1) && (f2<fs) && (fs<=300)
    % Cut-off Frequencies W1 and W2; and Stopband frequency at higher side Ws
        wa = 2*pi*f1;
        wb = 2*pi*f2;
        ws = 2*pi*fs;

% Selectivity
        S = (2*ws - wb - wa)/(wb-wa);
        % Displaying the selectivity
        Selectivity=num2str(S); % converting number to sder:tring
        set(handles.edit14_selectivity,'String',Selectivity); % Display the converted string

        % Degree
        Degree = (LA + LR + 6)/(20*log10(S + sqrt(S^2 - 1)));
        display('Degree of the filter:')
        Degree_rounded = round(Degree);
        if Degree_rounded<Degree
            N = Degree_rounded + 1;
        elseif Degree_rounded>=Degree
            N = Degree_rounded;
        else
        end
        N;
        % Displaying the filter order/degree
         filter_order=num2str(N); % converting number to sder:tring
         set(handles.FilterOrder,'String',filter_order); % Display the converted string

%     Reflection Coefficient
        Ref = 10^(-LR/20);
%     dB pass-band ripple
        Lar = -10*log10(1-Ref^2);
% Ripple level epsillon
        epsillon = sqrt(10^(0.1*Lar) - 1);
        %Displaying ripple value in GUI
        ripple_level=num2str(epsillon); % converting number to sder:tring
        set(handles.edit15_ripple,'String',ripple_level); % Display the converted string

        f = 0:0.01:30;              % Range of the frequency in GHz
        w = 0:(2*pi*0.01):(2*pi*30);   % Range of w = 2*pi*f.

        wo    = sqrt(wa*wb);
        fo    = wo/(2*pi);
        alpha = wo/(wb-wa);
```

```matlab
        for i=1:1:3001
            omega(i) = alpha*((w(i)./wo)-(wo./w(i)));

        if (omega(i)<=1)

            Deno = 1 + (epsillon*cos(N*acos(alpha*((w(i)./wo)-(wo./w(i)))))).^2;

        elseif (omega(i)>=1)

            Deno = 1 + (epsillon*cosh(N*acosh(alpha*((w(i)./wo)-(wo./w(i)))))).^2;
        else
        end

 % Computing S12 values for different values ABCD resulting from the variation in w.
    S1(i) = abs(1./Deno);
    S10(i) = 1 - abs(1./Deno);
     end

        S12 = 10*log10(S1);
        S11 = 10*log10(S10);

        plot(f,S12,'blue', f,S11,'red')
        graph = plot(f,S12,'blue', f,S11,'red');
        % Setting the line width or thickness for both curves
        set (graph, 'LineWidth',1.5);
        title('Chebyshev Response BP');
        ylabel('|S12|^2 and |S11|^2 in dB');
        xlabel('f in GHz');
        axis([0 (f1+fs) -80 5]);
        grid on;

    else
        helpdlg('Please check your input parameters. Make sure that f1s>=0.3GHz, f2>f1, fs>f2, f2<=300 GHz, LA>LR and LR>0 dB.');
    end
% |===================================================================|
% |                                                                   |
% |      Plotting Chebyshev HP Prototypes using A,B,C,D parameters:   |
% |                              4                                    |
%                    S12|^2 = -------------------                     |
% |                            |A+(B/Zo)+(C*Zo)+D|^2                  |
% |===================================================================|
elseif ChebyshevBP == 2   % Chebyshev BP Shunt Prototype to plot with its ABCD Parameters
% |*******************************************************************|
% |                  Plotting Chebyshev BP Shunt Prototype             |
% |*******************************************************************|
    % Generation an input dialog box for Chebyshev BP Shunt Prototype
    prompt={'INSERTION LOSS [dB]','RETURN LOSS [dB]','LOWER SIDE CUTOFF FREQUENCY f1 [GHz]','UPPER SIDE CUTOFF FREQUENCY f2 [GHz]','STOPBAND FREQUENCY AT HIGHER SIDE fs [GHz]','SYSTEM IMPEDANCE Zo [OHMS]'};
    title1='CHEBYSHEV BP SHUNT PROTOTYPE';
    numlines=1;
    defaultanswer={'','','','','',''};
    options.Resize='on';
    options.WindowStyle='normal';
    options.Interpreter='tex';
    input_array = inputdlg(prompt,title1,numlines,defaultanswer, options); % Complete input dialog box

    LA = str2num(input_array{1}); % Accepting first element of the input as LA
    LR = str2num(input_array{2}); % Accepting 2nd element of the input as LR
    f1 = str2num(input_array{3}); % Accepting 3rd element of the input as f1
    f2 = str2num(input_array{4});% Accepting 4th element of the input as f2
    fs = str2num(input_array{5});% Accepting 4th element of the input as fs
    Zo = str2num(input_array{6}); % Accepting last element of the input as Zo

    if (LA>LR) && (LR>=0) && (f1>=0.3) && (f2>f1) && (f2<fs) && (fs<=300)

        % Cut-off Frequencies W1 and W2; and Stopband frequency at higher side Ws
        wa = 2*pi*f1;
```

```matlab
            wb = 2*pi*f2;
            ws = 2*pi*fs;

% Selectivity
     S = (2*ws - wb - wa)/(wb-wa);
      % Displaying the selectivity
     Selectivity=num2str(S); % converting number to sder:tring
     set(handles.edit14_selectivity,'String',Selectivity); % Display the converted string

      % Degree
     Degree = (LA + LR + 6)/(20*log10(S + sqrt(S^2 - 1)));
     display('Degree of the filter:')
     Degree_rounded = round(Degree);
     if Degree_rounded<Degree
         N = Degree_rounded + 1;
     elseif Degree_rounded>=Degree
         N = Degree_rounded;
     else
     end
     N;
      % Displaying the filter order/degree
      filter_order=num2str(N); % converting number to sder:tring
      set(handles.FilterOrder,'String',filter_order); % Display the converted string

%    Reflection Coefficient
     Ref = 10^(-LR/20);
%    dB pass-band ripple
     Lar = -10*log10(1-Ref^2);
% Ripple level epsillon
     epsillon = sqrt(10^(0.1*Lar) - 1);
     %Displaying ripple value in GUI
     ripple_level=num2str(epsillon); % converting number to sder:tring
     set(handles.edit15_ripple,'String',ripple_level); % Display the converted string

      % Computation of element values,  Ref. G. Mattahaei MW Filters, Pg 99
R = coth(Lar/17.37);
beta = log(R);
gamma = sinh(beta/(2*N));
k = N+1;

for i=1:1:k
    a(i) = sin(((2*i-1).*pi)./(2*N));
    b(i) = gamma^2 + [sin((i.*pi)./N)]^2;

    if i==1

    g(1)= 2.*a(1)/gamma;

    elseif (i>1) && (i<=N)
        g(i) = (4.*a(i-1).*a(i))./(b(i-1).*g(i-1));

    else

        A = mod (N,2);
        if A==0
            g(k)=[coth(beta/4)]^2;

        elseif A==1

            g(k)=1;

        end

    end

end

%  Plotting with element values for different filter order
```

```matlab
% Generatng L and C elements for BP Prototype - Shunt
wo    = sqrt(wa*wb);
fo = wo/(2*pi);
alpha = wo/(wb-wa);
for i=1:1:N
    if i==1
        % Computing C(1) value
        Element(1) = g(1)/Zo;
        %Transforming Shunt C to Shunt LC circuit
        Inductor(1)= 1/(alpha*wo*Element(1));
        Capacitor(1)= (alpha*Element(1))/wo;
    elseif (i>1) && (i<=N)
        k=mod(i,2);
        if k==0
            % Computing L values for odd values of i
            Element(i) = g(i)*Zo;
            %Transforming series L to series LC circuit
            Inductor(i)= (alpha*Element(i))/wo;
            Capacitor(i)=1/(alpha*wo*Element(i));
        elseif k==1
            % Computing C values for even values of i
            Element(i) = g(i)/Zo;
            %Transforming Shunt C to Shunt LC circuit
            Inductor(i)= 1/(alpha*wo*Element(i));
            Capacitor(i)= (alpha*Element(i))/wo;
        else
        end
    else
    end
end
%Scaled reactive element values and scaled source Zs and load ZL impedances, Z = [Zs ZL]
Element;
Inductor(Inductor==0)=[];
Inductor;                      %L remains in nH.
Capacitor(Capacitor==0)=[];
Capacitor2 = 1000.*Capacitor;   %to convert from nF to pF.
Zs = 1*Zo;                     % Source impedance
ZL = g(N+1)*Zo;                % Load impedance in Ohm
Impedance = [Zs ZL];           %Z are in Ohms. in Ohm

% Displaying the inductor, capacitor and impedance values.
Inductor1 = mat2str(Inductor,4);
set(handles.InductorValues1,'String',Inductor1); % Display the converted string
Capacitor1 = mat2str(Capacitor2,4);
set(handles.CapacitorValues1,'String',Capacitor1); % Display the converted string
Impedance1 = mat2str(Impedance,4);
set(handles.ImpedanceValues1,'String',Impedance1); % Display the converted string

f = 0:0.01:30;                 % Range of the frequency in GHz
w = 0:(2*pi*0.01):(2*pi*30);   % Range of w = 2*pi*f.

 % Computing Z, Y and T = [A B;C D] values for different frequency values
 % This is for Chebyshev BP Prototype - Series

    for i=1:1:3001
        for t=1:1:N
            if t==1
               % Computing Z(C1) values
               Z(1) = j*Element(1).*((alpha*((w(i)./wo)-(wo./w(i)))));
               T = [1 0;Z(1) 1];
            elseif (t>1) && (t<=N)
                  k=mod(t,2);
               if k==0
                   % Computing Z(L) values
                   Z(t) = j*Element(t).*((alpha*((w(i)./wo)-(wo./w(i)))));
                   Tnew = T*[1 Z(t);0 1];
                   T = Tnew;
               elseif k==1
```

```matlab
                        % Computing Z(C) values
                        Z(t) = j*Element(t).*((alpha*((w(i)./wo)-(wo./w(i))))); 
                        Tnew = T*[1 0;Z(t) 1];
                        T = Tnew;
                    else
                    end
                else
                end
            end
            
        T;

% Getting ABCD values by identification using 2X2 Matrix T=[A B;C D] by using
    A = T(1,1);
    B = T(1,2);
    C = T(2,1);
    D = T(2,2);

 % Computing S12 values for different values ABCD resulting from the
 % variation in w.
    Sa(i) = abs(2./(A+(B./Zo)+(C.*Zo)+D));
    Sb(i) = 1 - abs((2./(A+(B./Zo)+(C.*Zo)+D)).^2);
   end

%                                        4
% Plotting with A,B,C,D parameters, |S12|^2 = --------------------
%                                       |A+(B/Zo)+(C*Zo)+D|^2

    S12 = 20*log10(Sa);  % Computing the dB values of |S12|^2
    S11 = 10*log10(Sb);

    plot(f,S12,'blue', f,S11,'red')
    graph = plot(f,S12,'blue', f,S11,'red');
    % Setting the line width or thickness for both curves
    set (graph, 'LineWidth',1.5);
    title('Chebyshev Response for BP Shunt Prototype');
    ylabel('|S12|^2 and |S11|^2 in dB');
    xlabel('f in GHz');
    axis([0 (f1+f2) -80 3]);
    grid on;

% |**********************************************************************|
% |                End Plotting Chebyshev BP shunt Prototype             |
% |**********************************************************************|

    else
        helpdlg('Please check your input parameters. Make sure that f1>=0.3GHz, f2>f1, fs>f2, f2<=300 GHz, LA>LR, LR>0 dB and it is recommended to choose Zo=50 Ohms');
    end

elseif ChebyshevBP == 3    % Chebyshev BP Series Prototype to plot with its ABCD Parameters
% |**********************************************************************|
% |                 Chebyshev - Series Band-Pass Prototype               |
% |**********************************************************************|
    % Generation an input dialog box for Chebyshev BP Series Prototype
    prompt={'INSERTION LOSS [dB]','RETURN LOSS [dB]','LOWER SIDE CUTOFF FREQUENCY f1 [GHz]','UPPER SIDE CUTOFF FREQUENCY f2 [GHz]','STOPBAND FREQUENCY AT HIGHER SIDE fs [GHz]','SYSTEM IMPEDANCE Zo [OHMS]'};
    title1='CHEBYSHEV BP SERIES PROTOTYPE';
    numlines=1;
    defaultanswer={'','','','','',''};
    options.Resize='on';
    options.WindowStyle='normal';
    options.Interpreter='tex';
    input_array = inputdlg(prompt,title1,numlines,defaultanswer, options); % Complete input dialog box

    LA = str2num(input_array{1}); % Accepting first element of the input as LA
    LR = str2num(input_array{2}); % Accepting 2nd element of the input as LR
    f1 = str2num(input_array{3}); % Accepting 3rd element of the input as f1
    f2 = str2num(input_array{4});% Accepting 4th element of the input as f2
```

```matlab
        fs = str2num(input_array{5});% Accepting 4th element of the input as fs
        Zo = str2num(input_array{6}); % Accepting last element of the input as Zo

    if (LA>LR) && (LR>0) && (f1>=0.3) && (f2>f1) && (f2<fs) && (fs<=300)
        % Cut-off Frequencies W1 and W2; and Stopband frequency at higher side Ws
        wa = 2*pi*f1;
        wb = 2*pi*f2;
        ws = 2*pi*fs;

% Selectivity
        S = (2*ws - wb - wa)/(wb-wa);
        % Displaying the selectivity
        Selectivity=num2str(S); % converting number to sder:tring
        set(handles.edit14_selectivity,'String',Selectivity); % Display the converted string

        % Degree
        Degree = (LA + LR + 6)/(20*log10(S + sqrt(S^2 - 1)));
        display('Degree of the filter:')
        Degree_rounded = round(Degree);
        if Degree_rounded<Degree
            N = Degree_rounded + 1;
        elseif Degree_rounded>=Degree
            N = Degree_rounded;
        else
        end
        N;
        % Displaying the filter order/degree
         filter_order=num2str(N); % converting number to sder:tring
         set(handles.FilterOrder,'String',filter_order); % Display the converted string

%     Reflection Coefficient
        Ref = 10^(-LR/20);
%     dB pass-band ripple
        Lar = -10*log10(1-Ref^2);
% Ripple level epsillon
       epsillon = sqrt(10^(0.1*Lar) - 1);
        %Displaying ripple value in GUI
        ripple_level=num2str(epsillon); % converting number to sder:tring
        set(handles.edit15_ripple,'String',ripple_level); % Display the converted string

     % Computation of element values,  Ref. G. Mattahaei MW Filters, Pg 99
R = coth(Lar/17.37);
beta = log(R);
gamma = sinh(beta/(2*N));
k = N+1;

for i=1:1:k
    a(i) = sin(((2*i-1).*pi)./(2*N));
    b(i) = gamma^2 + [sin((i.*pi)./N)]^2;

    if i==1
    g(1)= 2.*a(1)/gamma;
    elseif (i>1) && (i<=N)
        g(i) = (4.*a(i-1).*a(i))./(b(i-1).*g(i-1));
    else
        A = mod (N,2);
        if A==0
            g(k)=[coth(beta/4)]^2;
        elseif A==1
            g(k)=1;
        end
    end
end

%  Plotting with element values for different filter order

% Generatng L and C elements for BP Prototype - Series
wo    = sqrt(wa*wb);
fo = wo/(2*pi);
```

```matlab
alpha = wo/(wb-wa);

for i=1:1:N
    if i==1
        Element(1) = g(1)*Zo; % Computing L1 values
        %Transforming series L to series LC circuit
        Inductor(1)= (alpha*Element(1))/wo;
        Capacitor(1)=1/(alpha*wo*Element(1));
    elseif (i>1) && (i<=N)
        k=mod(i,2);
        if k==0
           Element(i) = g(i)/Zo; % Computing C values
           %Transforming Shunt C to Shunt LC circuit
           Inductor(i)= 1/(alpha*wo*Element(i));
           Capacitor(i)= (alpha*Element(i))/wo;
        elseif k==1
           Element(i) = g(i)*Zo; % Computing L values
           %Transforming series L to series LC circuit
           Inductor(i)= (alpha*Element(i))/wo;
           Capacitor(i)=1/(alpha*wo*Element(i));
        else
        end
    else
    end
end
%Scaled reactive element values and scaled source Zs and load ZL impedances, Z = [Zs ZL)
Element;
Inductor(Inductor==0)=[];
Inductor;                       %L remains in nH.
Capacitor(Capacitor==0)=[];
Capacitor2 = 1000.*Capacitor;   %to convert from nF to pF.
Zs = 1*Zo;                      % Source impedance
ZL = g(N+1)*Zo;                 % Load impedance in Ohm
Impedance = [Zs ZL];            %Z are in Ohms. in Ohm

% Displaying the inductor, capacitor and impedance values.
Inductor1 = mat2str(Inductor,4);
set(handles.InductorValues1,'String',Inductor1); % Display the converted string
Capacitor1 = mat2str(Capacitor2,4);
set(handles.CapacitorValues1,'String',Capacitor1); % Display the converted string
Impedance1 = mat2str(Impedance,4);
set(handles.ImpedanceValues1,'String',Impedance1); % Display the converted string

f = 0:0.01:30;                  % Range of the frequency in GHz
w = 0:(2*pi*0.01):(2*pi*30);    % Range of w = 2*pi*f.

 % Computing Z, Y and T = [A B;C D] values for different frequency values
 % This is for Chebyshev BP Prototype - Series

    for i=1:1:3001
        for t=1:1:N
            if t==1
                Z(1) = j*Element(1).*((alpha*((w(i)./wo)-(wo./w(i)))));  % Computing Z(L1) values
                T = [1 Z(1);0 1];
            elseif (t>1) && (t<=N)
                    k=mod(t,2);
                if k==0
                    Z(t) = j*Element(t).*((alpha*((w(i)./wo)-(wo./w(i))))); % Computing Zc values
                    Tnew = T*[1 0;Z(t) 1];
                    T = Tnew;
                elseif k==1
                    Z(t) = j*Element(t).*((alpha*((w(i)./wo)-(wo./w(i))))); % Computing Z(L) values
                    Tnew = T*[1 Z(t);0 1];
                    T = Tnew;
                else
                end
            else
            end
        end
```

```matlab
        T;

% Getting ABCD values by identification using 2X2 Matrix T=[A B;C D] by using
    A = T(1,1);
    B = T(1,2);
    C = T(2,1);
    D = T(2,2);

 % Computing S12 values for different values ABCD resulting from the
 % variation in w.
    Sa(i) = abs(2./(A+(B./Zo)+(C.*Zo)+D));
    Sb(i) = 1 - abs((2./(A+(B./Zo)+(C.*Zo)+D)).^2);
   end

%                                                  4
% Plotting with A,B,C,D parameters, |S12|^2 = -------------------
%                                             |A+(B/Zo)+(C*Zo)+D|^2

    S12 = 20*log10(Sa);  % Computing the dB values of |S12|^2
    S11 = 10*log10(Sb);

    plot(f,S12,'blue', f,S11,'red')
    graph = plot(f,S12,'blue', f,S11,'red');
    % Setting the line width or thickness for both curves
    set (graph, 'LineWidth',1.5);
    title('Chebyshev Response of BP Series Prototype');
    ylabel('|S12|^2 and |S11|^2 in dB');
    xlabel('f in GHz');
    axis([0 (f1+f2) -80 3]);
    grid on;

% |***********************************************************************|
% |                  End Plotting Chebyshev BP Series Prototype            |
% |***********************************************************************|

    else
        helpdlg('Please check your input parameters. Make sure that f1s>=0.3GHz, f2>f1, fs>f2, f2<=300 GHz, LA>LR, LR>0 dB and it is recommended to choose Zo=50 Ohms');
    end

elseif ChebyshevBP == 4   % Chebyshev BP Capacitively Coupled Filter

 % Generation an input dialog box for Chebyshev BP Capacitively Coupled Filter
    prompt={'INSERTION LOSS [dB]','RETURN LOSS [dB]','CENTER FREQUENCY fO [GHz]','PASSBAND BANDWIDTH [GHz]','STOPBAND BANDWIDTH [GHz]','SYSTEM IMPEDANCE Zo [OHMS]'};
    title1='CHEBYSHEV BP CAPACITIVELY COUPLED FILTER';
    numlines=1;
    defaultanswer={'','','','','',''};
    options.Resize='on';
    options.WindowStyle='normal';
    options.Interpreter='tex';
    input_array = inputdlg(prompt,title1,numlines,defaultanswer, options); % Complete input dialog box

    LA = str2num(input_array{1});  % Accepting first element of the input as LA
    LR = str2num(input_array{2});  % Accepting 2nd element of the input as LR
    fo = str2num(input_array{3});  % Accepting 3rd element of the input as fo
    BWp = str2num(input_array{4}); % Accepting 4th element of the input as BWp
    BWs = str2num(input_array{5}); % Accepting 4th element of the input as BWs
    Zo = str2num(input_array{6});  % Accepting last element of the input as Zo

    fp1 = fo - (BWp/2);
    fp2 = fo - (BWp/2);

    if (LA>LR) && (LR>0) && (BWp>0) && (BWs>BWp) && (fp1>=0.3) && (fp2<=300)
      % Selectivity
         S= BWs/BWp;
         % Displaying the selectivity
         Selectivity=num2str(S); % converting number to sder:tring
```

```matlab
    set(handles.edit14_selectivity,'String',Selectivity); % Display the converted string

 % Degree/Order
  Degree = (LA + LR + 6)/(20*log10(S + sqrt(S^2 - 1)));
  display('Degree of the filter:')
  Degree_rounded = round(Degree);
 if Degree_rounded<Degree
     N = Degree_rounded + 1;
 elseif Degree_rounded>=Degree
  N = Degree_rounded;
 else
 end
 N;
 % Displaying the filter order/degree
 filter_order=num2str(N); % converting number to sder:tring
 set(handles.FilterOrder,'String',filter_order); % Display the converted string

 %    Reflection Coefficient
 Ref = 10^(-LR/20);
 %    dB pass-band ripple
 Lar = -10*log10(1-Ref^2);
 % Ripple level epsillon
 epsillon = sqrt(10^(0.1*Lar) - 1);
 %Displaying ripple value in GUI
 ripple_level=num2str(epsillon); % converting number to sder:tring
 set(handles.edit15_ripple,'String',ripple_level); % Display the converted string

 f = 0:0.01:30;              % Range of the frequency in GHz
 w = 0:(2*pi*0.01):(2*pi*30);   % Range of w = 2*pi*f.

 % wo    = sqrt(wa*wb);
 wo = 2*pi*fo;
 % fo    = wo/(2*pi);
 alpha = fo/(BWp);

 % Computation of element values,  Ref. Ian Hunter Page 126 to 129.
  eta = sinh((1/N)*asinh(1/epsillon));

 % Generating Cr values
 for r=1:1:N
     C(r) = (2/eta)*sin(((2.*r-1).*pi)./(2*N));
 end

 % Generating Kr,r+1 values

 for r=1:1:(N-1)
     K(r) = (sqrt(eta^2 + (sin((pi.*r)./N)).^2))./eta;
 end
K(K==0)=[];

for i=1:1:(N+1)

    for k=2:1:(N+2)

    if (i==1) && (k==2)
        % C(01) Value
        Element2(1,2)= 1./(wo*sqrt(alpha-1));
        Element(1) = Element2(1,2);
    elseif (i==2) && (k==2)
        % C(11) Value
        Element2(2,2) = C(1)/wo - (sqrt(alpha-1)/(wo*alpha)) - K(1)/(wo*alpha);
        Element(i) = Element2(2,2);
        % L(1,1) values
         Element3(i) = 1./(wo*C(i-1));
    elseif (i==2) && (k==3)
        % C(1,2)
        Element2(2,3) = K(1)/(wo*alpha);
        Element(3) = Element2(2,3);
```

```matlab
            elseif (i>2) && (i<=N) && (k>=3) && (k<=(N+1))
                
                if i==k
                    h = (i+k-2)./2;
                     % C(r,r) values
                    Element2(i,k) = C(h)/(wo) - K(h-1)/(wo*alpha) - K(h)/(wo*alpha);
                    
                    
                    Element(i+k-2)=Element2(i,k);
                    % L(r,r) values
                    Element3(i,k) = 1./(wo*C(i-1));
                    
                elseif i<k
                    % C(r,r+1) Value
                    Element2(i,k)= K(i-1)./(alpha*wo);
                    Element(i+k-2) = Element2(i,k);
                else
                end
            elseif (i==N+1) && (k==N+1)
                % C(N,N)
                Element2(i,k) = C(N)/wo - (sqrt(alpha-1))/(wo*alpha) - K(N-1)/(wo*alpha);
                Element(i+k-2) = Element2(i,k);
                
                % L(r,r) values
                 Element3(i,k) = 1./(wo*C(i-1));
                
            elseif (i==N+1) && (k==N+2)
                % Last capacitive element C(N,N+1)
                 Element2(i,k) = 1./(wo*sqrt(alpha-1));
                 Element(i+k-2) = Element2(i,k);
            else
            end
            end
       
end
% Normalized Capacitive Elements
Element2;
Element;

%Normalized Inductive Elements
Element3(Element3==0)=[];
Element3;

format long;
% Scaled Capcitive Elements
Capacit  = Element./Zo;

% Scaled inductive Elements
Induct = Element3.*Zo;

%Scaled reactive element values and scaled source Zs and load ZL impedances, Z = [Zs ZL]
Inductor = Induct;          %L remains in nH.
Capacitor2 = 1000.*Capacit;  %to convert from nF to pF.
Impedance = [Zo Zo];        %Z are in Ohms.

% Displaying the inductor, capacitor and impedance values.
Inductor1 = mat2str(Inductor,4);
set(handles.InductorValues1,'String',Inductor1); % Display the converted string
Capacitor1 = mat2str(Capacitor2,4);
set(handles.CapacitorValues1,'String',Capacitor1); % Display the converted string
Impedance1 = mat2str(Impedance,4);
set(handles.ImpedanceValues1,'String',Impedance1); % Display the converted string

f = 0:0.001:30;              % Range of the frequency in GHz
w = 0:(2*pi*0.001):(2*pi*30);   % Range of w = 2*pi*f.
M = 2*N+1;

for i=1:1:30001
 for t=1:1:M
     if t==1
```

```matlab
                % Impedance for the first Series Element C(01)
                Z(1) = -j./(w(i).*Capacit(1));
                % T Matrix
                T = [1 Z(1);0 1];
            elseif (t>1) && (t<=M)
                    k=mod(t,2);
                if k==0
                    u = t/2;
                    % Equivalent Admittance for the the shunt Elements Lrr
                    % and Crr
                    Z(t) = j.*((w(i).*Capacit(t))-(1./(w(i).*Induct(u))));
                    % T Matrix
                    Tnew = T*[1 0;Z(t) 1];
                    T = Tnew;
                elseif k==1
                    % Impedance for the series elements Cr,r+1
                    Z(t) = -j./(w(i).*Capacit(t));
                    % T matrix
                    Tnew = T*[1 Z(t);0 1];
                    T = Tnew;
                else
                end
            else
            end
        end

    T;

% Getting ABCD values by identification using 2X2 Matrix T=[A B;C D] by using
    A = T(1,1);
    B = T(1,2);
    C = T(2,1);
    D = T(2,2);

 % Computing S12 values for different values ABCD resulting from the
 % variation in w.
    Sa(i) = abs(2./(A+(B./Zo)+(C.*Zo)+D));
    Sb(i) = abs(1 - (abs((2./(A+(B./Zo)+(C.*Zo)+D)))).^2);
    end

%                                              4
% Plotting with A,B,C,D parameters, |S12|^2 = -------------------
%                                             |A+(B/Zo)+(C*Zo)+D|^2

    S12 = 20*log10(Sa);  % Computing the dB values of |S12|^2
    S11 = 10*log10(Sb);

    plot(f,S12,'blue', f,S11,'red')
    graph = plot(f,S12,'blue', f,S11,'red');
    % Setting the line width or thickness for both curves
    set (graph, 'LineWidth',1.5);
    title('Response for Chebyshev Capacitively Coupled BPF');
    ylabel('|S12|^2 and |S11|^2 in dB');
    xlabel('f in GHz');
    axis([0 (2*fo) -80 5]);
    grid on;

    else
        helpdlg('Please check your input parameters. The BWp = fp2 - fp1, make sure that fp1>=0.3GHz, BWp>0,
BWs>BWp, fp2<=300 GHz, LA>LR, LR>0 dB and it is recommended to choose Zo=50 Ohms');
    end

elseif ChebyshevBP == 5   % Chebyshev BP Combline FIlter

    % Generation an input dialog box for Chebyshev BP Combline FIlter
    prompt={'RETURN LOSS [dB]','CENTER FREQUENCY fO [GHz]','PASSBAND BANDWIDTH [GHz]','FILTER ORDER','SYSTEM
IMPEDANCE Zo [OHMS]'};
    title1='CHEBYSHEV BP COMBLINE FILTER';
    numlines=1;
```

```matlab
    defaultanswer={'','','','','',''};
    options.Resize='on';
    options.WindowStyle='normal';
    options.Interpreter='tex';
    input_array = inputdlg(prompt,title1,numlines,defaultanswer, options); % Complete input dialog box

 LR = str2num(input_array{1}); % Accepting first element of the input as LA
 fo= str2num(input_array{2}); % Accepting 3rd element of the input as f1
 BW = str2num(input_array{3});% Accepting 4th element of the input as f2
 N = str2num(input_array{4});% Accepting 4th element of the input as fs
 Zo = str2num(input_array{5}); % Accepting last element of the input as Zo

 fs1 = fo - (BWs/2);
if (LR>=0) && (BW>0) && (fo>0) && (fo<=300) && (Zo==50)

helpdlg('For Combline filter, only the element values can be generated. No frequency response display.');

%Passband bandwidth in rad/s
    deltaw=2*pi*BW;
    %Center frequency in rad/s
    wo=2*pi*fo;

    %Theta0 is chosen to be 50degree for optimum response with BW=cte
    theta0=(50/180)*pi;              % i.e. 0.8726 radians approximately

    %    Reflection Coefficient
    Ref = 10^(-LR/20);
    %    dB pass-band ripple
    Lar = -10*log10(1-Ref^2);
    % Ripple level epsillon
    epsillon = sqrt(10^(0.1*Lar) - 1);
    % Ref. Ian Hunter Page 126 to 129.
    eta = sinh((1/N)*asinh(1/epsillon));
    %Bandwidth scaling factor alpha and the constant beta. Ref. Ian Hunter Page 191.
    alpha = 2*wo*tan(theta0)/(deltaw*(tan(theta0)+theta0*(1+(tan(theta0))^2)));
    beta = 1/(wo*tan(theta0));
    %Tranformer Capacitor to match input/output to the resonator
    %By choosing Yrr = 1, C = beta*Yrr thus:
    Cb = beta;
    % Scaled value of C is Cb/50 in nF
    C = Cb/50;

    %Generating CLr values.
    for r=1:1:N
        CL(r) = (2/eta)*sin(((2.*r-1).*pi)./(2*N));
    end

    % Generating Kr,r+1 values
    for r=1:1:(N-1)
        K(r) = (sqrt(eta^2 + (sin((pi.*r)./N)).^2))./eta;
    end
 K(K==0)=[];

 % Yrr is chosen to be 1 early in the calculation of C.
 Yrr = 1;
 % nr values
 for r=1:1:N
     n(r)=((alpha*CL(r)*tan(theta0))./Yrr).^0.5;
 end

 % Yr,r+1 values
 for r=1:1:(N-1)
     Yrr1(r) = K(r)*tan(theta0)/(n(r)*n(r+1));
 end

 %Yr values
 for r=2:1:(N-1)
     Y1(r) = Yrr - Yrr1(r-1)-Yrr1(r);
 end
```

```matlab
% Y0, Y1, YN and YN+1 values
Y1(1) = Yrr - Yrr1(1) + (1/(n(1))^2) - 1/(n(1)*cos(theta0));
Y1(N) = Y1(1);
Y0 = 1 - 1/(n(1)*cos(theta0));
Y1(N+1) = Y0;
%Y01 and YN,N+1
Y01 = 1/(n(1)*cos(theta0));
YNN1 = Y01;

%Putting all Yr elements together in one matrix (Admittance)
%Getting the corresponding scaled impedance
for r=1:1:(N+2)
    if r==1
        Y(r) = Y0;
        Zodd(r) = 50*(1/Y(r));
    elseif r>1
        Y(r) = Y1(r-1);
        Zodd(r) = 50*(1/Y(r));
    else
    end
end

  %Putting all Yr,r+1 elements together in one matrix (Admittance)
  %Getting the corresponding scaled impedance
for r=1:1:(N+1)
    if r==1
        Yr(r) = Y01;
        Zeven(r) = 50*(1/Yr(r));
    elseif (r>1) && (r<=N)
        Yr(r) = Yrr1(r-1);
        Zeven(r) = 50*(1/Yr(r));
    elseif r==(N+1)
        Yr(r) = YNN1;
        Zeven(r) = 50*(1/Yr(r));
    else
    end
end

 f = 0:0.001:30;                   % Range of the frequency in GHz
 w = 0:(2*pi*0.001):(2*pi*30);    % Range of w = 2*pi*f.
 M = 2*N+3;

 for i=1:1:30001
  for t=1:1:M
      if t==1
          % Impedance for the first Series Element C(01)
          Z(1) = -j/Zodd(1);
          % T Matrix
          T = [1 0;Z(1) 1];
      elseif (t>1) && (t<=M)
            k=mod(t,2);
          if k==0
              u = t/2;
              Z(t) = -j.*Zeven(u);
              % T Matrix
              Tnew = T*[1 Z(t);0 1];
              T = Tnew;
          elseif k==1
              v = round(t/2);
              Z(t) = j.*w(i).*C - j/Zodd(v);
              % T matrix
              Tnew = T*[1 0; Z(t) 1];
              T = Tnew;
          else
          end
      else
      end
  end
```

```matlab
    T;

% Getting ABCD values by identification using 2X2 Matrix T=[A B;C D] by using
    A = T(1,1);
    B = T(1,2);
    C = T(2,1);
    D = T(2,2);

 % Computing S12 values for different values ABCD resulting from the
 % variation in w.
    Sa(i) = abs(2./(A+(B./Zo)+(C.*Zo)+D));
    Sb(i) = abs(1 - (abs((2./(A+(B./Zo)+(C.*Zo)+D)))).^2);
   end

%                                                    4
% Plotting with A,B,C,D parameters, |S12|^2 = -------------------
%                                             |A+(B/Zo)+(C*Zo)+D|^2

    S12 = 20*log10(Sa);   % Computing the dB values of |S12|^2
    S11 = 10*log10(Sb);

    plot(f,S12,'blue', f,S11,'red')
    graph = plot(f,S12,'blue', f,S11,'red');
    % Setting the line width or thickness for both curves
    set (graph, 'LineWidth',1.5);
    title('Response for Chebyshev Combline Filter');
    ylabel('|S12|^2 and |S11|^2 in dB');
    xlabel('f in GHz');
    axis([0 (2*fo) -80 5]);
    grid on;

    else
        helpdlg('Please check your input parameters. The BWp = fp2 - fp1, make sure that fp1>=0.3GHz, BWp>0, BWs>BWp, fp2<=300 GHz, LA>LR, LR>0 dB and it is recommended to choose Zo=50 Ohms');
    end

elseif ChebyshevBP == 6   % 4th Order Chebyshev UWB BPF from its generalized equation

     % Generation an input dialog box for  4th Chebyshev UWB BPF
     prompt={'RETURN LOSS [dB]','LOWER CUTOFF FREQUENCY f1 [GHz]','UPPER CUTOFF FREQUENCY f2 [GHz]','FILTER ORDER'};
     title1='CHEBYSHEV UWB BPF';
     numlines=1;
     defaultanswer={'','','',''};
     options.Resize='on';
     options.WindowStyle='normal';
     options.Interpreter='tex';
     input_array = inputdlg(prompt,title1,numlines,defaultanswer, options); % Complete input dialog box

     LR = str2num(input_array{1}); % Accepting 2nd element of the input as LR
     LPF = str2num(input_array{2}); % Accepting 3rd element of the input as f1
     HPF = str2num(input_array{3});% Accepting 4th element of the input as f2
     order = str2num(input_array{4});% Accepting 4th element of the input as fs

     %    Reflection Coefficient
     Ref = 10^(-LR/20);
     %    dB pass-band ripple
     Lar = -10*log10(1-Ref^2);
     % Ripple level epsilon
     epsilon = sqrt(10^(0.1*Lar) - 1);
     %Displaying ripple value in GUI
     ripple_level=num2str(epsilon); % converting number to sder:tring
     set(handles.edit15_ripple,'String',ripple_level); % Display the converted string

     if (order==4)&&(LPF>=3.1)&&(HPF>LPF)&&(HPF<=10.6)

     % Computing the Bandwidth of the UWB Band Pass Filter in GHz
     BWD = HPF - LPF;
```

```matlab
        % Computing the center frequency fc in GHz
        fc = LPF + (BWD/2);

        % Converting the Bandwidth (GHz) value in radian
        Bandwidth = (pi/2)*(BWD/fc);

        % Computing alpha and sigma components
        alpha = (3/4)*(cos(Bandwidth/2) + (1/3))^2 - 4/3 ;
        sigma = 0.25*(sin(Bandwidth/2))^2*(1-cos(Bandwidth/2));

        % Checking for the design crieria on alpha & sigma values that satisfy
        % the following inequality:
        % alpha^2 - 4*sigma >0 and alpha + sigma + 1 >0
        cond1 = alpha^2 - 4*sigma;
        cond2 = alpha + sigma + 1;

        if (cond1>0)&&(cond2>0)

        % Computing the filtering coefficient A, B and C for 4th order UWB BPF
        A = epsilon/sigma;
        B = alpha*A;
        C = epsilon;

        % Computing the variables: theta and frequency ranges respectively
        f = 0:0.001:15;
        theta = (pi/2)*(f./fc);

        % Computing the UWB Bandpass Filtering Function
        T_num = A.*((cos(theta)).^4 + alpha*(cos(theta)).^2 + sigma);
        T_den = sin(theta);
        T_theta = T_num./T_den;

        % Computing the absolute value of T(theta)
        T = abs(T_theta);

        % Computing the dB values of S11 (squared) and S12 (squared)
        S11 = 20*log10(T) - 10*log10(1 + T.^2);
        S21 = - 10*log10(1 + T.^2);

        % Ploting the fequency responses of |S11|^2 and |S12|^2

        plot(f,S21,'blue', f,S11,'red');
        % Inserting legend for the plot
        graph = plot(f,S21,'blue', f,S11,'red');
        % Setting the line width or thickness for both curves
        set (graph, 'LineWidth',1.5);
        % Labelling x-axis, y-axis and giving title for the figure.
        title('Frequency Response of |S11|^2 and |S21|^2');
        xlabel('Frequency in GHz');
        ylabel('|S11|^2 and |S21|^2 in dB');
        % Setting min and max value of x-axis and y-axis
        axis([0 15 -90 10]);
        grid on;
        % Inserting legend for UWB BPF plot
        legend('S21','S11');
        % Moving the legend to the lower right corner of the graph
        legend('location','SouthEast');

        else
         helpdlg('Your design parameters violate the design criteria for 4th order UWB Filter Design.Please check your input parameters. Make sure that f1>=3.1 GHz; f2>f1; f2<=10.6 GHz, LR>=0 dB and it works for 4th order only.');
        end

        else
         helpdlg('Your design parameters violate the design criteria for 4th order UWB Filter Design.Please check your input parameters. Make sure that f1>=3.1 GHz; f2>f1; f2<=10.6 GHz, LR>=0 dB and it works for 4th order only.');
        end
```

```matlab
            else
            end

        % Chebyshev BS radio button is selected
        elseif selected_radio4 == 1
            if ChebyshevBS == 1     % Chebyshev BP from its Generalized Equation

                % Generation an input dialog box for Chebyshev BP Generalized Equation
                prompt={'INSERTION LOSS [dB]','RETURN LOSS [dB]','LOWER PASSBAND CUTOFF FREQUENCY f1 [GHz]','UPPER PASSBAND CUTOFF FREQUENCY f2 [GHz]','LOWER STOPBAND CUTOFF FREQUENCY fs1 [GHz]','UPPER STOPBAND CUTOFF FREQUENCY fs2 [GHz]'};
                title1='CHEBYSHEV BP GENERALIZED EQUATION';
                numlines=1;
                defaultanswer={'','','','','',''};
                options.Resize='on';
                options.WindowStyle='normal';
                options.Interpreter='tex';
                input_array = inputdlg(prompt,title1,numlines,defaultanswer, options); % Complete input dialog box

                LA = str2num(input_array{1}); % Accepting first element of the input as LA
                LR = str2num(input_array{2}); % Accepting 2nd element of the input as LR
                f1 = str2num(input_array{3}); % Accepting 3rd element of the input as f1
                f2 = str2num(input_array{4});% Accepting 4th element of the input as f2
                fs1= str2num(input_array{5});% Accepting 4th element of the input as fs1
                fs2 = str2num(input_array{6});% Accepting 4th element of the input as fs2

               if (LA>LR) && (LR>0) && (f1>=0.3) && (fs1>f1) && (fs2>fs1) && (f2>fs2) && (f2<=300)

                    % Cut-off Frequencies W1 and W2; and Stopband frequency at higher side Ws
                    wa = 2*pi*f1;
                    wb = 2*pi*f2;
                    ws1 = 2*pi*fs1;
                    ws2 = 2*pi*fs2;

                % Selectivity
                S = (wb-wa)/(ws2-ws1);
                % Displaying the selectivity
                Selectivity=num2str(S); % converting number to sder:tring
                set(handles.edit14_selectivity,'String',Selectivity); % Display the converted string

                    % Degree
                Degree = (LA + LR + 6)/(20*log10(S + sqrt(S^2 - 1)));
                display('Degree of the filter:')
                Degree_rounded = round(Degree);
                if Degree_rounded<Degree
                    N = Degree_rounded + 1;
                elseif Degree_rounded>=Degree
                    N = Degree_rounded;
                else
                end
                N;
                % Displaying the filter order/degree
                filter_order=num2str(N); % converting number to sder:tring
                set(handles.FilterOrder,'String',filter_order); % Display the converted string

%     Reflection Coefficient
                Ref = 10^(-LR/20);
%     dB pass-band ripple
                Lar = -10*log10(1-Ref^2);
% Ripple level epsillon
                epsillon = sqrt(10^(0.1*Lar) - 1);

                %Displaying ripple value in GUI
                ripple_level=num2str(epsillon); % converting number to sder:tring
                set(handles.edit15_ripple,'String',ripple_level); % Display the converted string

                    f = 0:0.001:30;              % Range of the frequency in GHz
```

```matlab
            w = 0:(2*pi*0.001):(2*pi*30);    % Range of w = 2*pi*f.
            %BS
            wo    = sqrt(wa*wb);
            fo    = wo/(2*pi);
            alpha = wo/(wb-wa);

    for i=1:1:30001
        omega(i) = abs((-1./(alpha*((w(i)./wo)-(wo./w(i))))));

       if (omega(i)<=1)

          Deno = 1 + (epsillon*cos(N*acos((-1./(alpha*((w(i)./wo)-(wo./w(i))))))).^2;

       elseif (omega(i)>=1)

           Deno = 1 + (epsillon*cosh(N*acosh((-1./(alpha*((w(i)./wo)-(wo./w(i))))))).^2;

       else
       end

 % Computing S12 values for different values ABCD resulting from the variation in w.
    S1(i) = abs(1./Deno);
    S10(i) = 1 - abs(1./Deno);
   end

         S12 = 10*log10(S1);
         S11 = 10*log10(S10);

         plot(f,S12,'blue', f,S11,'red')
         graph = plot(f,S12,'blue', f,S11,'red');
         % Setting the line width or thickness for both curves
         set (graph, 'LineWidth',1.5);
         title('Chebyshev Response BS');
         ylabel('|S12|^2 and |S11|^2 in dB');
         xlabel('f in GHz');
         axis([0 (2*fo) -80 5]);
         grid on;

    else
        helpdlg('Please check your input parameters. Make sure that f1>=0.3GHz, fs1>f1, fs2>fs1, f2>fs, f2<=300 GHz, LA>LR and LR>0 dB.');
    end
% |===================================================================|
% |                                                                   |
% | Plotting Chebyshev BS Prototype using A,B,C,D parameters:         |
% |                              4                                    |
% |               S12|^2 = -------------------                        |
% |                         |A+(B/Zo)+(C*Zo)+D|^2                     |
% |===================================================================|
elseif ChebyshevBS == 2    % Chebyshev BS Shunt Prototype to plot with its ABCD Parameters
% |*******************************************************************|
% |                 Plotting Chebyshev BS Shunt Prototype             |
% |*******************************************************************|
    % Generation an input dialog box for Chebyshev BP Shunt
    prompt={'INSERTION LOSS [dB]','RETURN LOSS [dB]','LOWER PASSBAND CUTOFF FREQUENCY f1 [GHz]','UPPER PASSBAND CUTOFF FREQUENCY f2 [GHz]','LOWER STOPBAND CUTOFF FREQUENCY fs1 [GHz]','UPPER STOPBAND CUTOFF FREQUENCY fs2 [GHz]','SYSTEM IMPEDANCE Zo [OHMS]'};
    title1='CHEBYSHEV BP SHUNT EQUATION';
    numlines=1;
    defaultanswer={'','','','','','',''};
    options.Resize='on';
    options.WindowStyle='normal';
    options.Interpreter='tex';
    input_array = inputdlg(prompt,title1,numlines,defaultanswer, options); % Complete input dialog box

    LA = str2num(input_array{1}); % Accepting first element of the input as LA
    LR = str2num(input_array{2}); % Accepting 2nd element of the input as LR
    f1 = str2num(input_array{3}); % Accepting 3rd element of the input as f1
    f2 = str2num(input_array{4});% Accepting 4th element of the input as f2
```

```matlab
         fs1= str2num(input_array{5});% Accepting 4th element of the input as fs1
          fs2 = str2num(input_array{6});% Accepting 4th element of the input as fs2
           Zo = str2num(input_array{7}); % Accepting last element of the input as Zo

        if (LA>LR) && (LR>0) && (f1>=0.3) && (fs1>f1) && (fs2>fs1) && (f2>fs2) && (f2<=300)

             % Cut-off Frequencies W1 and W2; and Stopband frequency at higher side Ws
             wa = 2*pi*f1;
             wb = 2*pi*f2;
             ws1 = 2*pi*fs1;
             ws2 = 2*pi*fs2;

        % Selectivity
         S = (wb-wa)/(ws2-ws1);
         % Displaying the selectivity
         Selectivity=num2str(S); % converting number to sder:tring
         set(handles.edit14_selectivity,'String',Selectivity); % Display the converted string

             % Degree
         Degree = (LA + LR + 6)/(20*log10(S + sqrt(S^2 - 1)));
         display('Degree of the filter:')
         Degree_rounded = round(Degree);
         if Degree_rounded<Degree
             N = Degree_rounded + 1;
         elseif Degree_rounded>=Degree
             N = Degree_rounded;
         else
         end
         N;
         % Displaying the filter order/degree
         filter_order=num2str(N); % converting number to sder:tring
         set(handles.FilterOrder,'String',filter_order); % Display the converted string

             %    Reflection Coefficient
         Ref = 10^(-LR/20);
%     dB pass-band ripple
         Lar = -10*log10(1-Ref^2);
% Ripple level epsillon
         epsillon = sqrt(10^(0.1*Lar) - 1);

         %Displaying ripple value in GUI
         ripple_level=num2str(epsillon); % converting number to sder:tring
         set(handles.edit15_ripple,'String',ripple_level); % Display the converted string

              % Computation of element values,  Ref. G. Mattahaei MW Filters, Pg 99
         R = coth(Lar/17.37);
         beta = log(R);
         gamma = sinh(beta/(2*N));
         k = N+1;

    for i=1:1:k
        a(i) = sin(((2*i-1).*pi)./(2*N));
        b(i) = gamma^2 + [sin((i.*pi)./N)]^2;

        if i==1

        g(1)= 2.*a(1)/gamma;

        elseif (i>1) && (i<=N)
            g(i) = (4.*a(i-1).*a(i))./(b(i-1).*g(i-1));

        else

            A = mod (N,2);
            if A==0
                g(k)=[coth(beta/4)]^2;

            elseif A==1
```

```matlab
            g(k)=1;

        end

    end

end

%  Plotting with element values for different filter order

% Generatng L and C elements for BS Prototype - Shunt
wo    = sqrt(wa*wb);
fo = wo/(2*pi);
alpha = wo/(wb-wa);

for i=1:1:N
    if i==1
        % Computing C(1) value
        Element(1) = g(1)/Zo;
        %Transforming shunt C to series LC circuit
        Inductor(1)= alpha/(wo*Element(1));
        Capacitor(1)=(Element(1))/(alpha*wo);
    elseif (i>1) && (i<=N)
        k=mod(i,2);
        if k==0
            % Computing L values for odd values of i
            Element(i) = g(i)*Zo;
            %Transforming series L to shunt LC circuit
            Inductor(i)= (Element(i))/(alpha*wo);
            Capacitor(i)=alpha/(wo*Element(i));
        elseif k==1
            % Computing C values for even values of i
            Element(i) = g(i)/Zo;
            %Transforming shunt C to series LC circuit
            Inductor(i)= alpha/(wo*Element(i));
            Capacitor(i)=(Element(i))/(alpha*wo);
        else
        end
    else
    end
end
%Scaled reactive element values and scaled source Zs and load ZL impedances, Z = [Zs ZL]
Element;
Inductor(Inductor==0)=[];
Inductor;                       %L remains in nH.
Capacitor(Capacitor==0)=[];
Capacitor2 = 1000.*Capacitor;   %to convert from nF to pF.
Zs = 1*Zo;                      % Source impedance go=1
ZL = g(N+1)*Zo;                 % Load impedance in Ohm
Impedance = [Zs ZL];            %Z are in Ohms. in Ohm

% Displaying the inductor, capacitor and impedance values.
Inductor1 = mat2str(Inductor,4);
set(handles.InductorValues1,'String',Inductor1); % Display the converted string
Capacitor1 = mat2str(Capacitor2,4);
set(handles.CapacitorValues1,'String',Capacitor1); % Display the converted string
Impedance1 = mat2str(Impedance,4);
set(handles.ImpedanceValues1,'String',Impedance1); % Display the converted string

f = 0:0.01:30;                  % Range of the frequency in GHz
w = 0:(2*pi*0.01):(2*pi*30);    % Range of w = 2*pi*f.

 % Computing Z, Y and T = [A B;C D] values for different frequency values
 % This is for Chebyshev BS Prototype - Series

    for i=1:1:3001
        for t=1:1:N
            if t==1
```

```matlab
                % Computing Z(C1) values
                Z(1) = j*Element(1).*(-1./(alpha*((w(i)./wo)-(wo./w(i)))));
                T = [1 0;Z(1) 1];
            elseif (t>1) && (t<=N)
                    k=mod(t,2);
                if k==0
                    % Computing Z(L) values
                    Z(t) = j*Element(t).*(-1./(alpha*((w(i)./wo)-(wo./w(i)))));
                    Tnew = T*[1 Z(t);0 1];
                    T = Tnew;
                elseif k==1
                    % Computing Z(C) values
                    Z(t) = j*Element(t).*(-1./(alpha*((w(i)./wo)-(wo./w(i)))));
                    Tnew = T*[1 0;Z(t) 1];
                    T = Tnew;
                else
                end
            else
            end
        end

    T;

% Getting ABCD values by identification using 2X2 Matrix T=[A B;C D] by using
    A = T(1,1);
    B = T(1,2);
    C = T(2,1);
    D = T(2,2);

 % Computing S12 values for different values ABCD resulting from the
 % variation in w.
    Sa(i) = abs(2./(A+(B./Zo)+(C.*Zo)+D));
    Sb(i) = 1 - abs((2./(A+(B./Zo)+(C.*Zo)+D)).^2);
   end

%                                                      4
% Plotting with A,B,C,D parameters, |S12|^2 = -------------------
%                                              |A+(B/Zo)+(C*Zo)+D|^2

    S12 = 20*log10(Sa);   % Computing the dB values of |S12|^2
    S11 = 10*log10(Sb);
    plot(f,S12,'blue', f,S11,'red')
    graph = plot(f,S12,'blue', f,S11,'red');
    % Setting the line width or thickness for both curves
    set (graph, 'LineWidth',1.5);
    title('Chebyshev Response for BS Shunt Prototype');
    ylabel('|S12|^2 and |S11|^2 in dB');
    xlabel('f in GHz');
    axis([0 2*fo -80 3]);
    grid on;

    else
        helpdlg('Please check your input parameters. Make sure that f1>=0.3GHz, fs1>f1, fs2>fs1, f2>fs, f2<=300 GHz, LA>LR, LR>0 dB and it is recommended to choose Zo=50 Ohms');
    end

elseif ChebyshevBS == 3   % Chebyshev BS Series Prototype to plot with its ABCD Parameters
% |*********************************************************************|
% |                 Chebyshev - Series Band-Stop Prototype               |
% |*********************************************************************|
    % Generation an input dialog box for Chebyshev BP Series
    prompt={'INSERTION LOSS [dB]','RETURN LOSS [dB]','LOWER PASSBAND CUTOFF FREQUENCY f1 [GHz]','UPPER PASSBAND CUTOFF FREQUENCY f2 [GHz]','LOWER STOPBAND CUTOFF FREQUENCY fs1 [GHz]','UPPER STOPBAND CUTOFF FREQUENCY fs2 [GHz]','SYSTEM IMPEDANCE Zo [OHMS]'};
    title1='CHEBYSHEV BP SERIES';
    numlines=1;
    defaultanswer={'','','','','','',''};
    options.Resize='on';
    options.WindowStyle='normal';
```

```matlab
    options.Interpreter='tex';
    input_array = inputdlg(prompt,title1,numlines,defaultanswer, options); % Complete input dialog box

     LA = str2num(input_array{1}); % Accepting first element of the input as LA
     LR = str2num(input_array{2}); % Accepting 2nd element of the input as LR
     f1 = str2num(input_array{3}); % Accepting 3rd element of the input as f1
     f2 = str2num(input_array{4});% Accepting 4th element of the input as f2
     fs1= str2num(input_array{5});% Accepting 4th element of the input as fs1
     fs2 = str2num(input_array{6});% Accepting 4th element of the input as fs2
     Zo = str2num(input_array{7}); % Accepting last element of the input as Zo

    if (LA>LR) && (LR>0) && (f1>=0.3) && (fs1>f1) && (fs2>fs1) && (f2>fs2) && (f2<=300)

        % Cut-off Frequencies W1 and W2; and Stopband frequency at higher side Ws
        wa = 2*pi*f1;
        wb = 2*pi*f2;
        ws1 = 2*pi*fs1;
        ws2 = 2*pi*fs2;

    % Selectivity
     S = (wb-wa)/(ws2-ws1);
     % Displaying the selectivity
     Selectivity=num2str(S); % converting number to sder:tring
     set(handles.edit14_selectivity,'String',Selectivity); % Display the converted string

        % Degree
     Degree = (LA + LR + 6)/(20*log10(S + sqrt(S^2 - 1)));
     display('Degree of the filter:')
     Degree_rounded = round(Degree);
     if Degree_rounded<Degree
         N = Degree_rounded + 1;
     elseif Degree_rounded>=Degree
         N = Degree_rounded;
     else
     end
     N;
     % Displaying the filter order/degree
     filter_order=num2str(N); % converting number to sder:tring
     set(handles.FilterOrder,'String',filter_order); % Display the converted string

     %   Reflection Coefficient
     Ref = 10^(-LR/20);
%    dB pass-band ripple
     Lar = -10*log10(1-Ref^2);
% Ripple level epsillon
    epsillon = sqrt(10^(0.1*Lar) - 1);

    %Displaying ripple value in GUI
    ripple_level=num2str(epsillon); % converting number to sder:tring
    set(handles.edit15_ripple,'String',ripple_level); % Display the converted string

    % Computation of element values,  Ref. G. Mattahaei MW Filters, Pg 99
R = coth(Lar/17.37);
beta = log(R);
gamma = sinh(beta/(2*N));
k = N+1;

for i=1:1:k
    a(i) = sin(((2*i-1).*pi)./(2*N));
    b(i) = gamma^2 + [sin((i.*pi)./N)]^2;

    if i==1

    g(1)= 2.*a(1)/gamma;

    elseif (i>1) && (i<=N)
        g(i) = (4.*a(i-1).*a(i))./(b(i-1).*g(i-1));

    else
```

```matlab
            A = mod (N,2);
            if A==0
                g(k)=[coth(beta/4)]^2;
            elseif A==1
                g(k)=1;
            end
        end
end

%  Plotting with element values for different filter order

% Generatng L and C elements for BS Prototype - Series
wo    = sqrt(wa*wb);
fo = wo/(2*pi);
alpha = wo/(wb-wa);
for i=1:1:N
    if i==1
        Element(1) = g(1)*Zo; % Computing L1 values
        %Transforming series L to shunt LC circuit
        Inductor(1)= (Element(1))/(alpha*wo);
        Capacitor(1)=alpha/(wo*Element(1));
    elseif (i>1) && (i<=N)
        k=mod(i,2);
        if k==0
            Element(i) = g(i)/Zo; % Computing C values
            %Transforming shunt C to series LC circuit
            Inductor(i)= alpha/(wo*Element(i));
            Capacitor(i)=(Element(i))/(alpha*wo);
        elseif k==1
            Element(i) = g(i)*Zo; % Computing L values
            %Transforming series L to shunt LC circuit
            Inductor(i)= (Element(i))/(alpha*wo);
            Capacitor(i)=alpha/(wo*Element(i));
        else
        end
    else
    end
end
%Scaled reactive element values and scaled source Zs and load ZL impedances, Z = [Zs ZL]
Element;
Inductor(Inductor==0)=[];
Inductor;                        %L remains in nH.
Capacitor(Capacitor==0)=[];
Capacitor2 = 1000.*Capacitor;   %to convert from nF to pF.
Zs = 1*Zo;                      % Source impedance
ZL = g(N+1)*Zo;                 % Load impedance in Ohm
Impedance = [Zs ZL];            %Z are in Ohms. in Ohm

% Displaying the inductor, capacitor and impedance values.
Inductor1 = mat2str(Inductor,4);
set(handles.InductorValues1,'String',Inductor1); % Display the converted string
Capacitor1 = mat2str(Capacitor2,4);
set(handles.CapacitorValues1,'String',Capacitor1); % Display the converted string
Impedance1 = mat2str(Impedance,4);
set(handles.ImpedanceValues1,'String',Impedance1); % Display the converted string

f = 0:0.01:30;                % Range of the frequency in GHz
w = 0:(2*pi*0.01):(2*pi*30);   % Range of w = 2*pi*f.
 % Computing Z, Y and T = [A B;C D] values for different frequency values
 % This is for Chebyshev BS Prototype - Series

    for i=1:1:3001
        for t=1:1:N
            if t==1
                Z(1) = j*Element(1).*(-1./(alpha*((w(i)./wo)-(wo./w(i)))));  % Computing Z(L1) values
                T = [1 Z(1);0 1];
            elseif (t>1) && (t<=N)
```

```matlab
                            k=mod(t,2);
                        if k==0
                            Z(t) = j*Element(t).*(-1./(alpha*((w(i)./wo)-(wo./w(i))))); % Computing Zc values
                            Tnew = T*[1 0;Z(t) 1];
                            T = Tnew;
                        elseif k==1
                            Z(t) = j*Element(t).*(-1./(alpha*((w(i)./wo)-(wo./w(i))))); % Computing Z(L) values
                            Tnew = T*[1 Z(t);0 1];
                            T = Tnew;
                        else
                        end
                    else
                    end
                end

        T;

    % Getting ABCD values by identification using 2X2 Matrix T=[A B;C D] by using
        A = T(1,1);
        B = T(1,2);
        C = T(2,1);
        D = T(2,2);

     % Computing S12 values for different values ABCD resulting from the
     % variation in w.
        Sa(i) = abs(2./(A+(B./Zo)+(C.*Zo)+D));
        Sb(i) = 1 - abs((2./(A+(B./Zo)+(C.*Zo)+D)).^2);
       end

    %                                                     4
    % Plotting with A,B,C,D parameters, |S12|^2 = -------------------
    %                                              |A+(B/Zo)+(C*Zo)+D|^2

        S12 = 20*log10(Sa);   % Computing the dB values of |S12|^2
        S11 = 10*log10(Sb);

        plot(f,S12,'blue', f,S11,'red')
        graph = plot(f,S12,'blue', f,S11,'red');
        % Setting the line width or thickness for both curves
        set (graph, 'LineWidth',1.5);
        title('Chebyshev Response of BS Series Prototype');
        ylabel('|S12|^2 and |S11|^2 in dB');
        xlabel('f in GHz');
        axis([0 2*fo -80 3]);
        grid on;

        else
            helpdlg('Please check your input parameters. Make sure that f1>=0.3GHz, fs1>f1, fs2>fs1, f2>fs, f2<=300 GHz, LA>LR, LR>0 dB and it is recommended to choose Zo=50 Ohms');
        end

        else
        end

        % Butterworth LP radio button is selected
    elseif selected_radio5 == 1

    if ButterworthLP == 1     % Butterworth LP from its USING ITS GENERALIZED Equation

        % Generation an input dialog box for Butterworth LP USING ITS GENERALIZED Equation
        prompt={'INSERTION LOSS [dB]','RETURN LOSS [dB]','PASSBAND FREQUENCY fp [GHz]','STOPBAND FREQUENCY fs [GHz]'};
        title1='BUTTWERWORTH LP GENERALIZED EQUATION';
        numlines=1;
        defaultanswer={'','','',''};
        options.Resize='on';
        options.WindowStyle='normal';
        options.Interpreter='tex';
        input_array = inputdlg(prompt,title1,numlines,defaultanswer, options); % Complete input dialog box
```

```matlab
    LA = str2num(input_array{1}); % Accepting first element of the input as LA
    LR = str2num(input_array{2}); % Accepting 2nd element of the input as LR
    fc = str2num(input_array{3}); % Accepting 3rd element of the input as fp
    fs = str2num(input_array{4});% Accepting 4th element of the input as fs

  if (LA>LR) && (LR>0) && (fc>=0.3) && (fs>fc) && (fs<=300) % Microwave frequence range: 300MHz - 300GHz.

% |===================================================================|
% |                                      1                            |
% | Plotting Butterworth LP Prototype from |S12|^2 = ----------------------|
% |                                      1 + [epsillon*T(w)]^2 |
% | Here: w <-------- w/wc                                            |
% |===================================================================|

% Cut-off frequency Wc and Stopband frequency Ws:
    wc = 2*pi*fc;
    ws = 2*pi*fs;

% Selectivity
    S = ws/wc;
    % Displaying the selectivity
    Selectivity=num2str(S); % converting number to sder:tring
    set(handles.edit14_selectivity,'String',Selectivity); % Display the converted string

% Degree for Butterworth Filter
    Degree = (LA + LR)/(20*log10(S));
    Degree_rounded = round(Degree);
    if Degree_rounded<Degree
        N = Degree_rounded + 1;
    elseif Degree_rounded>=Degree
        N = Degree_rounded;
    else
    end
    N;
% Displaying the filter order/degree
    filter_order=num2str(N); % converting number to sder:tring
    set(handles.FilterOrder,'String',filter_order); % Display the converted string

    %  Computation of S12 and its Plotting
    f = 0:0.01:30;
    w = 0:(2*pi*0.01):(30*2*pi);
    den = 1+(w./wc).^(2*N);

    S1 = abs(1./den);
    S10= 1-S1;

    S12 = 10*log10(S1);
    S11 = 10*log10(S10);

    plot(f,S12,'blue', f,S11,'red')
    graph = plot(f,S12,'blue', f,S11,'red');
    % Setting the line width or thickness for both curves
    set (graph, 'LineWidth',1.5);
    title('Maximally Flat Response of LP Prototype');
    ylabel('|S12|^2 and |S11|^2 in dB');
    xlabel('f in GHz');
    axis([0 2*fc -80 5]);
    grid on;

  else
      helpdlg('Please check your input parameters. Make sure that fp>=0.3GHz, fs>fp, fs<=300GHz, LA>LR and LR>0 dB.');
  end
% |===================================================================|
% |                                                                   |
% |      Plotting Butterworth LP Prototype using A,B,C,D parameters:    |
% |                                  4                                |
```

```matlab
%                         S12|^2 = -------------------            |
% |                               |A+(B/Zo)+(C*Zo)+D|^2           |
% |===================================================================|
elseif ButterworthLP == 2   % Butterworth LP Shunt Prototype to plot with its ABCD Parameters

    % Generation an input dialog box for Butterworth LP Shunt Prototype
    prompt={'INSERTION LOSS [dB]','RETURN LOSS [dB]','PASSBAND FREQUENCY fp [GHz]','STOPBAND FREQUENCY fs [GHz]','SYSTEM IMPEDANCE Zo [OHMS]'};
    title1='BUTTERWORTH LP SHUNT PROTOTYPE';
    numlines=1;
    defaultanswer={'','','','',''};
    options.Resize='on';
    options.WindowStyle='normal';
    options.Interpreter='tex';
    input_array = inputdlg(prompt,title1,numlines,defaultanswer, options); % Complete input dialog box

    LA = str2num(input_array{1}); % Accepting first element of the input as LA
    LR = str2num(input_array{2}); % Accepting 2nd element of the input as LR
    fc = str2num(input_array{3}); % Accepting 3rd element of the input as fp
    fs = str2num(input_array{4});% Accepting 4th element of the input as fs
    Zo = str2num(input_array{5}); % Accepting last element of the input as Zo

    if (LA>LR) && (LR>0) && (fc>=0.3) && (fs>fc) && (fs<=300)

% |************************************************************************|
% |                Butterworth - Shunt Low Pass Prototype                  |
% |************************************************************************|
% Cut-off frequency Wc and Stopband frequency Ws:
    wc = 2*pi*fc;
    ws = 2*pi*fs;

% Selectivity must be > 1.
    S = ws/wc;
    % Displaying the selectivity
    Selectivity=num2str(S); % converting number to sder:tring
    set(handles.edit14_selectivity,'String',Selectivity); % Display the converted string

% Degree for Butterworth Filter
    Degree = (LA + LR)/(20*log10(S));
    Degree_rounded = round(Degree);
    if Degree_rounded<Degree
        N = Degree_rounded + 1;
    elseif Degree_rounded>=Degree
        N = Degree_rounded;
    else
    end
    N;
    % Displaying the filter order/degree
        filter_order=num2str(N); % converting number to sder:tring
        set(handles.FilterOrder,'String',filter_order); % Display the converted string
     % Element Values for Maximally Flat LP prototype
    n = N;
    Lar = 3; % Attenuation in dB
    w1  = 1; % Frequency at which Lar is defined as the pass-band edge
    go  = 1; % is assumed to be 1

    for i=1:1:(n+1)

    if (i>=1)&&(i<=n)
    g(i)=2*sin((2*i-1).*pi./(2*n));
    else
    g(n+1) = 1;
    end
    end

%  Plotting with element values for different filter order

% Generatng L and C elements for LP Prototype - Shunt
```

```matlab
            for i=1:1:N
                if i==1
                    % Computing C(1) value
                    Element(1) = g(1)/(wc*Zo);
                    Capacitor(1)=Element(1);
                elseif (i>1) && (i<=N)
                    k=mod(i,2);
                    if k==0
                        % Computing L values for odd values of i
                        Element(i) = (g(i)*Zo)/wc;
                        Inductor(i)=Element(i);
                    elseif k==1
                        % Computing C values for even values of i
                        Element(i) = g(i)/(wc*Zo);
                        Capacitor(i)=Element(i);
                    else
                    end
                else
                end
            end
%
%Scaled reactive element values and scaled source Zs and load ZL impedances, Z = [Zs ZL)
Element;
Inductor(Inductor==0)=[];
Inductor;                         %L remains in nH.
Capacitor(Capacitor==0)=[];
Capacitor2 = 1000.*Capacitor;    %to convert from nF to pF.
Zs = 1*Zo;                        % Source impedance
ZL = g(N+1)*Zo;                   % Load impedance in Ohm
Impedance = [Zs ZL];              %Z are in Ohms. in Ohm

% Displaying the inductor, capacitor and impedance values.
Inductor1 = mat2str(Inductor,4);
set(handles.InductorValues1,'String',Inductor1); % Display the converted string
Capacitor1 = mat2str(Capacitor2,4);
set(handles.CapacitorValues1,'String',Capacitor1); % Display the converted string
Impedance1 = mat2str(Impedance,4);
set(handles.ImpedanceValues1,'String',Impedance1); % Display the converted string

f = 0:0.01:30;                   % Range of the frequency in GHz
w = 0:(2*pi*0.01):(2*pi*30);     % Range of w = 2*pi*f.

 % Computing Z, Y and T = [A B;C D] values for different frequency values
 % This is for Butterworth LP Prototype - Series

    for i=1:1:3001
        for t=1:1:N
            if t==1
                % Computing Z(C1) values
                Z(1) = j*Element(1).*w(i);
                T = [1 0;Z(1) 1];
            elseif (t>1) && (t<=N)
                    k=mod(t,2);
                if k==0
                    % Computing Z(L) values
                    Z(t) = j*Element(t).*w(i);
                    Tnew = T*[1 Z(t);0 1];
                    T = Tnew;
                elseif k==1
                    % Computing Z(C) values
                    Z(t) = j*Element(t).*w(i);
                    Tnew = T*[1 0;Z(t) 1];
                    T = Tnew;
                else
                end
            else
            end
        end
```

```matlab
        T;

% Getting ABCD values by identification using 2X2 Matrix T=[A B;C D] by using
    A = T(1,1);
    B = T(1,2);
    C = T(2,1);
    D = T(2,2);

 % Computing S12 values for different values ABCD resulting from the
 % variation in w.
    Sa(i) = abs(2./(A+(B./Zo)+(C.*Zo)+D));
    Sb(i)=abs(1-abs((2./(A+(B./Zo)+(C.*Zo)+D)).^2));
   end

%                                                   4
% Plotting with A,B,C,D parameters, |S12|^2 = -------------------
%                                             |A+(B/Zo)+(C*Zo)+D|^2

    S12 = 20*log10(Sa);  % Computing the dB values of |S12|^2
    S11 = 10*log10(Sb);
    plot(f,S12,'blue', f,S11,'red')
    graph = plot(f,S12,'blue', f,S11,'red');
    % Setting the line width or thickness for both curves
    set (graph, 'LineWidth',1.5);
    title('Maximally Flat Response for LP Shunt Prototype');
    ylabel('|S12|^2 and |S11|^2 in dB');
    xlabel('f in GHz');
    axis([0 2*fc -80 3]);
    grid on;

% |************************************************************************|
% |            End Plotting Butterworth LP shunt Prototype                  |
% |************************************************************************|
    else
        helpdlg('Please check your input parameters. Make sure that fp>=0.3GHz, fs>fp, fs<=300GHz, LA>LR and LR>0 dB. It is also advised to set Zo=50 Ohms');
    end

elseif ButterworthLP == 3   % Butterworth LP Series Prototype to plot with its ABCD Parameters

    % Generation an input dialog box for Butterworth LP Series Prototype
    prompt={'INSERTION LOSS [dB]','RETURN LOSS [dB]','PASSBAND FREQUENCY fp [GHz]','STOPBAND FREQUENCY fs [GHz]','SYSTEM IMPEDANCE Zo [OHMS]'};
    title1='BUTTERWORTH LP SERIES PROTOTYPE';
    numlines=1;
    defaultanswer={'','','','',''};
    options.Resize='on';
    options.WindowStyle='normal';
    options.Interpreter='tex';
    input_array = inputdlg(prompt,title1,numlines,defaultanswer, options); % Complete input dialog box

    LA = str2num(input_array{1}); % Accepting first element of the input as LA
    LR = str2num(input_array{2}); % Accepting 2nd element of the input as LR
    fc = str2num(input_array{3}); % Accepting 3rd element of the input as fp
    fs = str2num(input_array{4});% Accepting 4th element of the input as fs
    Zo = str2num(input_array{5}); % Accepting last element of the input as Zo

   if (LA>LR) && (LR>0) && (fc>=0.3) && (fs>fc) && (fs<=300)
% |************************************************************************|
% |                Butterworth - Shunt Low Pass Prototype                   |
% |************************************************************************|
% Cut-off frequency Wc and Stopband frequency Ws:
    wc = 2*pi*fc;
    ws = 2*pi*fs;

% Selectivity must be > 1.
    S = ws/wc;
    % Displaying the selectivity
    Selectivity=num2str(S); % converting number to sder:tring
```

```matlab
        set(handles.edit14_selectivity,'String',Selectivity); % Display the converted string

% Degree for Butterworth Filter
    Degree = (LA + LR)/(20*log10(S));
    Degree_rounded = round(Degree);
    if Degree_rounded<Degree
        N = Degree_rounded + 1;
    elseif Degree_rounded>=Degree
        N = Degree_rounded;
    else
    end
    N;
    % Displaying the filter order/degree
        filter_order=num2str(N); % converting number to sder:tring
        set(handles.FilterOrder,'String',filter_order); % Display the converted string
    % Element Values for Maximally Flat LP prototype
    n = N;
    Lar = 3; % Attenuation in dB
    w1  = 1; % Frequency at which Lar is defined as the pass-band edge
    go  = 1; % is assumed to be 1

    for i=1:1:(n+1)

    if (i>=1)&&(i<=n)
    g(i)=2*sin((2*i-1).*pi./(2*n));
    else
    g(n+1) = 1;
    end
    end

%  Plotting with element values for different filter order

% Generatng L and C elements for LP Prototype - Series

for i=1:1:N
    if i==1
        Element(1) = (g(1)*Zo)/wc; % Computing L1 values
        Inductor(1) = Element(1);
    elseif (i>1) && (i<=N)
        k=mod(i,2);
        if k==0
            Element(i) = g(i)/(wc*Zo); % Computing C values
            Capacitor(i)=Element(i);
        elseif k==1
            Element(i) = (g(i)*Zo)/wc; % Computing L values
            Inductor(i)=Element(i);
        else
        end
    else
    end
end
%Scaled reactive element values and scaled source Zs and load ZL impedances, Z = [Zs ZL)
Element;
Inductor(Inductor==0)=[];
Inductor;                        %L remains in nH.
Capacitor(Capacitor==0)=[];
Capacitor2 = 1000.*Capacitor;   %to convert from nF to pF.
Zs = 1*Zo;                       % Source impedance
ZL = g(N+1)*Zo;                  % Load impedance in Ohm
Impedance = [Zs ZL];             %Z are in Ohms. in Ohm

% Displaying the inductor, capacitor and impedance values.
Inductor1 = mat2str(Inductor,4);
set(handles.InductorValues1,'String',Inductor1); % Display the converted string
Capacitor1 = mat2str(Capacitor2,4);
set(handles.CapacitorValues1,'String',Capacitor1); % Display the converted string
Impedance1 = mat2str(Impedance,4);
set(handles.ImpedanceValues1,'String',Impedance1); % Display the converted string
```

```matlab
f = 0:0.01:30;               % Range of the frequency in GHz
w = 0:(2*pi*0.01):(2*pi*30);  % Range of w = 2*pi*f.

% Computing Z, Y and T = [A B;C D] values for different frequency values
% This is for Butterworth LP Prototype - Series

    for i=1:1:3001
        for t=1:1:N
            if t==1
                Z(1) = j*Element(1).*w(i);  % Computing Z(L1) values
                T = [1 Z(1);0 1];
            elseif (t>1) && (t<=N)
                    k=mod(t,2);
                if k==0
                    Z(t) = j*Element(t).*w(i); % Computing Zc values
                    Tnew = T*[1 0;Z(t) 1];
                    T = Tnew;
                elseif k==1
                    Z(t) = j*Element(t).*w(i); % Computing Z(L) values
                    Tnew = T*[1 Z(t);0 1];
                    T = Tnew;
                else
                end
            else
            end
        end

    T;

% Getting ABCD values by identification using 2X2 Matrix T=[A B;C D] by using
    A = T(1,1);
    B = T(1,2);
    C = T(2,1);
    D = T(2,2);

% Computing S12 values for different values ABCD resulting from the
% variation in w.
    Sa(i) = abs(2./(A+(B./Zo)+(C.*Zo)+D));
    Sb(i)= abs(1 - abs((2./(A+(B./Zo)+(C.*Zo)+D)).^2));
    end

%                                                    4
% Plotting with A,B,C,D parameters, |S12|^2 = -------------------
%                                             |A+(B/Zo)+(C*Zo)+D|^2

    S12 = 20*log10(Sa);   % Computing the dB values of |S12|^2
    S11 = 10*log10(Sb);

    plot(f,S12,'blue', f,S11,'red')
    graph = plot(f,S12,'blue', f,S11,'red');
    % Setting the line width or thickness for both curves
    set (graph, 'LineWidth',1.5);
    title('Maximally Flat Response of LP Series Prototype');
    ylabel('|S12|^2 and |S11|^2 in dB');
    xlabel('f in GHz');
    axis([0 2*fc -80 3]);
    grid on

% |*********************************************************************|
% |              End Plotting Butterworth LP Series Prototype            |
% |*********************************************************************|

    else
        helpdlg('Please check your input parameters. Make sure that fp>=0.3GHz, fs>fp, fs<=300GHz, LA>LR and LR>0 dB. It is also advised to set Zo=50 Ohms');
    end
else
end
```

```matlab
% |*********************************************************************|
% |                End Plotting Butterworth LP  Prototypes              |
% |*********************************************************************|

%==========================================================================

% |*********************************************************************|
% |                Begin Plotting Butterworth HP  Prototypes            |
% |*********************************************************************|

% Butterworth HP radio button is selected
elseif selected_radio6 == 1
if ButterworthHP == 1     % Butterworth HP from its Generalized Equation
% |=====================================================================|
% | Butterworth BP Prototype using LP to BP transformation:             |
% |  w <----- -wc/w   frequency substitution.                           |
% |                                            1                        |
% | Plotting Butterworth BP Prototype from   |S12|^2 = ----------------  |
% |                                              1 + w^2N               |
% |=====================================================================|
    % Generation an input dialog box for Butterworth HP USING ITS GENERALIZED Equation
    prompt={'INSERTION LOSS [dB]','RETURN LOSS [dB]','PASSBAND FREQUENCY fp [GHz]','STOPBAND FREQUENCY fs
[GHz]'};
    title1='BUTTERWORTH HP GENERALIZED EQUATION';
    numlines=1;
    defaultanswer={'','','',''};
    options.Resize='on';
    options.WindowStyle='normal';
    options.Interpreter='tex';
    input_array = inputdlg(prompt,title1,numlines,defaultanswer, options); % Complete input dialog box

    LA = str2num(input_array{1}); % Accepting first element of the input as LA
    LR = str2num(input_array{2}); % Accepting 2nd element of the input as LR
    fc = str2num(input_array{3}); % Accepting 3rd element of the input as fp
    fs = str2num(input_array{4});% Accepting 4th element of the input as fs

   if (LA>LR) && (LR>0) && (fs>=0.3) && (fc>fs) && (fc<=300)
    % Cut-off frequency Wc and Stopband frequency Ws:
    wc = 2*pi*fc;
    ws = 2*pi*fs;
% Selectivity
       S = wc/ws;
        % Displaying the selectivity
        Selectivity=num2str(S); % converting number to sder:tring
        set(handles.edit14_selectivity,'String',Selectivity); % Display the converted string

% Degree for Butterworth Filter
    Degree = (LA + LR)/(20*log10(S));
    Degree_rounded = round(Degree);
    if Degree_rounded<Degree
        N = Degree_rounded + 1;
    elseif Degree_rounded>=Degree
        N = Degree_rounded;
    else
    end
    N;
     % Displaying the filter order/degree
     filter_order=num2str(N); % converting number to sder:tring
     set(handles.FilterOrder,'String',filter_order); % Display the converted string

     f = 0:0.01:30;                  % Range of the frequency in GHz
     w = 0:(2*pi*0.01):(2*pi*30);    % Range of w = 2*pi*f.

     S12 = 10*log10(abs(1./(1+(-wc./w).^(2*N)))); % Computating |S12|^2 in dB
     S11 = 10*log10(abs(1 - abs(1./(1+(-wc./w).^(2*N)))));

% Plotting |S12|^2
        plot(f,S12,'blue', f,S11,'red')
        graph = plot(f,S12,'blue', f,S11,'red');
```

```matlab
        % Setting the line width or thickness for both curves
        set (graph, 'LineWidth',1.5);
        title('Maximally Flat Response HP');
        ylabel('|S12|^2 and |S11|^2 in dB');
        xlabel('f in GHz');
        axis([0 2*fc -80 5]);
        grid on;
% |**********************************************************************|
% |               End Plotting Butterworth HP Series Prototype           |
% |**********************************************************************|
    else
        helpdlg('Please check your input parameters. Make sure that fs>=0.3GHz, fs<fp, LA>LR and LR>0 dB.');
    end
% |**********************************************************************|
% |              Begin Plotting Butterworth HP Series Prototype          |
% |**********************************************************************|

% |====================================================================|
% |                                                                    |
% |    Plotting Butterworth HP Prototypes using A,B,C,D parameters:    |
% |                              4                                     |
% |              |S12|^2 = -------------------                         |
% |                       |A+(B/Zo)+(C*Zo)+D|^2                        |
% |====================================================================|
elseif ButterworthHP == 2   % Butterworth HP Shunt Prototype to plot with its ABCD Parameters

    % Generation an input dialog box for Butterworth HP Shunt Prototype
    prompt={'INSERTION LOSS [dB]','RETURN LOSS [dB]','PASSBAND FREQUENCY fp [GHz]','STOPBAND FREQUENCY fs [GHz]','SYSTEM IMPEDANCE Zo [OHMS]'};
    title1='BUTTERWORTH HP SHUNT PROTOTYPE';
    numlines=1;
    defaultanswer={'','','','',''};
    options.Resize='on';
    options.WindowStyle='normal';
    options.Interpreter='tex';
    input_array = inputdlg(prompt,title1,numlines,defaultanswer, options); % Complete input dialog box

    LA = str2num(input_array{1}); % Accepting first element of the input as LA
    LR = str2num(input_array{2}); % Accepting 2nd element of the input as LR
    fc = str2num(input_array{3}); % Accepting 3rd element of the input as fp
    fs = str2num(input_array{4}); % Accepting 4th element of the input as fs
    Zo = str2num(input_array{5}); % Accepting last element of the input as Zo

    if (LA>LR) && (LR>=0) && (fs>=0.3) && (fs<fc) && (fc<=300) && (Zo==50)
    % Cut-off frequency Wc and Stopband frequency Ws:
    wc = 2*pi*fc;
    ws = 2*pi*fs;
% Selectivity
        S = wc/ws;
        % Displaying the selectivity
        Selectivity=num2str(S); % converting number to sder:tring
        set(handles.edit14_selectivity,'String',Selectivity); % Display the converted string

% Degree for Butterworth Filter
    Degree = (LA + LR)/(20*log10(S));
    Degree_rounded = round(Degree);
    if Degree_rounded<Degree
        N = Degree_rounded + 1;
    elseif Degree_rounded>=Degree
        N = Degree_rounded;
    else
    end
    N;
    % Displaying the filter order/degree
    filter_order=num2str(N); % converting number to sder:tring
    set(handles.FilterOrder,'String',filter_order); % Display the converted string

    % Element Values for Maximally Flat HP prototype
    n = N;
```

```matlab
    Lar = 3; % Attenuation in dB
    w1  = 1; % Frequency at which Lar is defined as the pass-band edge
    go  = 1; % is assumed to be 1

    for i=1:1:(n+1)

    if (i>=1)&&(i<=n)
    g(i)=2*sin((2*i-1).*pi./(2*n));
    else
    g(n+1) = 1;
    end
    end

% Plotting with element values for different filter order

% Generatng L and C elements for HP Prototype - Series

for i=1:1:N
    if i==1
        Element(1) = (g(1)*Zo)*wc; % Computing L1 values
        %LP to HP transformation for element value
        Capacitor(1)=1/Element(1);
    elseif (i>1) && (i<=N)
        k=mod(i,2);
        if k==0
            Element(i) = (g(i)*wc)/Zo; % Computing C values
            %LP to HP transformation for element values
            Inductor(i)=1./Element(i);
        elseif k==1
            Element(i) = (g(i)*Zo)*wc; % Computing L values
            %LP to HP transformation for element values
            Capacitor(i)=1./Element(i);
        else
        end
    else
    end
end
%Scaled reactive element values and scaled source Zs and load ZL impedances, Z = [Zs ZL]
Element;
Inductor(Inductor==0)=[];
Inductor;                       %L remains in nH.
Capacitor(Capacitor==0)=[];
Capacitor2 = 1000.*Capacitor;   %to convert from nF to pF.
Zs = 1*Zo;                      % Source impedance
ZL = g(N+1)*Zo;                 % Load impedance in Ohm
Impedance = [Zs ZL];            %Z are in Ohms. in Ohm

% Displaying the inductor, capacitor and impedance values.
Inductor1 = mat2str(Inductor,4);
set(handles.InductorValues1,'String',Inductor1); % Display the converted string
Capacitor1 = mat2str(Capacitor2,4);
set(handles.CapacitorValues1,'String',Capacitor1); % Display the converted string
Impedance1 = mat2str(Impedance,4);
set(handles.ImpedanceValues1,'String',Impedance1); % Display the converted string

f = 0:0.01:30;               % Range of the frequency in GHz
w = 0:(2*pi*0.01):(2*pi*30); % Range of w = 2*pi*f.

 % Computing Z, Y and T = [A B;C D] values for different frequency values
 % This is for Butterworth HP Prototype - Series

    for i=1:1:3001
        for t=1:1:N
            if t==1
                Z(1) = -j*Element(1)./w(i);  % Computing Z(L1) values
                T = [1 Z(1);0 1];
            elseif (t>1) && (t<=N)
                k=mod(t,2);
```

```matlab
                if k==0
                    Z(t) = -j*Element(t)./w(i); % Computing Zc values
                    Tnew = T*[1 0;Z(t) 1];
                    T = Tnew;
                elseif k==1
                    Z(t) = -j*Element(t)./w(i); % Computing Z(L) values
                    Tnew = T*[1 Z(t);0 1];
                    T = Tnew;
                else
                end
            else
            end
        end

    T;

% Getting ABCD values by identification using 2X2 Matrix T=[A B;C D] by using
    A = T(1,1);
    B = T(1,2);
    C = T(2,1);
    D = T(2,2);

 % Computing S12 values for different values ABCD resulting from the
 % variation in w.
    Sa(i) = abs(2./(A+(B./Zo)+(C.*Zo)+D));
    Sb(i) = abs(1 - abs((2./(A+(B./Zo)+(C.*Zo)+D)).^2));
   end

%                                                  4
% Plotting with A,B,C,D parameters, |S12|^2 = -------------------
%                                             |A+(B/Zo)+(C*Zo)+D|^2

    S12 = 20*log10(Sa);  % Computing the dB values of |S12|^2
    S11 = 10*log10(Sb);

    plot(f,S12,'blue', f,S11,'red')
    graph = plot(f,S12,'blue', f,S11,'red');
    % Setting the line width or thickness for both curves
    set (graph, 'LineWidth',1.5);
    title('Maximally Flat Response of HP Shunt Prototype');
    ylabel('|S12|^2 and |S11|^2 in dB');
    xlabel('f in GHz');
    axis([0 2*fc -80 3]);
    grid on;

% |************************************************************|
% |            End Plotting Butterworth HP shunt Prototype     |
% |************************************************************|

    else
        helpdlg('Please check your input parameters. Make sure that fs>=0.3GHz, fs<fp, LA>LR, LR>0 dB and it is recommended to keep Zo=50 Ohms');
    end

elseif ButterworthHP == 3   % Butterworth HP Series Prototype to plot with its ABCD Parameters

    % Generation an input dialog box for Butterworth HP Series Prototype
    prompt={'INSERTION LOSS [dB]','RETURN LOSS [dB]','PASSBAND FREQUENCY fp [GHz]','STOPBAND FREQUENCY fs [GHz]','SYSTEM IMPEDANCE Zo [OHMS]'};
    title1='BUTTERWORTH HP SERIES PROTOTYPE';
    numlines=1;
    defaultanswer={'','','','',''};
    options.Resize='on';
    options.WindowStyle='normal';
    options.Interpreter='tex';
    input_array = inputdlg(prompt,title1,numlines,defaultanswer, options); % Complete input dialog box

    LA = str2num(input_array{1}); % Accepting first element of the input as LA
    LR = str2num(input_array{2}); % Accepting 2nd element of the input as LR
```

```matlab
    fc = str2num(input_array{3}); % Accepting 3rd element of the input as fp
    fs = str2num(input_array{4});% Accepting 4th element of the input as fs
    Zo = str2num(input_array{5}); % Accepting last element of the input as Zo

   if (LA>LR) && (LR>=0) && (fs>=0.3) && (fs<fc) && (fc<=300) && (Zo==50)
   % Cut-off frequency Wc and Stopband frequency Ws:
    wc = 2*pi*fc;
    ws = 2*pi*fs;
% Selectivity
        S = wc/ws;
        % Displaying the selectivity
        Selectivity=num2str(S); % converting number to sder:tring
        set(handles.edit14_selectivity,'String',Selectivity); % Display the converted string

% Degree for Butterworth Filter
    Degree = (LA + LR)/(20*log10(S));
    Degree_rounded = round(Degree);
    if Degree_rounded<Degree
        N = Degree_rounded + 1;
    elseif Degree_rounded>=Degree
        N = Degree_rounded;
    else
    end
    N;
    % Displaying the filter order/degree
    filter_order=num2str(N); % converting number to sder:tring
    set(handles.FilterOrder,'String',filter_order); % Display the converted string
   % Element Values for Maximally Flat HP prototype
   n = N;
   Lar = 3; % Attenuation in dB
   w1  = 1; % Frequency at which Lar is defined as the pass-band edge
   go  = 1; % is assumed to be 1

   for i=1:1:(n+1)

   if (i>=1)&&(i<=n)
   g(i)=2*sin((2*i-1).*pi./(2*n));
   else
   g(n+1) = 1;
   end
   end

% Plotting with element values for different filter order

% Generatng L and C elements for HP Prototype - Shunt

for i=1:1:N
    if i==1
        % Computing C(1) value
        Element(1) = (g(1)*wc)/Zo;
        %LP to HP transformation
        Inductor(1)=1/Element(1);
    elseif (i>1) && (i<=N)
        k=mod(i,2);
        if k==0
            % Computing L values for odd values of i
            Element(i) = (g(i)*Zo)*wc;
            %LP to HP transformation
            Capacitor(i)=1./Element(i);
        elseif k==1
            % Computing C values for even values of i
            Element(i) = (g(i)*wc)/Zo;
            %LP to HP transformation
            Inductor(i)=1./Element(i);
        else
        end
    else
    end
end
```

```matlab
%Scaled reactive element values and scaled source Zs and load ZL impedances, Z = [Zs ZL]
Element;
Inductor(Inductor==0)=[];
Inductor;                       %L remains in nH.
Capacitor(Capacitor==0)=[];
Capacitor2 = 1000.*Capacitor;   %to convert from nF to pF.
Zs = 1*Zo;                      % Source impedance
ZL = g(N+1)*Zo;                 % Load impedance in Ohm
Impedance = [Zs ZL];            %Z are in Ohms. in Ohm

% Displaying the inductor, capacitor and impedance values.
Inductor1 = mat2str(Inductor,4);
set(handles.InductorValues1,'String',Inductor1); % Display the converted string
Capacitor1 = mat2str(Capacitor2,4);
set(handles.CapacitorValues1,'String',Capacitor1); % Display the converted string
Impedance1 = mat2str(Impedance,4);
set(handles.ImpedanceValues1,'String',Impedance1); % Display the converted string

f = 0:0.01:30;                  % Range of the frequency in GHz
w = 0:(2*pi*0.01):(2*pi*30);    % Range of w = 2*pi*f.

% Computing Z, Y and T = [A B;C D] values for different frequency values
% This is for Butterworth HP Prototype - Series

    for i=1:1:3001
        for t=1:1:N
            if t==1
                % Computing Z(C1) values
                Z(1) = -j*Element(1)./w(i);
                T = [1 0;Z(1) 1];
            elseif (t>1) && (t<=N)
                    k=mod(t,2);
                if k==0
                    % Computing Z(L) values
                    Z(t) = -j*Element(t)./w(i);
                    Tnew = T*[1 Z(t);0 1];
                    T = Tnew;
                elseif k==1
                    % Computing Z(C) values
                    Z(t) = -j*Element(t)./w(i);
                    Tnew = T*[1 0;Z(t) 1];
                    T = Tnew;
                else
                end
            else
            end
        end

    T;

% Getting ABCD values by identification using 2X2 Matrix T=[A B;C D] by using
    A = T(1,1);
    B = T(1,2);
    C = T(2,1);
    D = T(2,2);

% Computing S12 values for different values ABCD resulting from the
% variation in w.
    Sa(i) = abs(2./(A+(B./Zo)+(C.*Zo)+D));
    Sb(i) = abs(1-abs((2./(A+(B./Zo)+(C.*Zo)+D)).^2));
    end

%                                                            4
% Plotting with A,B,C,D parameters, |S12|^2 = -------------------
%                                              |A+(B/Zo)+(C*Zo)+D|^2

    S12 = 20*log10(Sa);  % Computing the dB values of |S12|^2
    S11 = 10*log10(Sb);
```

```matlab
        plot(f,S12,'blue', f,S11,'red')
        graph = plot(f,S12,'blue', f,S11,'red');
        % Setting the line width or thickness for both curves
        set (graph, 'LineWidth',1.5);
        title('Maximally Flat Response for HP Series Prototype');
        ylabel('|S12|^2 and |S11|^2 in dB');
        xlabel('f in GHz');
        axis([0 2*fc -80 3]);
        grid on;

% |************************************************************************|
% |             End Plotting Butterworth HP Series Prototype               |
% |************************************************************************|

    else
        helpdlg('Please check your input parameters. Make sure that fp>=0.3 GHz, fs<fp, LA>LR, LR>=0 dB and  it is
advised to keep Zo=50 Ohms');
    end
else
end

% |************************************************************************|
% |              End Plotting Butterworth HP   Prototypes                  |
% |************************************************************************|

%===========================================================================

% |************************************************************************|
% |              Begin Plotting Butterworth BP   Prototypes                |
% |************************************************************************|

% Butterworth BP radio button is selected
elseif selected_radio7 == 1

if ButterworthBP == 1     % Butterworth BP from its Generalized Equation
% |========================================================================|
% | Butterworth BP Prototype using LP to BP transformation:                |
% |  w <----- [wo/(w1-w1)]*[(w./wo)-(wo./w)]   with wo = sqrt(w1*w2)       |
% |                                              1                        |
% | Plotting Butterworth BP Prototype from   |S12|^2 = ----------------    |
% |                                              1 + w^2N                 |
% |========================================================================|
    % Generation an input dialog box for Butterworth BP USING ITS GENERALIZED Equation
    prompt={'INSERTION LOSS [dB]','RETURN LOSS [dB]','LOWER SIDE CUTOFF FREQUENCY f1 [GHz]','UPPER SIDE CUTOFF
FREQUENCY f2 [GHz]','STOPBAND FREQUENCY AT HIGHER SIDE fs [GHz]'};
    title1='BUTTERWORTH BP GENERALIZED EQUATION';
    numlines=1;
    defaultanswer={'','','','',''};
    options.Resize='on';
    options.WindowStyle='normal';
    options.Interpreter='tex';
    input_array = inputdlg(prompt,title1,numlines,defaultanswer, options); % Complete input dialog box

    LA = str2num(input_array{1}); % Accepting first element of the input as LA
    LR = str2num(input_array{2}); % Accepting 2nd element of the input as LR
    f1 = str2num(input_array{3}); % Accepting 3rd element of the input as f1
    f2 = str2num(input_array{4});% Accepting 4th element of the input as f2
    fs = str2num(input_array{5});% Accepting 4th element of the input as fs

    if (LA>LR) && (LR>0) && (f1>=0.3) && (f2>f1) && (f2<fs) && (fs<=300)
    % Cut-off Frequencies W1 and W2; and Stopband frequency at higher side Ws
        wa = 2*pi*f1;
        wb = 2*pi*f2;
        ws = 2*pi*fs;

% Selectivity
    S = (2*ws - wb - wa)/(wb-wa);
    % Displaying the selectivity
    Selectivity=num2str(S); % converting number to sder:tring
```

```matlab
        set(handles.edit14_selectivity,'String',Selectivity); % Display the converted string

        % Degree
        Degree = (LA + LR)/(20*log10(S));
        Degree_rounded = round(Degree);
        if Degree_rounded<Degree
            N = Degree_rounded + 1;
        elseif Degree_rounded>=Degree
            N = Degree_rounded;
        else
        end
        N;
        % Displaying the filter order/degree
         filter_order=num2str(N); % converting number to sder:tring
         set(handles.FilterOrder,'String',filter_order); % Display the converted string

          f = 0:0.01:30;             % Range of the frequency in GHz
          w = 0:(2*pi*0.01):(2*pi*30);   % Range of w = 2*pi*f.
          %BP
          wo    = sqrt(wa*wb);
          fo    = wo/(2*pi);
          alpha = wo/(wb-wa);

          S12  = 10*log10(1./(1+(alpha*((w./wo)-(wo./w))).^(2*N)));
          S11 = 10*log10(1 - abs(1./(1+(alpha*((w./wo)-(wo./w))).^(2*N))));

          plot(f,S12,'blue', f,S11,'red')
          graph = plot(f,S12,'blue', f,S11,'red');
          % Setting the line width or thickness for both curves
          set (graph, 'LineWidth',1.5);
          title('Maximally Flat Response BP');
          ylabel('|S12|^2 and |S11|^2 in dB');
          xlabel('f in GHz');
          axis([0 (f1+f2) -80 5]);
          grid on;

    else
        helpdlg('Please check your input parameters. Make sure that f1s>=0.3GHz, f2>f1, fs>f2, f2<=300 GHz, LA>LR and LR>0 dB.');
    end
% |=====================================================================|
% |                                                                     |
% |       Plotting Butterworth HP Prototypes using A,B,C,D parameters:   |
% |                              4                                       |
%                   S12|^2 = -------------------                         |
% |                          |A+(B/Zo)+(C*Zo)+D|^2                       |
% |=====================================================================|
elseif ButterworthBP == 2   % Butterworth BP Shunt Prototype to plot with its ABCD Parameters
% |*********************************************************************|
% |                  Plotting Butterworth BP Shunt Prototype             |
% |*********************************************************************|
    % Generation an input dialog box for Butterworth BP Shunt Prototype
    prompt={'INSERTION LOSS [dB]','RETURN LOSS [dB]','LOWER SIDE CUTOFF FREQUENCY f1 [GHz]','UPPER SIDE CUTOFF FREQUENCY f2 [GHz]','STOPBAND FREQUENCY AT HIGHER SIDE fs [GHz]','SYSTEM IMPEDANCE Zo [OHMS]'};
    title1='BUTTERWORTH BP SHUNT PROTOTYPE';
    numlines=1;
    defaultanswer={'','','','','',''};
    options.Resize='on';
    options.WindowStyle='normal';
    options.Interpreter='tex';
    input_array = inputdlg(prompt,title1,numlines,defaultanswer, options); % Complete input dialog box

    LA = str2num(input_array{1}); % Accepting first element of the input as LA
    LR = str2num(input_array{2}); % Accepting 2nd element of the input as LR
    f1 = str2num(input_array{3}); % Accepting 3rd element of the input as f1
    f2 = str2num(input_array{4});% Accepting 4th element of the input as f2
    fs = str2num(input_array{5});% Accepting 4th element of the input as fs
    Zo = str2num(input_array{6}); % Accepting last element of the input as Zo
```

```matlab
        if (LA>LR) && (LR>=0) && (f1>=0.3) && (f2>f1) && (f2<fs) && (fs<=300)

            % Cut-off Frequencies W1 and W2; and Stopband frequency at higher side Ws
            wa = 2*pi*f1;
            wb = 2*pi*f2;
            ws = 2*pi*fs;

% Selectivity
        S = (2*ws - wb - wa)/(wb-wa);
        % Displaying the selectivity
        Selectivity=num2str(S); % converting number to sder:tring
        set(handles.edit14_selectivity,'String',Selectivity); % Display the converted string

        % Degree
        Degree = (LA + LR)/(20*log10(S));
        Degree_rounded = round(Degree);
        if Degree_rounded<Degree
            N = Degree_rounded + 1;
        elseif Degree_rounded>=Degree
            N = Degree_rounded;
        else
        end
        N;
        % Displaying the filter order/degree
         filter_order=num2str(N); % converting number to sder:tring
         set(handles.FilterOrder,'String',filter_order); % Display the converted string
        % Element Values for Maximally Flat BP prototype
        n = N;
        Lar = 3; % Attenuation in dB
        w1  = 1; % Frequency at which Lar is defined as the pass-band edge
        go  = 1; % is assumed to be 1

        for i=1:1:(n+1)

        if (i>=1)&&(i<=n)
        g(i)=2*sin((2*i-1).*pi./(2*n));
        else
        g(n+1) = 1;
        end
        end

%  Plotting with element values for different filter order

% Generatng L and C elements for BP Prototype - Shunt
wo    = sqrt(wa*wb);
fo = wo/(2*pi);
alpha = wo/(wb-wa);
for i=1:1:N
    if i==1
        % Computing C(1) value
        Element(1) = g(1)/Zo;
        %Transforming Shunt C to Shunt LC circuit
        Inductor(1)= 1/(alpha*wo*Element(1));
        Capacitor(1)= (alpha*Element(1))/wo;
    elseif (i>1) && (i<=N)
        k=mod(i,2);
        if k==0
            % Computing L values for odd values of i
            Element(i) = g(i)*Zo;
            %Transforming series L to series LC circuit
            Inductor(i)= (alpha*Element(i))/wo;
            Capacitor(i)=1/(alpha*wo*Element(i));
        elseif k==1
            % Computing C values for even values of i
            Element(i) = g(i)/Zo;
            %Transforming Shunt C to Shunt LC circuit
            Inductor(i)= 1/(alpha*wo*Element(i));
            Capacitor(i)= (alpha*Element(i))/wo;
        else
```

```matlab
        end
    else
    end
end
%Scaled reactive element values and scaled source Zs and load ZL impedances, Z = [Zs ZL]
Element;
Inductor(Inductor==0)=[];
Inductor;                       %L remains in nH.
Capacitor(Capacitor==0)=[];
Capacitor2 = 1000.*Capacitor;   %to convert from nF to pF.
Zs = 1*Zo;                      % Source impedance
ZL = g(N+1)*Zo;                 % Load impedance in Ohm
Impedance = [Zs ZL];            %Z are in Ohms. in Ohm

% Displaying the inductor, capacitor and impedance values.
Inductor1 = mat2str(Inductor,4);
set(handles.InductorValues1,'String',Inductor1); % Display the converted string
Capacitor1 = mat2str(Capacitor2,4);
set(handles.CapacitorValues1,'String',Capacitor1); % Display the converted string
Impedance1 = mat2str(Impedance,4);
set(handles.ImpedanceValues1,'String',Impedance1); % Display the converted string

f = 0:0.01:30;                  % Range of the frequency in GHz
w = 0:(2*pi*0.01):(2*pi*30);    % Range of w = 2*pi*f.
 % Computing Z, Y and T = [A B;C D] values for different frequency values
 % This is for Butterworth BP Prototype - Series

    for i=1:1:3001
        for t=1:1:N
            if t==1
                % Computing Z(C1) values
                Z(1) = j*Element(1).*((alpha*((w(i)./wo)-(wo./w(i)))));
                T = [1 0;Z(1) 1];
            elseif (t>1) && (t<=N)
                    k=mod(t,2);
                if k==0
                    % Computing Z(L) values
                    Z(t) = j*Element(t).*((alpha*((w(i)./wo)-(wo./w(i)))));
                    Tnew = T*[1 Z(t);0 1];
                    T = Tnew;
                elseif k==1
                    % Computing Z(C) values
                    Z(t) = j*Element(t).*((alpha*((w(i)./wo)-(wo./w(i)))));
                    Tnew = T*[1 0;Z(t) 1];
                    T = Tnew;
                else
                end
            else
            end
        end

    T;

% Getting ABCD values by identification using 2X2 Matrix T=[A B;C D] by using
    A = T(1,1);
    B = T(1,2);
    C = T(2,1);
    D = T(2,2);

 % Computing S12 values for different values ABCD resulting from the
 % variation in w.
    Sa(i) = abs(2./(A+(B./Zo)+(C.*Zo)+D));
    Sb(i) = abs(1 - abs((2./(A+(B./Zo)+(C.*Zo)+D)).^2));
    end

%                                          4
% Plotting with A,B,C,D parameters, |S12|^2 = -------------------
%                                         |A+(B/Zo)+(C*Zo)+D|^2
```

```matlab
        S12 = 20*log10(Sa);  % Computing the dB values of |S12|^2
        S11 = 10*log10(Sb);

        plot(f,S12,'blue', f,S11,'red')
        graph = plot(f,S12,'blue', f,S11,'red');
        % Setting the line width or thickness for both curves
        set (graph, 'LineWidth',1.5);
        title('Maximally Flat Response for BP Shunt Prototype');
        ylabel('|S12|^2 and |S11|^2 in dB');
        xlabel('f in GHz');
        axis([0 (f1+f2) -80 3]);
        grid on;

% |************************************************************************|
% |                End Plotting Butterworth BP shunt Prototype             |
% |************************************************************************|

    else
        helpdlg('Please check your input parameters. Make sure that f1>=0.3GHz, f2>f1, fs>f2, f2<=300 GHz, LA>LR, LR>0 dB and it is recommended to choose Zo=50 Ohms');
    end

elseif ButterworthBP == 3   % Butterworth BP Series Prototype to plot with its ABCD Parameters
% |************************************************************************|
% |                Butterworth - Series Band-Pass Prototype                |
% |************************************************************************|
    % Generation an input dialog box for Butterworth BP Series Prototype
    prompt={'INSERTION LOSS [dB]','RETURN LOSS [dB]','LOWER SIDE CUTOFF FREQUENCY f1 [GHz]','UPPER SIDE CUTOFF FREQUENCY f2 [GHz]','STOPBAND FREQUENCY AT HIGHER SIDE fs [GHz]','SYSTEM IMPEDANCE Zo [OHMS]'};
    title1='BUTTERWORTH BP SERIES PROTOTYPE';
    numlines=1;
    defaultanswer={'','','','','',''};
    options.Resize='on';
    options.WindowStyle='normal';
    options.Interpreter='tex';
    input_array = inputdlg(prompt,title1,numlines,defaultanswer, options); % Complete input dialog box

    LA = str2num(input_array{1}); % Accepting first element of the input as LA
    LR = str2num(input_array{2}); % Accepting 2nd element of the input as LR
    f1 = str2num(input_array{3}); % Accepting 3rd element of the input as f1
    f2 = str2num(input_array{4});% Accepting 4th element of the input as f2
    fs = str2num(input_array{5});% Accepting 4th element of the input as fs
    Zo = str2num(input_array{6}); % Accepting last element of the input as Zo

    if (LA>LR) && (LR>0) && (f1>=0.3) && (f2>f1) && (f2<fs) && (fs<=300)
        % Cut-off Frequencies W1 and W2; and Stopband frequency at higher side Ws
        wa = 2*pi*f1;
        wb = 2*pi*f2;
        ws = 2*pi*fs;

% Selectivity
        S = (2*ws - wb - wa)/(wb-wa);
        % Displaying the selectivity
        Selectivity=num2str(S); % converting number to sder:tring
        set(handles.edit14_selectivity,'String',Selectivity); % Display the converted string

        % Degree
        Degree = (LA + LR)/(20*log10(S));
        Degree_rounded = round(Degree);
        if Degree_rounded<Degree
            N = Degree_rounded + 1;
        elseif Degree_rounded>=Degree
            N = Degree_rounded;
        else
        end
        N;
        % Displaying the filter order/degree
         filter_order=num2str(N); % converting number to sder:tring
         set(handles.FilterOrder,'String',filter_order); % Display the converted string
```

```matlab
        % Element Values for Maximally Flat BP prototype
       n = N;
       Lar = 3; % Attenuation in dB
       w1  = 1; % Frequency at which Lar is defined as the pass-band edge
       go  = 1; % is assumed to be 1

       for i=1:1:(n+1)

       if (i>=1)&&(i<=n)
       g(i)=2*sin((2*i-1).*pi./(2*n));
       else
       g(n+1) = 1;
       end
       end

%  Plotting with element values for different filter order

% Generatng L and C elements for BP Prototype - Series
wo    = sqrt(wa*wb);
fo = wo/(2*pi);
alpha = wo/(wb-wa);
for i=1:1:N
    if i==1
        Element(1) = g(1)*Zo; % Computing L1 values
        %Transforming series L to series LC circuit
        Inductor(1)= (alpha*Element(1))/wo;
        Capacitor(1)=1/(alpha*wo*Element(1));
    elseif (i>1) && (i<=N)
        k=mod(i,2);
        if k==0
            Element(i) = g(i)/Zo; % Computing C values
            %Transforming Shunt C to Shunt LC circuit
            Inductor(i)= 1/(alpha*wo*Element(i));
            Capacitor(i)= (alpha*Element(i))/wo;
        elseif k==1
            Element(i) = g(i)*Zo; % Computing L values
            %Transforming series L to series LC circuit
            Inductor(i)= (alpha*Element(i))/wo;
            Capacitor(i)=1/(alpha*wo*Element(i));
        else
        end
    else
    end
end
%Scaled reactive element values and scaled source Zs and load ZL impedances, Z = [Zs ZL]
Element;
Inductor(Inductor==0)=[];
Inductor;                      %L remains in nH.
Capacitor(Capacitor==0)=[];
Capacitor2 = 1000.*Capacitor;  %to convert from nF to pF.
Zs = 1*Zo;                     % Source impedance
ZL = g(N+1)*Zo;                % Load impedance in Ohm
Impedance = [Zs ZL];           %Z are in Ohms. in Ohm

% Displaying the inductor, capacitor and impedance values.
Inductor1 = mat2str(Inductor,4);
set(handles.InductorValues1,'String',Inductor1); % Display the converted string
Capacitor1 = mat2str(Capacitor2,4);
set(handles.CapacitorValues1,'String',Capacitor1); % Display the converted string
Impedance1 = mat2str(Impedance,4);
set(handles.ImpedanceValues1,'String',Impedance1); % Display the converted string

f = 0:0.01:30;                 % Range of the frequency in GHz
w = 0:(2*pi*0.01):(2*pi*30);   % Range of w = 2*pi*f.
 % Computing Z, Y and T = [A B;C D] values for different frequency values
 % This is for Butterworth BP Prototype - Series

    for i=1:1:3001
```

```matlab
        for t=1:1:N
            if t==1
                Z(1) = j*Element(1).*((alpha*((w(i)./wo)-(wo./w(i)))));   % Computing Z(L1) values
                T = [1 Z(1);0 1];
            elseif (t>1) && (t<=N)
                    k=mod(t,2);
                if k==0
                    Z(t) = j*Element(t).*((alpha*((w(i)./wo)-(wo./w(i))))); % Computing Zc values
                    Tnew = T*[1 0;Z(t) 1];
                    T = Tnew;
                elseif k==1
                    Z(t) = j*Element(t).*((alpha*((w(i)./wo)-(wo./w(i))))); % Computing Z(L) values
                    Tnew = T*[1 Z(t);0 1];
                    T = Tnew;
                else
                end
            else
            end
        end

    T;

% Getting ABCD values by identification using 2X2 Matrix T=[A B;C D] by using
    A = T(1,1);
    B = T(1,2);
    C = T(2,1);
    D = T(2,2);

 % Computing S12 values for different values ABCD resulting from the
 % variation in w.
    Sa(i) = abs(2./(A+(B./Zo)+(C.*Zo)+D));
    Sb(i) = abs(1 -abs((2./(A+(B./Zo)+(C.*Zo)+D)).^2));
   end

%                                                   4
% Plotting with A,B,C,D parameters, |S12|^2 = -------------------
%                                              |A+(B/Zo)+(C*Zo)+D|^2

    S12 = 20*log10(Sa);   % Computing the dB values of |S12|^2
    S11 = 10*log10(Sb);

    plot(f,S12,'blue', f,S11,'red')
    graph = plot(f,S12,'blue', f,S11,'red');
    % Setting the line width or thickness for both curves
    set (graph, 'LineWidth',1.5);
    title('Maximally Flat Response of BP Series Prototype');
    ylabel('|S12|^2 and |S11|^2 in dB');
    xlabel('f in GHz');
    axis([0 (f1+f2) -80 3]);
    grid on;

% |********************************************************************|
% |                 End Plotting Butterworth BP Series Prototype       |
% |********************************************************************|

    else
        helpdlg('Please check your input parameters. Make sure that f1s>=0.3GHz, f2>f1, fs>f2, f2<=300 GHz, LA>LR, LR>0 dB and it is recommended to choose Zo=50 Ohms');
    end

else
end

    % Butterworth BS radio button is selected
elseif selected_radio8 == 1
    if ButterworthBS == 1     % Butterworth BP from its Generalized Equation

    % Generation an input dialog box for Butterworth BP USING ITS GENERALIZED Equation
```

```matlab
    prompt={'INSERTION LOSS [dB]','RETURN LOSS [dB]','LOWER PASSBAND CUTOFF FREQUENCY f1 [GHz]','UPPER PASSBAND CUTOFF FREQUENCY f2 [GHz]','LOWER STOPBAND CUTOFF FREQUENCY fs1 [GHz]','UPPER STOPBAND CUTOFF FREQUENCY fs2 [GHz]'};
    title1='BUTTERWORTH BP GENERALIZED EQUATION';
    numlines=1;
    defaultanswer={'','','','','',''};
    options.Resize='on';
    options.WindowStyle='normal';
    options.Interpreter='tex';
    input_array = inputdlg(prompt,title1,numlines,defaultanswer, options); % Complete input dialog box
    
    LA = str2num(input_array{1}); % Accepting first element of the input as LA
    LR = str2num(input_array{2}); % Accepting 2nd element of the input as LR
    f1 = str2num(input_array{3}); % Accepting 3rd element of the input as f1
    f2 = str2num(input_array{4});% Accepting 4th element of the input as f2
    fs1= str2num(input_array{5});% Accepting 4th element of the input as fs1
    fs2 = str2num(input_array{6});% Accepting 4th element of the input as fs2
    
    
if (LA>LR) && (LR>0) && (f1>=0.3) && (fs1>f1) && (fs2>fs1) && (f2>fs2) && (f2<=300)
    
    % Cut-off Frequencies W1 and W2; and Stopband frequency at higher side Ws
    wa = 2*pi*f1;
    wb = 2*pi*f2;
    ws1 = 2*pi*fs1;
    ws2 = 2*pi*fs2;
    
% Selectivity
 S = (wb-wa)/(ws2-ws1);
% Displaying the selectivity
Selectivity=num2str(S); % converting number to sder:tring
set(handles.edit14_selectivity,'String',Selectivity); % Display the converted string

    % Degree
Degree = (LA + LR)/(20*log10(S));
Degree_rounded = round(Degree);
if Degree_rounded<Degree
    N = Degree_rounded + 1;
elseif Degree_rounded>=Degree
    N = Degree_rounded;
else
end
N;
% Displaying the filter order/degree
filter_order=num2str(N); % converting number to sder:tring
set(handles.FilterOrder,'String',filter_order); % Display the converted string
% Maxi Flat BS

    f = 0:0.01:30;              % Range of the frequency in GHz
    w = 0:(2*pi*0.01):(2*pi*30);   % Range of w = 2*pi*f.
    %BS
    wo    = sqrt(wa*wb);
    fo    = wo/(2*pi);
    alpha = wo/(wb-wa);
    
    S12 = 10*log10(1./(1+(-1./(alpha*((w./wo)-(wo./w)))).^(2*N)));
    S11 = 10*log10(1 - abs(1./(1+(-1./(alpha*((w./wo)-(wo./w)))).^(2*N))));
    
    plot(f,S12,'blue', f,S11,'red')
    graph = plot(f,S12,'blue', f,S11,'red');
    % Setting the line width or thickness for both curves
    set (graph, 'LineWidth',1.5);
    title('Maximally Flat Response BS');
    ylabel('|S12|^2 and |S11|^2 in dB');
    xlabel('f in GHz');
    axis([0 (f1+fs2) -80 5]);
    grid on;

else
```

```matlab
        helpdlg('Please check your input parameters. Make sure that f1>=0.3GHz, fs1>f1, fs2>fs1, f2>fs, f2<=300 GHz, LA>LR and LR>0 dB.');
    end

% |===================================================================|
% |                                                                   |
% | Plotting Butterworth BS Prototype using A,B,C,D parameters:       |
% |                              4                                    |
% |            S12|^2 = -------------------                           |
% |                        |A+(B/Zo)+(C*Zo)+D|^2                      |
% |===================================================================|
elseif ButterworthBS == 2   % Butterworth BS Shunt Prototype to plot with its ABCD Parameters
% |*******************************************************************|
% |                 Plotting Butterworth BS Shunt Prototype           |
% |*******************************************************************|
    % Generation an input dialog box for Butterworth BP Shunt prototype
    prompt={'INSERTION LOSS [dB]','RETURN LOSS [dB]','LOWER PASSBAND CUTOFF FREQUENCY f1 [GHz]','UPPER PASSBAND CUTOFF FREQUENCY f2 [GHz]','LOWER STOPBAND CUTOFF FREQUENCY fs1 [GHz]','UPPER STOPBAND CUTOFF FREQUENCY fs2 [GHz]','SYSTEM IMPEDANCE Zo [OHMS]'};
    title1='BUTWERWORTH BP SHUNT PROTOTYPE';
    numlines=1;
    defaultanswer={'','','','','','',''};
    options.Resize='on';
    options.WindowStyle='normal';
    options.Interpreter='tex';
    input_array = inputdlg(prompt,title1,numlines,defaultanswer, options); % Complete input dialog box

    LA = str2num(input_array{1}); % Accepting first element of the input as LA
    LR = str2num(input_array{2}); % Accepting 2nd element of the input as LR
    f1 = str2num(input_array{3}); % Accepting 3rd element of the input as f1
    f2 = str2num(input_array{4});% Accepting 4th element of the input as f2
    fs1= str2num(input_array{5});% Accepting 4th element of the input as fs1
    fs2 = str2num(input_array{6});% Accepting 4th element of the input as fs2
    Zo = str2num(input_array{7}); % Accepting last element of the input as Zo

    if (LA>LR) && (LR>0) && (f1>=0.3) && (fs1>f1) && (fs2>fs1) && (f2>fs2) && (f2<=300)

        % Cut-off Frequencies W1 and W2; and Stopband frequency at higher side Ws
        wa = 2*pi*f1;
        wb = 2*pi*f2;
        ws1 = 2*pi*fs1;
        ws2 = 2*pi*fs2;

    % Selectivity
    S = (wb-wa)/(ws2-ws1);
    % Displaying the selectivity
    Selectivity=num2str(S); % converting number to sder:tring
    set(handles.edit14_selectivity,'String',Selectivity); % Display the converted string

        % Degree
    Degree = (LA + LR)/(20*log10(S));
    Degree_rounded = round(Degree);
    if Degree_rounded<Degree
        N = Degree_rounded + 1;
    elseif Degree_rounded>=Degree
        N = Degree_rounded;
    else
    end
    N;
    % Displaying the filter order/degree
    filter_order=num2str(N); % converting number to sder:tring
    set(handles.FilterOrder,'String',filter_order); % Display the converted string

    % Element Values for Maximally Flat BS prototype
    n = N;
    Lar = 3; % Attenuation in dB
    w1  = 1; % Frequency at which Lar is defined as the pass-band edge
```

```matlab
        go  = 1; % is assumed to be 1

        for i=1:1:(n+1)

        if (i>=1)&&(i<=n)
        g(i)=2*sin((2*i-1).*pi./(2*n));
        else
        g(n+1) = 1;
        end
        end

% Plotting with element values for different filter order

% Generatng L and C elements for BS Prototype - Shunt
wo    = sqrt(wa*wb);
fo = wo/(2*pi);
alpha = wo/(wb-wa);

for i=1:1:N
    if i==1
        % Computing C(1) value
        Element(1) = g(1)/Zo;
        %Transforming shunt C to series LC circuit
        Inductor(1)= alpha/(wo*Element(1));
        Capacitor(1)=(Element(1))/(alpha*wo);
    elseif (i>1) && (i<=N)
        k=mod(i,2);
        if k==0
            % Computing L values for odd values of i
            Element(i) = g(i)*Zo;
            %Transforming series L to shunt LC circuit
            Inductor(i)= (Element(i))/(alpha*wo);
            Capacitor(i)=alpha/(wo*Element(i));
        elseif k==1
            % Computing C values for even values of i
            Element(i) = g(i)/Zo;
            %Transforming shunt C to series LC circuit
            Inductor(i)= alpha/(wo*Element(i));
            Capacitor(i)=(Element(i))/(alpha*wo);
        else
        end
    else
    end
end
%Scaled reactive element values and scaled source Zs and load ZL impedances, Z = [Zs ZL]
Element;
Inductor(Inductor==0)=[];
Inductor;                          %L remains in nH.
Capacitor(Capacitor==0)=[];
Capacitor2 = 1000.*Capacitor;   %to convert from nF to pF.
Zs = 1*Zo;                      % Source impedance go=1
ZL = g(N+1)*Zo;                 % Load impedance in Ohm
Impedance = [Zs ZL];            %Z are in Ohms. in Ohm

% Displaying the inductor, capacitor and impedance values.
Inductor1 = mat2str(Inductor,4);
set(handles.InductorValues1,'String',Inductor1); % Display the converted string
Capacitor1 = mat2str(Capacitor2,4);
set(handles.CapacitorValues1,'String',Capacitor1); % Display the converted string
Impedance1 = mat2str(Impedance,4);
set(handles.ImpedanceValues1,'String',Impedance1); % Display the converted string

f = 0:0.01:30;                  % Range of the frequency in GHz
w = 0:(2*pi*0.01):(2*pi*30);    % Range of w = 2*pi*f.

 % Computing Z, Y and T = [A B;C D] values for different frequency values
 % This is for Butterworth BS Prototype - Series

    for i=1:1:3001
```

```matlab
        for t=1:1:N
            if t==1
                % Computing Z(C1) values
                Z(1) = j*Element(1).*(-1./(alpha*((w(i)./wo)-(wo./w(i)))));
                T = [1 0;Z(1) 1];
            elseif (t>1) && (t<=N)
                    k=mod(t,2);
                if k==0
                    % Computing Z(L) values
                    Z(t) = j*Element(t).*(-1./(alpha*((w(i)./wo)-(wo./w(i)))));
                    Tnew = T*[1 Z(t);0 1];
                    T = Tnew;
                elseif k==1
                    % Computing Z(C) values
                    Z(t) = j*Element(t).*(-1./(alpha*((w(i)./wo)-(wo./w(i)))));
                    Tnew = T*[1 0;Z(t) 1];
                    T = Tnew;
                else
                end
            else
            end
        end

    T;

% Getting ABCD values by identification using 2X2 Matrix T=[A B;C D] by using
    A = T(1,1);
    B = T(1,2);
    C = T(2,1);
    D = T(2,2);

 % Computing S12 values for different values ABCD resulting from the
 % variation in w.
    Sa(i) = abs(2./(A+(B./Zo)+(C.*Zo)+D));
    Sb(i) = abs(1 - abs((2./(A+(B./Zo)+(C.*Zo)+D)).^2));
   end
%                                                  4
% Plotting with A,B,C,D parameters, |S12|^2 = -------------------
%                                              |A+(B/Zo)+(C*Zo)+D|^2

    S12 = 20*log10(Sa);  % Computing the dB values of |S12|^2
    S11 = 10*log10(Sb);

    plot(f,S12,'blue', f,S11,'red')
    graph = plot(f,S12,'blue', f,S11,'red');
    % Setting the line width or thickness for both curves
    set (graph, 'LineWidth',1.5);
    title('Maximally Flat Response for BS Shunt Prototype');
    ylabel('|S12|^2 and |S11|^2 in dB');
    xlabel('f in GHz');
    axis([0 (f1+fs2) -80 3]);
    grid on;

    else
        helpdlg('Please check your input parameters. Make sure that f1>=0.3GHz, fs1>f1, fs2>fs1, f2>fs, f2<=300 GHz, LA>LR, LR>0 dB and it is recommended to choose Zo=50 Ohms');
    end

elseif ButterworthBS == 3   % Butterworth BS Series Prototype to plot with its ABCD Parameters
% |********************************************************************|
% |                 Butterworth - Series Band-Stop Prototype             |
% |********************************************************************|
    % Generation an input dialog box for Butterworth BP Series
    prompt={'INSERTION LOSS [dB]','RETURN LOSS [dB]','LOWER PASSBAND CUTOFF FREQUENCY f1 [GHz]','UPPER PASSBAND CUTOFF FREQUENCY f2 [GHz]','LOWER STOPBAND CUTOFF FREQUENCY fs1 [GHz]','UPPER STOPBAND CUTOFF FREQUENCY fs2 [GHz]','SYSTEM IMPEDANCE Zo [OHMS]'};
    title1='BUTTWERWORTH BS SERIES PROTOTYPE';
    numlines=1;
    defaultanswer={'','','','','','',''};
```

```matlab
    options.Resize='on';
    options.WindowStyle='normal';
    options.Interpreter='tex';
    input_array = inputdlg(prompt,title1,numlines,defaultanswer, options); % Complete input dialog box

    LA = str2num(input_array{1}); % Accepting first element of the input as LA
    LR = str2num(input_array{2}); % Accepting 2nd element of the input as LR
    f1 = str2num(input_array{3}); % Accepting 3rd element of the input as f1
    f2 = str2num(input_array{4});% Accepting 4th element of the input as f2
    fs1= str2num(input_array{5});% Accepting 4th element of the input as fs1
    fs2 = str2num(input_array{6});% Accepting 4th element of the input as fs2
    Zo = str2num(input_array{7}); % Accepting last element of the input as Zo

if (LA>LR) && (LR>0) && (f1>=0.3) && (fs1>f1) && (fs2>fs1) && (f2>fs2) && (f2<=300)

    % Cut-off Frequencies W1 and W2; and Stopband frequency at higher side Ws
    wa = 2*pi*f1;
    wb = 2*pi*f2;
    ws1 = 2*pi*fs1;
    ws2 = 2*pi*fs2;

% Selectivity
S = (wb-wa)/(ws2-ws1);
% Displaying the selectivity
Selectivity=num2str(S); % converting number to sder:tring
set(handles.edit14_selectivity,'String',Selectivity); % Display the converted string

    % Degree
Degree = (LA + LR)/(20*log10(S));
Degree_rounded = round(Degree);
if Degree_rounded<Degree
    N = Degree_rounded + 1;
elseif Degree_rounded>=Degree
    N = Degree_rounded;
else
end
N;
% Displaying the filter order/degree
filter_order=num2str(N); % converting number to sder:tring
set(handles.FilterOrder,'String',filter_order); % Display the converted string

% Element Values for Maximally Flat BS prototype
n = N;
Lar = 3; % Attenuation in dB
w1  = 1; % Frequency at which Lar is defined as the pass-band edge
go  = 1; % is assumed to be 1

for i=1:1:(n+1)

if (i>=1)&&(i<=n)
g(i)=2*sin((2*i-1).*pi./(2*n));
else
g(n+1) = 1;
end
end

%  Plotting with element values for different filter order

% Generatng L and C elements for BS Prototype - Series
wo    = sqrt(wa*wb);
fo = wo/(2*pi);
alpha = wo/(wb-wa);

for i=1:1:N
    if i==1
        Element(1) = g(1)*Zo; % Computing L1 values
        %Transforming series L to shunt LC circuit
        Inductor(1)= (Element(1))/(alpha*wo);
```

```matlab
            Capacitor(1)=alpha/(wo*Element(1));
        elseif (i>1) && (i<=N)
            k=mod(i,2);
            if k==0
                Element(i) = g(i)/Zo; % Computing C values
                %Transforming shunt C to series LC circuit
                Inductor(i)= alpha/(wo*Element(i));
                Capacitor(i)=(Element(i))/(alpha*wo);
            elseif k==1
                Element(i) = g(i)*Zo; % Computing L values
                %Transforming series L to shunt LC circuit
                Inductor(i)= (Element(i))/(alpha*wo);
                Capacitor(i)=alpha/(wo*Element(i));
            else
            end
        else
        end
end
%Scaled reactive element values and scaled source Zs and load ZL impedances, Z = [Zs ZL)
Element;
Inductor(Inductor==0)=[];
Inductor;                      %L remains in nH.
Capacitor(Capacitor==0)=[];
Capacitor2 = 1000.*Capacitor;  %to convert from nF to pF.
Zs = 1*Zo;                     % Source impedance
ZL = g(N+1)*Zo;                % Load impedance in Ohm
Impedance = [Zs ZL];           %Z are in Ohms. in Ohm

% Displaying the inductor, capacitor and impedance values.
Inductor1 = mat2str(Inductor,4);
set(handles.InductorValues1,'String',Inductor1); % Display the converted string
Capacitor1 = mat2str(Capacitor2,4);
set(handles.CapacitorValues1,'String',Capacitor1); % Display the converted string
Impedance1 = mat2str(Impedance,4);
set(handles.ImpedanceValues1,'String',Impedance1); % Display the converted string

f = 0:0.01:30;                 % Range of the frequency in GHz
w = 0:(2*pi*0.01):(2*pi*30);   % Range of w = 2*pi*f.
% Computing Z, Y and T = [A B;C D] values for different frequency values
% This is for Butterworth BS Prototype - Series

    for i=1:1:3001
        for t=1:1:N
            if t==1
                Z(1) = j*Element(1).*(-1./(alpha*((w(i)./wo)-(wo./w(i)))));  % Computing Z(L1) values
                T = [1 Z(1);0 1];
            elseif (t>1) && (t<=N)
                    k=mod(t,2);
                if k==0
                    Z(t) = j*Element(t).*(-1./(alpha*((w(i)./wo)-(wo./w(i))))); % Computing Zc values
                    Tnew = T*[1 0;Z(t) 1];
                    T = Tnew;
                elseif k==1
                    Z(t) = j*Element(t).*(-1./(alpha*((w(i)./wo)-(wo./w(i))))); % Computing Z(L) values
                    Tnew = T*[1 Z(t);0 1];
                    T = Tnew;
                else
                end
            else
            end
        end

    T;
% Getting ABCD values by identification using 2X2 Matrix T=[A B;C D] by using
    A = T(1,1);
    B = T(1,2);
    C = T(2,1);
    D = T(2,2);
```

```matlab
 % Computing S12 values for different values ABCD resulting from the
 % variation in w.
    Sa(i) = abs(2./(A+(B./Zo)+(C.*Zo)+D));
    Sb(i) = abs(1 - abs((2./(A+(B./Zo)+(C.*Zo)+D)).^2));
   end

%                                          4
% Plotting with A,B,C,D parameters, |S12|^2 = -------------------
%                                          |A+(B/Zo)+(C*Zo)+D|^2

    S12 = 20*log10(Sa);   % Computing the dB values of |S12|^2
    S11 = 10*log10(Sb);

    plot(f,S12,'blue', f,S11,'red')
    graph = plot(f,S12,'blue', f,S11,'red');
    % Setting the line width or thickness for both curves
    set (graph, 'LineWidth',1.5);
    title('Maximally Flat Response of BS Series Prototype');
    ylabel('|S12|^2 and |S11|^2 in dB');
    xlabel('f in GHz');
    axis([0 (f1+fs2) -80 3]);
    grid on;

    else
        helpdlg('Please check your input parameters. Make sure that f1>=0.3GHz, fs1>f1, fs2>fs1, f2>fs, f2<=300 GHz, LA>LR, LR>0 dB and it is recommended to choose Zo=50 Ohms');
    end

    else
    end
else
    helpdlg('Please select filter type and class from one of the radio buttons at the top left corner (under Chebyshev and Butterworth Filters) in order to proceed.');
end

function ImpedanceValues1_Callback(hObject, eventdata, handles)
% hObject    handle to ImpedanceValues1 (see GCBO)
% eventdata  reserved - to be defined in a future version of MATLAB
% handles    structure with handles and user data (see GUIDATA)
% Hints: get(hObject,'String') returns contents of ImpedanceValues1 as text
% str2double(get(hObject,'String')) returns contents of ImpedanceValues1 as a double

% --- Executes during object creation, after setting all properties.
function ImpedanceValues1_CreateFcn(hObject, eventdata, handles)
if ispc && isequal(get(hObject,'BackgroundColor'), get(0,'defaultUicontrolBackgroundColor'))
    set(hObject,'BackgroundColor','white');
end

% |===================================================================|
% |===================================================================|
% |                              END                                  |
% |===================================================================|
% |===================================================================|
```